\documentclass{article}

\usepackage[dvipdfmx]{graphicx}
\usepackage{amsmath}        
\usepackage{amssymb}        
\usepackage{cite}
\usepackage[dvips]{color}
\usepackage{fancyhdr}

\def\erase#1{{}}
\def\EqArrerase#1{{}}

\makeatletter
 \renewcommand{\theequation}{%
 \thesection.\arabic{equation}}
 \@addtoreset{equation}{section}
\makeatother

\def\T{{\rm T}}

\def\GL{{G\kern-.12em L\kern-.04em}}
\def\OSp{{O\kern-.11em S\kern-.04em p}}
\def\IOSp{{I\kern-.06em O\kern-.11em S\kern-.04em p}}
\def\MN{{M\kern-.14em N}}
\def\NM{{N\kern-.14em M}}
\def\NL{{N\kern-.14em L}}
\def\LN{{L\kern-.11em N}}
\def\ML{{M\kern-.14em L}}
\def\LM{{L\kern-.11em M}}
\def\RN{{R\kern-.11em N}}
\def\NR{{N\kern-.14em R}}
\def\RM{{R\kern-.11em M}}
\def\MR{{M\kern-.14em R}}
\def\RL{{R\kern-.11em L}}
\def\LR{{L\kern-.11em R}}
\def\RS{{R\kern-.11em S}}
\def\SR{{S\kern-.11em R}}
\def\SN{{S\kern-.11em N}}
\def\NS{{N\kern-.11em S}}
\def\SM{{S\kern-.11em M}}
\def\MS{{M\kern-.11em S}}
\def\SL{{S\kern-.11em L}}
\def\LS{{L\kern-.11em S}}
\def\sqr#1#2{{\vcenter{\hrule height.#2pt
      \hbox{\vrule width.#2pt height#1pt \kern#1pt
          \vrule width.#2pt}
      \hrule height.#2pt}}}
\def\bra0{\langle0|}
\def\ket0{|0\rangle}
\def\soeji#1_#2#3{#1_{#2}\cdots#1_{#3}}
\def\longgLRarrow{\longleftarrow\kern-3pt\relbar\kern-3pt\relbar\kern-3pt%
\longrightarrow}
\def\longLRarrow{\longleftarrow\kern-3pt\relbar\kern-3pt\longrightarrow}
\def\longLarrow{\longleftarrow\kern-3pt\relbar\kern-3pt\relbar\kern-3pt\relbar}
\def\longRarrow{\relbar\kern-3pt\relbar\kern-3pt\relbar\kern-3pt\longrightarrow}
\def\bothDer#1#2#3{%
\overset{\kern-.7em\stackrel{#1}{#2}}{\partial_{#3}}}
 
\makeatletter
 \renewcommand{\theequation}{%
 \thesection.\arabic{equation}}
 \@addtoreset{equation}{section}
\makeatother
\begin{document}
\thispagestyle{fancy}

\title{Quantum Theory of Weyl Invariant Scalar-tensor Gravity}

\author{Ichiro Oda
\footnote{Electronic address: ioda@sci.u-ryukyu.ac.jp}
\\
{\it\small
\begin{tabular}{c}
Department of Physics, Faculty of Science, University of the 
           Ryukyus,\\
           Nishihara, Okinawa 903-0213, Japan\\      
\end{tabular}
}
}
\date{}

\maketitle

\thispagestyle{fancy}

\begin{abstract}

We perform a manifestly covariant quantization of a Weyl invariant, i.e., a locally scale invariant, scalar-tensor
gravity in the extended de Donder gauge condition (or harmonic gauge condition) for general coordinate 
invariance and a new scalar gauge for Weyl invariance within the framework of BRST formalism. It is shown 
that choral symmetry, which is a Poincar${\rm{\acute{e}}}$-like $\IOSp(8|8)$ supersymmetry in case of Einstein 
gravity, is extended to a Poincar${\rm{\acute{e}}}$-like $\IOSp(10|10)$ supersymmetry. We point out that 
there is a gravitational conformal symmetry in quantum gravity and account for how conventional conformal symmetry 
in a flat Minkowski space-time is related to the gravitational conformal symmetry. Moreover, we examine 
the mechanism of spontaneous symmetry breakdown of the choral symmetry, and show that the gravitational
conformal symmetry is spontaneously broken to the Poincar\'e symmetry and the corresponding massless 
Nambu-Goldstone bosons are the graviton and the dilaton. We also prove the unitarity of the physical S-matrix 
on the basis of the BRST quartet mechanism. 

\end{abstract}

\section{Introduction}

There is no question that symmetry plays the central role in both elementary-particle physics 
and quantum gravity. For instance, in the Yang-Mills theory it has been found that we have 
the non-abelian gauge symmetry and that this symmetry gives rise to physically significant effects, 
such as the asymptotic freedom and the quark confinement. 

It is well known that there are two kinds of symmetries in nature: global symmetry and gauge
symmetry. In order to understand the nature more deeply, it is necessary to understand the 
meaning of the both symmetries.  The meaning of the global symmetry is clear in the sense that
it operates physical observables in a direct manner and shows the real symmetry of a physical
system. On the other hand, the meaning of the gauge symmetry is more elusive than that of the
global symmetry since it does not operate on physical observables directly. To treat the gauge
symmetry properly in quantum field theory, it is essential to fix the gauge symmetry by a suitable
gauge condition and consequently physical observables are defined as BRST invariant operators.
Thus, it is sometimes said that the gauge symmetry is redundancy in the mathematical description 
of a physical system rather than the property of the system itself.

Another important property of the symmetries is that many of the global symmetry are not exact 
but only approximate whereas the gauge symmetry is an exact one. For instance, there is 
a clear prediction of violation of baryon and lepton numbers by a quantum anomaly in the
standard model. This fact is also supported by a theory of quantum gravity. In particular,
when a black hole evaporates at the quantum level, the baryon and lepton numbers are
not conserved whereas gauge quantum numbers such as electric and magnetic charges are
precisely conserved since they are measured by the flux integrals at infinity.  

Thus, if a certain global symmetry plays a critical role in physics, it must be promoted to 
the gauge symmetry. This statement particularly holds in constructing theories involving
quantum gravity.  In our previous work \cite{Oda-Q}, we have presented a quantum theory of a globally scale 
invariant gravity with a real scalar field, which is equivalent to the well-known Brans-Dicke gravity \cite{Brans}, 
by constructing its manifestly covariant BRST formalism.  Since many of studies of the Brans-Dicke 
gravity have been so far limited to a classical analysis, our theory has provided us with some useful 
information on quantum aspects of the Brans-Dicke gravity. Indeed, based on this quantum gravity 
we have elucidated a mechanism of how a scale invariance is spontaneously broken 
and consequently exactly massless ``dilaton'' emerges thanks to the Nambu-Goldstone theorem 
in quantum gravity \cite{Oda-Q, Oda-SI}. Then, it is natural to generalize our formulation to a case of 
a locally scale invariant, or equivalently Weyl invariant, scalar-tensor gravity and ask if we can get some 
useful knowledge about quantum aspects of the theory.   

In this article, we perform a manifestly covariant BRST quantization of a Weyl invariant scalar-tensor
gravity with a real scalar field in addition to the metric tensor field, investigate the remaining global
symmetries and their spontaneous symmetry breakdown, prove the unitarity of the S-matrix, and then
elucidate that there exists a gravitational analog of conformal symmetry in our theory.     
Long ago, in a pioneering work by Nakanishi \cite{Nakanishi, N-O-text}, on the basis of the Einstein-Hilbert 
action in the de Donder gauge (harmonic gauge) for general coordinate transformation (GCT), it has been 
shown that there remains a huge residual symmetry, which is a Poincar\'e-like $ISOp(8|8)$ 
supersymmetry, called ``choral symmetry'', including the BRST symmetry and $GL(4)$ symmetry etc. 
In our present formulation, adopting the extended de Donder gauge condition for the GCT and a new scalar gauge
condition for the Weyl transformation, the choral symmetry is extended to a Poincar\'e-like 
$ISOp(10|10)$ supersymmetry, which includes the scale symmetry and the gravitational special conformal
symmetry. It is of interest that as in a flat Minkowski space-time, both the scale symmetry and the special
conformal symmetry are spontaneously broken, and the dilation is not only a Nambu-Goldstone boson for
the scale symmetry but also its derivative provides a Nambu-Goldstone boson for the special conformal transformation.  

The paper is organized as follows. In Section 2, we discuss a general gravitational theory 
for which there are two local symmetries, those are, the general coordinate invariance and the Weyl symmetry. 
It is pointed out that in such a theory, we must choose a gauge fixing condition for GCT carefully in such a way 
that it does not violate the Weyl symmetry, and similarly a gauge fixing condition for the Weyl transformation 
should be selected in order not to break the GCT.  In Section 3, beginning with a Weyl invariant scalar-tensor 
gravity \cite{Fujii}, we fix the GCT and the Weyl transformation by the extended de Donder gauge and
the new scalar gauge conditions, and construct a gauge-fixed, BRST-invariant quantum Lagrangian.
In Section 4, we calculate various equal-time (anti-)commutation relations (ETCRs) among the fundamental
fields, in particular, the Nakanishi-Lautrup auxiliary field, the Faddeev-Popov (FP) ghosts. In Section 5, we derive 
the ETCRs involving the gravitational field. In Section 6, we prove the unitarity of the physical S-matrix
by means of the BRST quartet mechanism. In Section 7, we show that there is a choral symmetry, which is an
$IOSp(10|10)$ supersymmetry, in our theory. In Section 8, we point out the existence of a gravitational
conformal symmetry even in quantum gravity, and we investigate its spontaneous symmetry breaking 
in Section 9. The final section is devoted to discussion. 

Two appendices are put for technical details. In Appendix A, a derivation of the equation for the $b_\rho$ field is given, 
and in Appendix B we have accounted for the relationship between the gravitational conformal symmetry and 
conventional conformal symmetry.

\section{Consistency between two BRST symmetries}

We wish to perform a manifestly covariant BRST quantization of a gravitational theory which
is invariant under both general coordinate transformation (GCT) and Weyl transformation, or equivalently
local scale transformation. To take a more general theory into consideration, without specifying the concrete 
expression of the gravitational Lagrangian density, we will start with the following classical 
Lagrangian density:\footnote{We follow the notation and conventions of MTW textbook \cite{MTW}. Greek 
little letters $\mu, \nu, \cdots$ and Latin ones $i, j, \cdots$ are used for space-time and spatial indices, 
respectively; for instance, $\mu= 0, 1, 2, 3$ and $i = 1, 2, 3$. The Riemann curvature tensor 
and the Ricci tensor are respectively defined by $R^\rho{}_{\sigma\mu\nu} = \partial_\mu \Gamma^\rho_{\sigma\nu} 
- \partial_\nu \Gamma^\rho_{\sigma\mu} + \Gamma^\rho_{\lambda\mu} \Gamma^\lambda_{\sigma\nu} 
- \Gamma^\rho_{\lambda\nu} \Gamma^\lambda_{\sigma\mu}$ and $R_{\mu\nu} = R^\rho{}_{\mu\rho\nu}$. 
The Minkowski metric tensor is denoted by $\eta_{\mu\nu}$; $\eta_{00} = - \eta_{11} = - \eta_{22} 
= - \eta_{33} = -1$ and $\eta_{\mu\nu} = 0$ for $\mu \neq \nu$.} 
%**   Lc   %%%%%%%%%%%%%%%%%%%%%%%%%%%%%%%%%%%%%%%%%%%%%%%%%%%%%%%%%
\begin{eqnarray}
{\cal L}_c = {\cal L}_c ( g_{\mu\nu}, \phi ),
\label{Lc}  
\end{eqnarray}
%%%%%%%%%%%%%%%%%%%%%%%%%%%%%%%%%%%%%%%%%%%%%%%%%%%%%%%%%%%%%%%%%%% 
which includes the metric tensor field $g_{\mu\nu}$ and a scalar field $\phi$ as dynamical variables.\footnote{It is 
straightforward to add the other fields such as gauge fields and spinors.} 
We assume that ${\cal{L}}_c$ does not involve more than first order derivatives of the metric and matter fields.

We have a physical situation in mind that we fix the general coordinate symmetry and the Weyl symmetry by suitable 
gauge conditions. It is a familiar fact that after introducing the gauge conditions, instead of such the two local gauge 
symmetries, we are left with two kinds of global symmetries named as the BRST symmetries.  The BRST transformation, 
which is denoted as $\delta_B$, corresponding to the GCT is defined as
%**   GCT-BRST   %%%%%%%%%%%%%%%%%%%%%%%%%%%%%%%%%%%%%%%%%%%%%%%%%%%%%%%%%
\begin{eqnarray}
\delta_B g_{\mu\nu} &=& - ( \nabla_\mu c_\nu+ \nabla_\nu c_\mu)
\nonumber\\
&=& - ( c^\alpha\partial_\alpha g_{\mu\nu} + \partial_\mu c^\alpha g_{\alpha\nu} 
+ \partial_\nu c^\alpha g_{\mu\alpha} ),
\nonumber\\
\delta_B \tilde g^{\mu\nu} &=& h ( \nabla^\mu c^\nu+ \nabla^\nu c^\mu 
- g^{\mu\nu} \nabla_\rho c^\rho),
\nonumber\\
\delta_B \phi &=& - c^\lambda \partial_\lambda \phi, \quad 
\delta_B c^\rho = - c^\lambda\partial_\lambda c^\rho, 
\nonumber\\
\delta_B \bar c_\rho &=& i B_\rho, \quad 
\delta_B B_\rho = 0, 
\label{GCT-BRST}  
\end{eqnarray}
%%%%%%%%%%%%%%%%%%%%%%%%%%%%%%%%%%%%%%%%%%%%%%%%%%%%%%%%%%%%%%%%%%% 
where $c^\rho$ and $\bar c_\rho$ are respectively the Faddeev-Popov (FP) ghost and anti-ghost, 
$B_\rho$ is the Nakanishi-Lautrup (NL) field, and we have defined $\tilde g^{\mu\nu} \equiv \sqrt{-g} 
g^{\mu\nu} \equiv h g^{\mu\nu}$.  For later convenience, in place of the NL field $B_\rho$ we will 
introduce a new NL field defined as
%**   b-rho-field   %%%%%%%%%%%%%%%%%%%%%%%%%%%%%%%%%%%%%%%%%%%%%%%%%%%%%%%%%
\begin{eqnarray}
b_\rho= B_\rho- i c^\lambda\partial_\lambda\bar c_\rho,
\label{b-rho-field}  
\end{eqnarray}
%%%%%%%%%%%%%%%%%%%%%%%%%%%%%%%%%%%%%%%%%%%%%%%%%%%%%%%%%%%%%%%%%%% 
and its BRST transformation reads
%**   b-BRST   %%%%%%%%%%%%%%%%%%%%%%%%%%%%%%%%%%%%%%%%%%%%%%%%%%%%%%%%%
\begin{eqnarray}
\delta_B b_\rho= - c^\lambda\partial_\lambda b_\rho.
\label{b-BRST}  
\end{eqnarray}
%%%%%%%%%%%%%%%%%%%%%%%%%%%%%%%%%%%%%%%%%%%%%%%%%%%%%%%%%%%%%%%%%%%
The other BRST transformation, which is denoted as $\bar \delta_B$, 
corresponding to the Weyl transformation is defined as
%**   Weyl-BRST   %%%%%%%%%%%%%%%%%%%%%%%%%%%%%%%%%%%%%%%%%%%%%%%%%%%%%%%%%
\begin{eqnarray}
\bar \delta_B g_{\mu\nu} &=& 2 c g_{\mu\nu}, \quad
\bar \delta_B \tilde g^{\mu\nu} = 2 c \tilde g_{\mu\nu},
\nonumber\\
\bar \delta_B \phi &=& - c \phi, \quad 
\bar \delta_B \bar c = i B, \quad 
\bar \delta_B c = \bar \delta_B B = 0, 
\label{Weyl-BRST}  
\end{eqnarray}
%%%%%%%%%%%%%%%%%%%%%%%%%%%%%%%%%%%%%%%%%%%%%%%%%%%%%%%%%%%%%%%%%%% 
where $c$ and $\bar c$ are respectively the FP ghost and FP anti-ghost, 
$B$ is the NL field. Note that the two BRST transformations are nilpotent, i.e.,
%**   Nilpotent   %%%%%%%%%%%%%%%%%%%%%%%%%%%%%%%%%%%%%%%%%%%%%%%%%%%%%%%%%
\begin{eqnarray}
\delta_B^2 = \bar \delta_B^2 = 0.   
\label{Nilpotent}  
\end{eqnarray}
%%%%%%%%%%%%%%%%%%%%%%%%%%%%%%%%%%%%%%%%%%%%%%%%%%%%%%%%%%%%%%%%%%%  

To complete the two BRST transformations, we have to fix not only the GCT BRST transformation
$\delta_B$ on $c, \bar c$ and $B$ but also the Weyl BRST transformation $\delta_B$ on
$c^\rho, \bar c_\rho$ and $b_\rho$. It is easy to determine the former BRST transformation since the
fields $c, \bar c$ and $B$ are all scalar fields so their BRST transformations should take the form:
%**   GCT-BRST2   %%%%%%%%%%%%%%%%%%%%%%%%%%%%%%%%%%%%%%%%%%%%%%%%%%%%%%%%%
\begin{eqnarray}
\delta_B B = - c^\lambda\partial_\lambda B, \quad
\delta_B c = - c^\lambda\partial_\lambda c, \quad
\delta_B \bar c = - c^\lambda\partial_\lambda \bar c.   
\label{GCT-BRST2}  
\end{eqnarray}
%%%%%%%%%%%%%%%%%%%%%%%%%%%%%%%%%%%%%%%%%%%%%%%%%%%%%%%%%%%%%%%%%%%  
On the other hand, there is an ambiguity in fixing the latter BRST transformation, but we would like to
propose a recipe for achieving this goal. The recipe is to just assume that the two BRST transformations 
are anti-commute with each other, that is,   
%**   GCT-Weyl-BRST   %%%%%%%%%%%%%%%%%%%%%%%%%%%%%%%%%%%%%%%%%%%%%%%%%%%%%%%%%
\begin{eqnarray}
\{ \delta_B, \bar \delta_B \} \equiv \delta_B \bar \delta_B + \bar \delta_B \delta_B = 0,
\label{GCT-Weyl-BRST}  
\end{eqnarray}
%%%%%%%%%%%%%%%%%%%%%%%%%%%%%%%%%%%%%%%%%%%%%%%%%%%%%%%%%%%%%%%%%%%  
which requires us to take 
%**   Weyl-BRST2   %%%%%%%%%%%%%%%%%%%%%%%%%%%%%%%%%%%%%%%%%%%%%%%%%%%%%%%%%
\begin{eqnarray}
\bar \delta_B b_\rho = \bar \delta_B c^\rho = \bar \delta_B \bar c_\rho = 0.   
\label{Weyl-BRST2}  
\end{eqnarray}
%%%%%%%%%%%%%%%%%%%%%%%%%%%%%%%%%%%%%%%%%%%%%%%%%%%%%%%%%%%%%%%%%%%  

Now we would like to explain an important point, which is sometimes missed in the theoretical physics
literature, when two BRST transformations coexist in a theory. Suppose that we fix the GCT 
by a gauge condition $F^\alpha ( g_{\mu\nu}, \phi ) = 0$ and the Weyl transformation by a gauge conditon 
$F ( g_{\mu\nu}, \phi ) = 0$. Then, the gauge-fixed and BRST invariant Lagrangian density is given by
%**   Lq   %%%%%%%%%%%%%%%%%%%%%%%%%%%%%%%%%%%%%%%%%%%%%%%%%%%%%%%%%
\begin{eqnarray}
{\cal L}_q = {\cal L}_c + \delta_B ( \bar c_\alpha F^\alpha ) + \bar \delta_B ( \bar c F ),
\label{Lq}  
\end{eqnarray}
%%%%%%%%%%%%%%%%%%%%%%%%%%%%%%%%%%%%%%%%%%%%%%%%%%%%%%%%%%%%%%%%%%% 
where the first term is the classical Lagrangian density (\ref{Lc}). Under this situation,
a natural question arises about the gauge-fixing conditions: Can we take any gauge-fixing
conditions if they fix gauge symmetries anyway? If not, what gauge conditions are suitable for $F^\alpha$ and $F$?

In order to answer the questions, let us take the two BRST transformations separately and check whether
the quantum Lagrangian density (\ref{Lq}) is really invariant under the BRST transformations up to surface
terms. First, taking the Weyl BRST transformation leads to
%**   W-BRST-Lq   %%%%%%%%%%%%%%%%%%%%%%%%%%%%%%%%%%%%%%%%%%%%%%%%%%%%%%%%%
\begin{eqnarray}
\bar \delta_B {\cal{L}}_q &=& \bar \delta_B \delta_B ( \bar c_\alpha F^\alpha ) 
= - \delta_B \bar \delta_B ( \bar c_\alpha F^\alpha ) 
\nonumber\\
&=& - \delta_B \left[ ( \bar \delta_B \bar c_\alpha ) F^\alpha - \bar c_\alpha \bar \delta_B F^\alpha \right],
\label{W-BRST-Lq}  
\end{eqnarray}
%%%%%%%%%%%%%%%%%%%%%%%%%%%%%%%%%%%%%%%%%%%%%%%%%%%%%%%%%%%%%%%%%%% 
where we have used $\bar \delta_B {\cal{L}}_c = 0$, and Eqs. (\ref{Nilpotent}) and (\ref{GCT-Weyl-BRST}). 
This equation clearly shows that the conditions
%**   Cond-GCT-gauge   %%%%%%%%%%%%%%%%%%%%%%%%%%%%%%%%%%%%%%%%%%%%%%%%%%%%%%%%%
\begin{eqnarray}
\bar \delta_B \bar c_\alpha = 0, \quad
\bar \delta_B F^\alpha = 0,
\label{Cond-GCT-gauge}  
\end{eqnarray}
%%%%%%%%%%%%%%%%%%%%%%%%%%%%%%%%%%%%%%%%%%%%%%%%%%%%%%%%%%%%%%%%%%% 
are sufficient conditions such that the Lagrangian density (\ref{Lq}) is invariant under the Weyl BRST 
transformation.

It is of interest to notice that the former condition in (\ref{Cond-GCT-gauge}) leads to two remaining equations 
in (\ref{Weyl-BRST2}). To see this fact, let us take the GCT BRST transformation of the former equation  
as follows:
%**   Cond-GCT-gauge1   %%%%%%%%%%%%%%%%%%%%%%%%%%%%%%%%%%%%%%%%%%%%%%%%%%%%%%%%%
\begin{eqnarray}
0 &=& \delta_B \bar \delta_B \bar c_\alpha = - \bar \delta_B \delta_B \bar c_\alpha = - i \bar \delta_B B_\alpha
\nonumber\\
&=& - i \left[  \bar \delta_B b_\alpha + i ( \bar \delta_B c^\lambda ) \partial_\lambda \bar c_\alpha \right],
\label{Cond-GCT-gauge1}  
\end{eqnarray}
%%%%%%%%%%%%%%%%%%%%%%%%%%%%%%%%%%%%%%%%%%%%%%%%%%%%%%%%%%%%%%%%%%% 
which implies $\bar \delta_B b_\alpha = \bar \delta_B c^\lambda = 0$, which coincide with the remaining 
two equations in (\ref{Weyl-BRST2}).

On the other hand, the latter condition in (\ref{Cond-GCT-gauge}) gives rise to important information 
on the gauge condition for the GCT: The gauge-fixing condition for the GCT must be
invariant under the Weyl transformation. Thus, for instance, the conventional de Donder gauge condition
(or harmonic gauge condition) 
%**   Donder   %%%%%%%%%%%%%%%%%%%%%%%%%%%%%%%%%%%%%%%%%%%%%%%%%%%%%%%%%
\begin{eqnarray}
\partial_\mu\tilde g^{\mu\nu} = 0,
\label{Donder}  
\end{eqnarray}
%%%%%%%%%%%%%%%%%%%%%%%%%%%%%%%%%%%%%%%%%%%%%%%%%%%%%%%%%%%%%%%%%%% 
is not suitable when there is the Weyl invariance.\footnote{In two space-time dimensions the de Donder
condition is Weyl invariant so it is allowed to use it as the gauge-fixing condition for the GCT.}

Next, let us operate the GCT BRST transformation on ${\cal L}_q$. To do that, since the Lagrangian
density is in general a quantity with density, it is more convenient to write as ${\cal L}_q \equiv \sqrt{-g} 
{\cal L}_q^\prime$ and $F = \sqrt{-g} F^\prime$ where $F$ and $F^\prime$ are scalars. 
Then, taking the GCT BRST variation leads to 
%**   GCT-BRST-Lq   %%%%%%%%%%%%%%%%%%%%%%%%%%%%%%%%%%%%%%%%%%%%%%%%%%%%%%%%%
\begin{eqnarray}
\delta_B {\cal L}_q &=& \delta_B ( \sqrt{-g} {\cal L}_q^\prime )
= \delta_B \bar \delta_B ( \sqrt{-g} \bar c F^\prime )
= - \bar \delta_B \delta_B ( \sqrt{-g} \bar c F^\prime )
\nonumber\\
&=& - \bar \delta_B \left[ - \sqrt{-g} \nabla_\rho c^\rho \bar c F^\prime
+ \sqrt{-g} ( - c^\rho \partial_\rho \bar c ) F^\prime - \sqrt{-g} \bar c ( - c^\rho \partial_\rho F^\prime ) \right]
\nonumber\\
&=& \partial_\rho \bar \delta_B ( c^\rho \bar c F ),
\label{GCT-BRST-Lq}  
\end{eqnarray}
%%%%%%%%%%%%%%%%%%%%%%%%%%%%%%%%%%%%%%%%%%%%%%%%%%%%%%%%%%%%%%%%%%% 
which means that ${\cal L}_q$ is indeed invariant under the GCT BRST transformation up to
a surface term. In obtaining this result, we have assumed 
%**   GCT-transf-F   %%%%%%%%%%%%%%%%%%%%%%%%%%%%%%%%%%%%%%%%%%%%%%%%%%%%%%%%%
\begin{eqnarray}
\delta_B F^\prime = - c^\rho \partial_\rho F^\prime,
\label{GCT-transf-F}  
\end{eqnarray}
%%%%%%%%%%%%%%%%%%%%%%%%%%%%%%%%%%%%%%%%%%%%%%%%%%%%%%%%%%%%%%%%%%% 
which is nothing but the requirement that the quantity $F^\prime$ should be a scalar under the GCT. Thus,
only a scalar function $F^\prime$, or equivalently a scalar density $F$, makes sense as a gauge-fixing
condition for the Weyl invariance. Of course, this scalar function must break the Weyl invariance.
As suitable gauge-fixing conditions, in this paper we will choose $F^\nu = \partial_\mu ( \tilde g^{\mu\nu}
\phi^2 )$ and $F = \partial_\mu ( \tilde g^{\mu\nu} \phi \partial_\nu \phi )$.

\section{Quantum Weyl invariant scalar-tensor gravity}

In this section, as a classical Lagrangian\footnote{For simplicity, we henceforth call a Lagrangian density
a Lagrangian.}, we will take a Weyl invariant scalar-tensor gravity whose Lagrangian is of form \cite{Fujii} 
%**   WIST-gravity   %%%%%%%%%%%%%%%%%%%%%%%%%%%%%%%%%%%%%%%%%%%%%%%%%%%%%%%%%
\begin{eqnarray}
{\cal L}_c = \sqrt{-g} \left( \frac{1}{12} \phi^2 R + \frac{1}{2} g^{\mu\nu} \partial_\mu \phi \partial_\nu \phi \right),
\label{WIST-gravity}  
\end{eqnarray}
%%%%%%%%%%%%%%%%%%%%%%%%%%%%%%%%%%%%%%%%%%%%%%%%%%%%%%%%%%%%%%%%%%% 
where $\phi$ is a real scalar field with a ghost-like kinetic term, and $R$ the scalar curvature.
In addition to the invariance under the general coordinate transformation (GCT), this Lagrangian is also invariant 
under the Weyl transformation (or the local scale transformation) defined as
%**   Weyl-transf   %%%%%%%%%%%%%%%%%%%%%%%%%%%%%%%%%%%%%%%%%%%%%%%%%%%%%%%%%
\begin{eqnarray}
g_{\mu\nu} \rightarrow g^\prime_{\mu\nu} = \Omega^2 (x) g_{\mu\nu}, \qquad 
\phi \rightarrow \phi^\prime = \Omega^{-1}(x) \phi.
\label{Weyl-transf}  
\end{eqnarray}
%%%%%%%%%%%%%%%%%%%%%%%%%%%%%%%%%%%%%%%%%%%%%%%%%%%%%%%%%%%%%%%%%%% 
Recall that in order to prove the invariance, we need to use the following transformation of the scalar curvature 
under (\ref{Weyl-transf}):
%**   Weyl-R   %%%%%%%%%%%%%%%%%%%%%%%%%%%%%%%%%%%%%%%%%%%%%%%%%%%%%%%%%
\begin{eqnarray}
R \rightarrow R^\prime = \Omega^{-2} ( R - 6 \Omega^{-1} \Box \Omega ),
\label{Weyl-R}  
\end{eqnarray}
%%%%%%%%%%%%%%%%%%%%%%%%%%%%%%%%%%%%%%%%%%%%%%%%%%%%%%%%%%%%%%%%%%% 
where $\Box \Omega \equiv h^{-1} \partial_\mu ( \tilde g^{\mu\nu} \partial_\nu \Omega )$.

As explained in the previous section, we have to pay attention to what gauge-fixing conditions 
should be chosen for the GCT and the Weyl transformation in a consistent manner. For instance, taking the de Donder 
condition as a gauge condition for GCT is not allowed since it breaks the Weyl symmetry in four space-time dimensions.
There are several interesting choices of suitable gauge conditions for the GCT, but we shall refer to only two 
representative examples: The first gauge condition for the GCT is a Weyl invariant version of the de Donder gauge:
%**   Mod-de-Donder   %%%%%%%%%%%%%%%%%%%%%%%%%%%%%%%%%%%%%%%%%%%%%%%%%%%%%%%%%
\begin{eqnarray}
\partial_\mu \left( ( - g )^{\frac{1}{4}} g^{\mu\nu} \right) = 0. 
\label{Mod-de-Donder}  
\end{eqnarray}
%%%%%%%%%%%%%%%%%%%%%%%%%%%%%%%%%%%%%%%%%%%%%%%%%%%%%%%%%%%%%%%%%%% 
This gauge choice is invariant under the Weyl transformation (\ref{Weyl-transf}) and is physically interesting 
in the sense that it makes use of only the metric tensor field. However, some fields such as the Nakanishi-Lautrup 
field become not a normal vector field but a vector field with density, which makes several formulas ugly, 
so we will not adopt (\ref{Mod-de-Donder}) as a gauge condition for the GCT.  The second gauge condition, 
which we will take in this article and call it ``the extended de Donder gauge'', is given by
%**   Ext-de-Donder   %%%%%%%%%%%%%%%%%%%%%%%%%%%%%%%%%%%%%%%%%%%%%%%%%%%%%%%%%
\begin{eqnarray}
\partial_\mu ( \tilde g^{\mu\nu} \phi^2 ) = 0,
\label{Ext-de-Donder}  
\end{eqnarray}
%%%%%%%%%%%%%%%%%%%%%%%%%%%%%%%%%%%%%%%%%%%%%%%%%%%%%%%%%%%%%%%%%%% 
which is also invariant under the Weyl transformation (\ref{Weyl-transf}). 

Next, let us consider a gauge-fixing condition for the Weyl transformation. From the consistency discussed
in Section 2, an appropriate gauge condition must obey the condition that it is invariant under the GCT, that is, 
a scalar quantity. Since there are many of scalars constructed out of the real scalar field $\phi$ and the 
Riemannian tensors, we might be left in the dark on this issue. However, surprisingly enough, if we impose 
the requirement such that the FP ghost's Lagrangian should have a Weyl invariant metric $\tilde g^{\mu\nu} \phi^2$ 
instead of the standard metric $\tilde g^{\mu\nu}$, the suitable gauge condition for the GCT can be uniquely picked up. 
Such the gauge condition, we will call ``the scalar gauge condition'', reads 
%**   Scalar-gauge   %%%%%%%%%%%%%%%%%%%%%%%%%%%%%%%%%%%%%%%%%%%%%%%%%%%%%%%%%
\begin{eqnarray}
\partial_\mu ( \tilde g^{\mu\nu} \phi \partial_\nu \phi ) = 0,
\label{Scalar-gauge}  
\end{eqnarray}
%%%%%%%%%%%%%%%%%%%%%%%%%%%%%%%%%%%%%%%%%%%%%%%%%%%%%%%%%%%%%%%%%%% 
which can be alternatively written as 
%**   Alt-Scalar-gauge   %%%%%%%%%%%%%%%%%%%%%%%%%%%%%%%%%%%%%%%%%%%%%%%%%%%%%%%%%
\begin{eqnarray}
\Box \, \phi^2 = 0.
\label{Alt-Scalar-gauge}  
\end{eqnarray}
%%%%%%%%%%%%%%%%%%%%%%%%%%%%%%%%%%%%%%%%%%%%%%%%%%%%%%%%%%%%%%%%%%% 
Incidentally, the unitary gauge $\phi = \rm{constant}$ is often taken to show that the Weyl invariant scalar-tensor
gravity (\ref{WIST-gravity}) is equivalent to the Einstein-Hilbert term, but this gauge choice is not so
interesting since there remains no conformal symmetry behind.   

After taking the extended de Donder gauge condition (\ref{Ext-de-Donder}) for the GCT and the scalar gauge condition
(\ref{Scalar-gauge}) for the Weyl transformation, the gauge-fixed and BRST invariant quantum Lagrangian is given by
%**   ST-q-Lag   %%%%%%%%%%%%%%%%%%%%%%%%%%%%%%%%%%%%%%%%%%%%%%%%%%%%%%%%%
\begin{eqnarray}
{\cal L}_q &=& {\cal L}_c + {\cal L}_{GF+FP} + \bar {\cal L}_{GF+FP}
\nonumber\\
&=& {\cal L}_c + i \delta_B ( \tilde g^{\mu\nu} \phi^2 \partial_\mu \bar c_\nu )
+ i \bar \delta_B \left[ \bar c \partial_\mu ( \tilde g^{\mu\nu} \phi \partial_\nu \phi ) \right] 
\nonumber\\
&=& \sqrt{-g} \left( \frac{1}{12} \phi^2 R + \frac{1}{2} g^{\mu\nu} \partial_\mu \phi \partial_\nu \phi \right)
- \tilde g^{\mu\nu} \phi^2 ( \partial_\mu b_\nu + i \partial_\mu \bar c_\lambda  \partial_\nu c^\lambda )
\nonumber\\
&+& \tilde g^{\mu\nu} \phi \partial_\mu B \partial_\nu \phi - i \tilde g^{\mu\nu} \phi^2 \partial_\mu \bar c 
\partial_\nu c,
\label{ST-q-Lag}  
\end{eqnarray}
%%%%%%%%%%%%%%%%%%%%%%%%%%%%%%%%%%%%%%%%%%%%%%%%%%%%%%%%%%%%%%%%%%% 
where surface terms are dropped. Note that the last term, which is the FP ghost's term for the Weyl transformation,
certainly involves the Weyl invariant metric $\tilde g^{\mu\nu} \phi^2$. Let us rewrite this Lagrangian concisely as
%**   ST-q-Lag2   %%%%%%%%%%%%%%%%%%%%%%%%%%%%%%%%%%%%%%%%%%%%%%%%%%%%%%%%%
\begin{eqnarray}
{\cal L}_q = \sqrt{-g} \frac{1}{12} \phi^2 R - \frac{1}{2} \tilde g^{\mu\nu} E_{\mu\nu},
\label{ST-q-Lag2}  
\end{eqnarray}
%%%%%%%%%%%%%%%%%%%%%%%%%%%%%%%%%%%%%%%%%%%%%%%%%%%%%%%%%%%%%%%%%%% 
where we have defined  
%**   E   %%%%%%%%%%%%%%%%%%%%%%%%%%%%%%%%%%%%%%%%%%%%%%%%%%%%%%%%%
\begin{eqnarray}
E_{\mu\nu} &\equiv& - \frac{1}{2} \partial_\mu \phi \partial_\nu \phi + \phi^2 ( \partial_\mu b_\nu 
+ i \partial_\mu \bar c_\lambda  \partial_\nu c^\lambda )
\nonumber\\
&-& \phi \partial_\mu B \partial_\nu \phi + i \phi^2 \partial_\mu \bar c \partial_\nu c
+ ( \mu \leftrightarrow \nu ). 
\label{E}  
\end{eqnarray}
%%%%%%%%%%%%%%%%%%%%%%%%%%%%%%%%%%%%%%%%%%%%%%%%%%%%%%%%%%%%%%%%%%% 
Moreover, it is sometimes more convenient to introduce the dilaton $\sigma (x)$ by defining
%**   Dilaton   %%%%%%%%%%%%%%%%%%%%%%%%%%%%%%%%%%%%%%%%%%%%%%%%%%%%%%%%%
\begin{eqnarray}
\phi (x) \equiv e^{\sigma (x)},
\label{Dilaton}  
\end{eqnarray}
%%%%%%%%%%%%%%%%%%%%%%%%%%%%%%%%%%%%%%%%%%%%%%%%%%%%%%%%%%%%%%%%%%% 
and rewrite (\ref{ST-q-Lag2}) further into the form  
%**   ST-q-Lag3   %%%%%%%%%%%%%%%%%%%%%%%%%%%%%%%%%%%%%%%%%%%%%%%%%%%%%%%%%
\begin{eqnarray}
{\cal L}_q = e^{2 \sigma (x)} \left( \sqrt{-g} \frac{1}{12} R - \frac{1}{2} \tilde g^{\mu\nu} 
\hat E_{\mu\nu} \right),
\label{ST-q-Lag3}  
\end{eqnarray}
%%%%%%%%%%%%%%%%%%%%%%%%%%%%%%%%%%%%%%%%%%%%%%%%%%%%%%%%%%%%%%%%%%% 
where we have defined  
%**   hat-E   %%%%%%%%%%%%%%%%%%%%%%%%%%%%%%%%%%%%%%%%%%%%%%%%%%%%%%%%%
\begin{eqnarray}
\hat E_{\mu\nu} &\equiv& - \frac{1}{2} \partial_\mu \sigma \partial_\nu \sigma + \partial_\mu b_\nu 
+ i \partial_\mu \bar c_\lambda  \partial_\nu c^\lambda
\nonumber\\
&-& \partial_\mu B \partial_\nu \sigma + i \partial_\mu \bar c \partial_\nu c
+ ( \mu \leftrightarrow \nu ). 
\label{hat-E}  
\end{eqnarray}
%%%%%%%%%%%%%%%%%%%%%%%%%%%%%%%%%%%%%%%%%%%%%%%%%%%%%%%%%%%%%%%%%%% 
Note that the relation between $E_{\mu\nu}$ and $\hat E_{\mu\nu}$ is given by
%**   E vs h-E   %%%%%%%%%%%%%%%%%%%%%%%%%%%%%%%%%%%%%%%%%%%%%%%%%%%%%%%%%
\begin{eqnarray}
E_{\mu\nu} = \phi^2 \hat E_{\mu\nu} = e^{2 \sigma } \hat E_{\mu\nu}.
\label{E vs h-E}  
\end{eqnarray}
%%%%%%%%%%%%%%%%%%%%%%%%%%%%%%%%%%%%%%%%%%%%%%%%%%%%%%%%%%%%%%%%%%% 

From the Lagrangian ${\cal L}_q$, it is straightforward to derive the field equations by taking 
the variation with respect to $g_{\mu\nu}$, $\phi$ (or $\sigma$), $b_\nu$, $B$, $c^\rho$, 
$\bar c_\rho$, $c$ and $\bar c$ in order:
%**   q-field-eq   %%%%%%%%%%%%%%%%%%%%%%%%%%%%%%%%%%%%%%%%%%%%%%%%%%%%%%%%%
\begin{eqnarray}
&{}& \frac{1}{12} \phi^2 G_{\mu\nu} - \frac{1}{12} ( \nabla_\mu \nabla_\nu - g_{\mu\nu} \Box ) \phi^2  
- \frac{1}{2} ( E_{\mu\nu} - \frac{1}{2} g_{\mu\nu} E ) = 0, 
\nonumber\\
&{}& \frac{1}{6} \phi^2 R - E - 2 g^{\mu\nu} \phi \partial_\mu B \partial_\nu \phi - \phi^2 \Box B = 0,
\nonumber\\
&{}& \partial_\mu ( \tilde g^{\mu\nu} \phi^2 ) = 0, \qquad
\partial_\mu ( \tilde g^{\mu\nu} \phi \partial_\nu \phi ) = 0, 
\nonumber\\
&{}& g^{\mu\nu} \partial_\mu \partial_\nu\bar c_\rho = g^{\mu\nu} \partial_\mu \partial_\nu c^\rho 
= g^{\mu\nu} \partial_\mu \partial_\nu\bar c = g^{\mu\nu} \partial_\mu \partial_\nu c = 0.
\label{q-field-eq}  
\end{eqnarray}
%%%%%%%%%%%%%%%%%%%%%%%%%%%%%%%%%%%%%%%%%%%%%%%%%%%%%%%%%%%%%%%%%%% 
where we have defined the Einstein tensor $G_{\mu\nu} \equiv R_{\mu\nu} - \frac{1}{2} g_{\mu\nu} R$
and $E \equiv g^{\mu\nu} E_{\mu\nu}$. The two gauge-fixing conditions in (\ref{q-field-eq})
lead to a very simple equation for the dilaton:
%**   Dilaton-eq   %%%%%%%%%%%%%%%%%%%%%%%%%%%%%%%%%%%%%%%%%%%%%%%%%%%%%%%%%
\begin{eqnarray}
g^{\mu\nu} \partial_\mu \partial_\nu \sigma = 0.
\label{Dilaton-eq}  
\end{eqnarray}
%%%%%%%%%%%%%%%%%%%%%%%%%%%%%%%%%%%%%%%%%%%%%%%%%%%%%%%%%%%%%%%%%%% 
It is worthwhile to notice that it is not the scalar field $\phi$ but the dilaton $\sigma$ that satisfies 
this type of equation. Furthermore, the trace part of the Einstein equation, i.e., the first field equation 
in (\ref{q-field-eq}) and the field equation for $\phi$ also give us the equation for $B$:
%**   B-eq   %%%%%%%%%%%%%%%%%%%%%%%%%%%%%%%%%%%%%%%%%%%%%%%%%%%%%%%%%
\begin{eqnarray}
g^{\mu\nu} \partial_\mu \partial_\nu B = 0.
\label{B-eq}  
\end{eqnarray}
%%%%%%%%%%%%%%%%%%%%%%%%%%%%%%%%%%%%%%%%%%%%%%%%%%%%%%%%%%%%%%%%%%% 
Finally, using the field equations obtained thus far, after some calculations, we can also derive 
the equation for $b_\rho$:\footnote{The detail of the calculation is presented in Appendix A.}
%**   b-rho-eq   %%%%%%%%%%%%%%%%%%%%%%%%%%%%%%%%%%%%%%%%%%%%%%%%%%%%%%%%%
\begin{eqnarray}
g^{\mu\nu} \partial_\mu \partial_\nu b_\rho = 0.
\label{b-rho-eq}  
\end{eqnarray}
%%%%%%%%%%%%%%%%%%%%%%%%%%%%%%%%%%%%%%%%%%%%%%%%%%%%%%%%%%%%%%%%%%% 
In other words, setting $X^M = \{ x^\mu, b_\mu, \sigma, B, c^\mu, \bar c_\mu, c, \bar c \}$, $X^M$
turns out to obey the very simple equation:
%**   X-M-eq   %%%%%%%%%%%%%%%%%%%%%%%%%%%%%%%%%%%%%%%%%%%%%%%%%%%%%%%%%
\begin{eqnarray}
g^{\mu\nu} \partial_\mu \partial_\nu X^M = 0.
\label{X-M-eq}  
\end{eqnarray}
%%%%%%%%%%%%%%%%%%%%%%%%%%%%%%%%%%%%%%%%%%%%%%%%%%%%%%%%%%%%%%%%%%% 
This fact, together with the gauge condition $\partial_\mu ( \tilde g^{\mu\nu} \phi^2 ) = 0$,
produces the two kinds of conserved currents:
%**   Cons-currents   %%%%%%%%%%%%%%%%%%%%%%%%%%%%%%%%%%%%%%%%%%%%%%%%%%%%%%%%%
\begin{eqnarray}
{\cal P}^{\mu M} &\equiv& \tilde g^{\mu\nu} \phi^2 \partial_\nu X^M 
= \tilde g^{\mu\nu} \phi^2 \bigl( 1 \overset{\leftrightarrow}{\partial}_\nu X^M \bigr)
\nonumber\\
{\cal M}^{\mu M N} &\equiv& \tilde g^{\mu\nu} \phi^2 \bigl( X^M 
\overset{\leftrightarrow}{\partial}_\nu Y^N \bigr),
\label{Cons-currents}  
\end{eqnarray}
%%%%%%%%%%%%%%%%%%%%%%%%%%%%%%%%%%%%%%%%%%%%%%%%%%%%%%%%%%%%%%%%%%% 
where we have defined $X^M \overset{\leftrightarrow}{\partial}_\mu Y^N \equiv
X^M \partial_\mu Y^N - ( \partial_\mu X^M ) Y^N$.

\section{Canonical quantization and equal-time commutation relations}

In this section, after introducing the canonical commutation relations (CCRs), we will evaluate various equal-time 
commutation relations (ETCRs) among fundamental variables. To simplify various expressions, we will obey 
the following abbreviations adopted in the textbook of Nakanishi and Ojima \cite{N-O-text}:
%**   abbreviation   %%%%%%%%%%%%%%%%%%%%%%%%%%%%%%%%%%%%%%%%%%%%%%%%%%%%%%%%%
\begin{eqnarray}
[ A, B^\prime ] &=& [ A(x), B(x^\prime) ] |_{x^0 = x^{\prime 0}},
\qquad \delta^3 = \delta(\vec{x} - \vec{x}^\prime), 
\nonumber\\
\tilde f &=& \frac{1}{\tilde g^{00}} = \frac{1}{\sqrt{-g} g^{00}} = \frac{1}{h g^{00}},
\label{abbreviation}  
\end{eqnarray}
%%%%%%%%%%%%%%%%%%%%%%%%%%%%%%%%%%%%%%%%%%%%%%%%%%%%%%%%%%%%%%%%%%% 
where we assume that $\tilde g^{00}$ is invertible. 

Now let us set up the canonical (anti-)commutation relations: 
%**   CCRs   %%%%%%%%%%%%%%%%%%%%%%%%%%%%%%%%%%%%%%%%%%%%%%%%%%%%%%%%%
\begin{eqnarray}
&{}& [ g_{\mu\nu}, \pi_g^{\rho\lambda\prime} ] = i \frac{1}{2} ( \delta_\mu^\rho\delta_\nu^\lambda 
+ \delta_\mu^\lambda\delta_\nu^\rho) \delta^3,  \quad 
[ \phi, \pi_\phi^\prime ] = + i \delta^3,  \quad
[ B, \pi_B^\prime ] = + i \delta^3,
\nonumber\\
&{}& \{ c^\sigma, \pi_{c \lambda}^\prime,  \} = \{ \bar c_\lambda, \pi_{\bar c}^{\sigma\prime} \}
= + i \delta_\lambda^\sigma \delta^3,  \quad
\{ c, \pi_c^\prime  \} = \{ \bar c, \pi_{\bar c}^\prime \} = + i \delta^3,
\label{CCRs}  
\end{eqnarray}
%%%%%%%%%%%%%%%%%%%%%%%%%%%%%%%%%%%%%%%%%%%%%%%%%%%%%%%%%%%%%%%%%%% 
where the other (anti-)commutation relations vanish.
Here the canonical variables are $g_{\mu\nu}, \phi, B, c^\rho, \bar c_\rho, c, \bar c$ and the corresponding canonical
conjugate momenta are $\pi_g^{\mu\nu}, \pi_\phi, \pi_B, \pi_{c \rho}, \pi_{\bar c}^\rho, \pi_c, \pi_{\bar c}$, respectively 
and the $b_\mu$ field is regarded as not a canonical variable but a conjugate momentum of $\tilde g^{0 \mu}$. 

To remove second order derivatives of the metric involved in $R$, we perform the integration by parts once and
rewrite the Lagrangian (\ref{ST-q-Lag}) as
%**   Mod-ST-q-Lag   %%%%%%%%%%%%%%%%%%%%%%%%%%%%%%%%%%%%%%%%%%%%%%%%%%%%%%%%%
\begin{eqnarray}
{\cal L}_q &=& - \frac{1}{12} \tilde g^{\mu\nu} \phi^2 ( \Gamma^\sigma_{\mu\nu} \Gamma^\alpha_{\sigma\alpha}  
-  \Gamma^\sigma_{\mu\alpha} \Gamma^\alpha_{\sigma\nu} ) 
- \frac{1}{6} \phi \partial_\mu \phi ( \tilde g^{\alpha\beta} \Gamma^\mu_{\alpha\beta}  
- \tilde g^{\mu\nu} \Gamma^\alpha_{\nu\alpha} ) 
\nonumber\\
&+& \frac{1}{2} \tilde g^{\mu\nu} \partial_\mu \phi \partial_\nu \phi 
+ \partial_\mu ( \tilde g^{\mu\nu} \phi^2 ) b_\nu 
- i \tilde g^{\mu\nu} \phi^2 \partial_\mu \bar c_\rho \partial_\nu c^\rho 
+ \tilde g^{\mu\nu} \partial_\mu B \phi \partial_\nu \phi
\nonumber\\
&-& i \tilde g^{\mu\nu} \phi^2 \partial_\mu \bar c \partial_\nu c
+ \partial_\mu {\cal{V}}^\mu,
\label{Mod-ST-q-Lag}  
\end{eqnarray}
%%%%%%%%%%%%%%%%%%%%%%%%%%%%%%%%%%%%%%%%%%%%%%%%%%%%%%%%%%%%%%%%%%% 
where a surface term ${\cal{V}}^\mu$ is defined as
%**   surface   %%%%%%%%%%%%%%%%%%%%%%%%%%%%%%%%%%%%%%%%%%%%%%%%%%%%%%%%%
\begin{eqnarray}
{\cal{V}}^\mu =  \frac{1}{12} \phi^2 ( \tilde g^{\alpha\beta} \Gamma^\mu_{\alpha\beta} 
- \tilde g^{\mu\nu} \Gamma^\alpha_{\nu\alpha} ) - \tilde g^{\mu\nu} \phi^2 b_\nu.
\label{surface}  
\end{eqnarray}
%%%%%%%%%%%%%%%%%%%%%%%%%%%%%%%%%%%%%%%%%%%%%%%%%%%%%%%%%%%%%%%%%%% 
Using this Lagrangian, the concrete expressions for canonical conjugate momenta become
%**   CCM   %%%%%%%%%%%%%%%%%%%%%%%%%%%%%%%%%%%%%%%%%%%%%%%%%%%%%%%%%
\begin{eqnarray}
\pi_g^{\mu\nu} &=& \frac{\partial {\cal L}_q}{\partial \dot g_{\mu\nu}} 
\nonumber\\
&=& - \frac{1}{24} \sqrt{-g} \phi^2 \Bigl[ - g^{0 \lambda} g^{\mu\nu} g^{\sigma\tau} - g^{0 \tau} g^{\mu\lambda} g^{\nu\sigma}
- g^{0 \sigma} g^{\mu\tau} g^{\nu\lambda} + g^{0 \lambda} g^{\mu\tau} g^{\nu\sigma} 
\nonumber\\
&+& g^{0 \tau} g^{\mu\nu} g^{\lambda\sigma}
+ \frac{1}{2} ( g^{0 \mu} g^{\nu\lambda} + g^{0 \nu} g^{\mu\lambda} ) g^{\sigma\tau} \Bigr] \partial_\lambda g_{\sigma\tau}
\nonumber\\
&-& \frac{1}{6} \sqrt{-g} \Bigl[ \frac{1}{2} ( g^{0 \mu} g^{\rho\nu} + g^{0 \nu} g^{\rho\mu} ) 
- g^{\mu\nu} g^{\rho 0} \Bigr] \phi \partial_\rho \phi 
\nonumber\\
&-& \frac{1}{2} \sqrt{-g} ( g^{0 \mu} g^{\nu\rho} + g^{0 \nu} g^{\mu\rho} - g^{0 \rho} g^{\mu\nu} ) \phi^2 b_\rho,
\nonumber\\
\pi_\phi &=& \frac{\partial {\cal L}_q}{\partial \dot \phi} = \tilde g^{0 \mu} \partial_\mu \phi
+ 2 \tilde g^{0 \mu} \phi b_\mu + \frac{1}{6} \phi ( - \tilde g^{\alpha\beta} \Gamma^0_{\alpha\beta} 
+ \tilde g^{0 \alpha} \Gamma^\beta_{\alpha\beta} ) + \tilde g^{0 \mu} \partial_\mu B \phi,
\nonumber\\
\pi_B &=& \frac{\partial {\cal L}_q}{\partial \dot B} = \tilde g^{0 \mu} \phi \partial_\mu \phi,
\nonumber\\
\pi_{c \sigma} &=& \frac{\partial {\cal L}_q}{\partial \dot c^\sigma} = - i \tilde g^{0 \mu} \phi^2 \partial_\mu \bar c_\sigma,
\nonumber\\
\pi_{\bar c}^\sigma &=& \frac{\partial {\cal L}_q}{\partial \dot {\bar c}_\sigma} = i \tilde g^{0 \mu} \phi^2 \partial_\mu c^\sigma,
\nonumber\\
\pi_c &=& \frac{\partial {\cal L}_q}{\partial \dot c} = - i \tilde g^{0 \mu} \phi^2 \partial_\mu \bar c,
\nonumber\\
\pi_{\bar c} &=& \frac{\partial {\cal L}_q}{\partial \dot {\bar c}} = i \tilde g^{0 \mu} \phi^2 \partial_\mu c,
\label{CCM}  
\end{eqnarray}
%%%%%%%%%%%%%%%%%%%%%%%%%%%%%%%%%%%%%%%%%%%%%%%%%%%%%%%%%%%%%%%%%%% 
where we have defined the time derivative such as $\dot g_{\mu\nu} \equiv \frac{\partial g_{\mu\nu}}{\partial t}
\equiv \partial_0 g_{\mu\nu}$, and differentiation of ghosts is taken from the right. 

From now on, we would like to evaluate various nontrivial equal-time commutation relations (ETCRs) in order.
Let us first work with the ETCR in Eq. (\ref{CCRs}):
%**   pi(a0)-g   %%%%%%%%%%%%%%%%%%%%%%%%%%%%%%%%%%%%%%%%%%%%%%%%%%%%%%%%%
\begin{eqnarray}
[ \pi_g^{\alpha 0}, g_{\mu\nu}^\prime ] = - i \frac{1}{2} ( \delta_\mu^\alpha \delta_\nu^0 
+ \delta_\mu^0 \delta_\nu^\alpha) \delta^3.
\label{pi(a0)-g}  
\end{eqnarray}
%%%%%%%%%%%%%%%%%%%%%%%%%%%%%%%%%%%%%%%%%%%%%%%%%%%%%%%%%%%%%%%%%%% 
The canonical conjugate momentum $\pi_g^{\alpha0}$ has a structure
%**   pi(a0)   %%%%%%%%%%%%%%%%%%%%%%%%%%%%%%%%%%%%%%%%%%%%%%%%%%%%%%%%%
\begin{eqnarray}
\pi_g^{\alpha 0} = A^\alpha+ B^{\alpha\beta} \partial_\beta \phi+ C^{\alpha\beta} b_\beta,
\label{pi(a0)}  
\end{eqnarray}
%%%%%%%%%%%%%%%%%%%%%%%%%%%%%%%%%%%%%%%%%%%%%%%%%%%%%%%%%%%%%%%%%%% 
where $A^\alpha, B^{\alpha\beta}$ and $C^{\alpha\beta} \equiv - \frac{1}{2} \tilde g^{00} g^{\alpha\beta} \phi^2$ 
have no $\dot g_{\mu\nu}$, and $B^{\alpha\beta} \partial_\beta \phi$ does not have $\dot \phi$ 
since $\pi_g^{\alpha 0}$ does not include the dynamics of the metric and the scalar fields. Then, we find 
that Eq. (\ref{pi(a0)-g}) produces
%**   g-b   %%%%%%%%%%%%%%%%%%%%%%%%%%%%%%%%%%%%%%%%%%%%%%%%%%%%%%%%%
\begin{eqnarray}
[ g_{\mu\nu}, b_\rho^\prime ] = - i \tilde f \phi^{-2} ( \delta_\mu^0 g_{\rho\nu} + \delta_\nu^0 g_{\rho\mu} ) \delta^3.
\label{g-b}  
\end{eqnarray}
%%%%%%%%%%%%%%%%%%%%%%%%%%%%%%%%%%%%%%%%%%%%%%%%%%%%%%%%%%%%%%%%%%% 
From this ETCR, we can easily derive ETCRs:
%**   3-g-b   %%%%%%%%%%%%%%%%%%%%%%%%%%%%%%%%%%%%%%%%%%%%%%%%%%%%%%%%%
\begin{eqnarray}
&{}& [ g^{\mu\nu}, b_\rho^\prime ] = i \tilde f \phi^{-2} ( g^{\mu0} \delta_\rho^\nu+ g^{\nu0} \delta_\rho^\mu) \delta^3,
\nonumber\\
&{}& [ \tilde g^{\mu\nu}, b_\rho^\prime ] = i \tilde f \phi^{-2} ( \tilde g^{\mu0} \delta_\rho^\nu+ \tilde g^{\nu0} \delta_\rho^\mu 
- \tilde g^{\mu\nu} \delta_\rho^0 ) \delta^3.
\label{3-g-b}  
\end{eqnarray}
%%%%%%%%%%%%%%%%%%%%%%%%%%%%%%%%%%%%%%%%%%%%%%%%%%%%%%%%%%%%%%%%%%% 
Here we have used the following fact; since a commutator works as a derivation, we can have formulae:
%**   Simple formulae   %%%%%%%%%%%%%%%%%%%%%%%%%%%%%%%%%%%%%%%%%%%%%%%%%%%%%%%%%
\begin{eqnarray}
&{}& [ g^{\mu\nu}, \Phi^\prime ] = - g^{\mu\alpha} g^{\nu\beta} [ g_{\alpha\beta}, \Phi^\prime ],
\nonumber\\
&{}& [ \tilde g^{\mu\nu}, \Phi^\prime ] = - \left( \tilde g^{\mu\alpha} g^{\nu\beta} - \frac{1}{2} \tilde g^{\mu\nu} 
g^{\alpha\beta} \right) [ g_{\alpha\beta}, \Phi^\prime ],
\label{Simple formulae}  
\end{eqnarray}
%%%%%%%%%%%%%%%%%%%%%%%%%%%%%%%%%%%%%%%%%%%%%%%%%%%%%%%%%%%%%%%%%%% 
where $\Phi$ is a generic field. Similarly, the ETCR, $[ \pi_g^{\alpha 0}, \phi^\prime ] = 0$ yields
%**   phi-b   %%%%%%%%%%%%%%%%%%%%%%%%%%%%%%%%%%%%%%%%%%%%%%%%%%%%%%%%%
\begin{eqnarray}
[ \phi, b_\rho^\prime ] = 0.
\label{phi-b}  
\end{eqnarray}
%%%%%%%%%%%%%%%%%%%%%%%%%%%%%%%%%%%%%%%%%%%%%%%%%%%%%%%%%%%%%%%%%%% 
The ETCR, $[ \pi_g^{\alpha 0}, B^\prime ] = 0$ yields
%**   B-b   %%%%%%%%%%%%%%%%%%%%%%%%%%%%%%%%%%%%%%%%%%%%%%%%%%%%%%%%%
\begin{eqnarray}
[ B, b_\rho^\prime ] = 0.
\label{B-b}  
\end{eqnarray}
%%%%%%%%%%%%%%%%%%%%%%%%%%%%%%%%%%%%%%%%%%%%%%%%%%%%%%%%%%%%%%%%%%% 
Moreover, the ETCRs, $[ \pi_B, \phi^\prime ] = 0$ and $[ \pi_B, B^\prime ] = - i \delta^3$ respectively produce
%**   2-phi   %%%%%%%%%%%%%%%%%%%%%%%%%%%%%%%%%%%%%%%%%%%%%%%%%%%%%%%%%
\begin{eqnarray}
[ \dot \phi, \phi^\prime ] = 0, \qquad
[ \dot \phi, B^\prime ] = - i \tilde f \phi^{-1} \delta^3.
\label{2-phi}  
\end{eqnarray}
%%%%%%%%%%%%%%%%%%%%%%%%%%%%%%%%%%%%%%%%%%%%%%%%%%%%%%%%%%%%%%%%%%% 

As for the ETCRs involving FP ghosts, let us first consider the anti-ETCRs, $\{ \pi_{c \lambda}, c^{\sigma\prime} \}
= \{ \pi_{\bar c}^\sigma, \bar c_\lambda^\prime \} = i \delta_\lambda^\sigma \delta^3$. These anti-ETCRs lead to
the same anti-ETCR: 
%**   gh-antigh   %%%%%%%%%%%%%%%%%%%%%%%%%%%%%%%%%%%%%%%%%%%%%%%%%%%%%%%%%
\begin{eqnarray}
\{ \dot{\bar c}_\lambda, c^{\sigma\prime} \} = - \{ \dot c^\sigma, \bar c^\prime_\lambda \} 
= - \tilde f \phi^{-2} \delta_\lambda^\sigma \delta^3,
\label{gh-antigh} 
\end{eqnarray} 
%%%%%%%%%%%%%%%%%%%%%%%%%%%%%%%%%%%%%%%%%%%%%%%%%%%%%%%%%%%%%%%%%%% 
where we have used a useful identity for generic variables $\Phi$ and $\Psi$: 
%**   identity   %%%%%%%%%%%%%%%%%%%%%%%%%%%%%%%%%%%%%%%%%%%%%%%%%%%%%%%%%
\begin{eqnarray}
[ \Phi, \dot \Psi^\prime] = \partial_0 [ \Phi, \Psi^\prime] - [ \dot \Phi, \Psi^\prime],
\label{identity}  
\end{eqnarray}
%%%%%%%%%%%%%%%%%%%%%%%%%%%%%%%%%%%%%%%%%%%%%%%%%%%%%%%%%%%%%%%%%%% 
which holds for the anti-commutation relation as well. In a similar way, the anti-ETCRs, 
$\{ \pi_c, c^\prime \} = \{ \pi_{\bar c}, \bar c^\prime \} = i \delta^3$ yield
%**   W-gh-antigh   %%%%%%%%%%%%%%%%%%%%%%%%%%%%%%%%%%%%%%%%%%%%%%%%%%%%%%%%%
\begin{eqnarray}
\{ \dot{\bar c}, c^\prime \} = - \{ \dot c, \bar c^\prime \} 
= - \tilde f \phi^{-2} \delta^3.
\label{W-gh-antigh} 
\end{eqnarray} 
%%%%%%%%%%%%%%%%%%%%%%%%%%%%%%%%%%%%%%%%%%%%%%%%%%%%%%%%%%%%%%%%%%% 
Moreover, $[ \pi_g^{\alpha 0}, c^{\sigma\prime} ] = [ \pi_g^{\alpha 0}, \bar c_\lambda^\prime ] = 0$
give us the ETCRs:
%**   b-FP-ghosts   %%%%%%%%%%%%%%%%%%%%%%%%%%%%%%%%%%%%%%%%%%%%%%%%%%%%%%%%%
\begin{eqnarray}
[ b_\rho, c^{\sigma\prime} ] = [ b_\rho, \bar c^\prime_\lambda ] = 0, 
\label{b-FP-ghosts} 
\end{eqnarray} 
%%%%%%%%%%%%%%%%%%%%%%%%%%%%%%%%%%%%%%%%%%%%%%%%%%%%%%%%%%%%%%%%%%% 
and similarly $[ \pi_g^{\alpha 0}, c^\prime ] = [ \pi_g^{\alpha 0}, \bar c^\prime ] = 0$ produce
%**   b-W-FP-ghosts   %%%%%%%%%%%%%%%%%%%%%%%%%%%%%%%%%%%%%%%%%%%%%%%%%%%%%%%%%
\begin{eqnarray}
[ b_\rho, c^\prime ] = [ b_\rho, \bar c^\prime ] = 0.
\label{b-W-FP-ghosts} 
\end{eqnarray} 
%%%%%%%%%%%%%%%%%%%%%%%%%%%%%%%%%%%%%%%%%%%%%%%%%%%%%%%%%%%%%%%%%%% 
To calculate the ETCRs between $B$ and the FP-ghosts, it is necessary to utilize the
ETCRs, $[ B, \pi_{c \lambda}^\prime ] = [ B, \pi_{\bar c}^{\sigma\prime} ] = [ B, \pi_c^\prime ] 
= [ B, \pi_{\bar c}^\prime ] = 0$, and consequently we have
%**   B-FP   %%%%%%%%%%%%%%%%%%%%%%%%%%%%%%%%%%%%%%%%%%%%%%%%%%%%%%%%%
\begin{eqnarray}
[ B, \dot{\bar c}^\prime_\lambda ] = [ B, \dot c^{\sigma\prime} ] = [ B, \dot{\bar c}^\prime ] 
= [ B, \dot c^\prime ] =0.
\label{B-FP} 
\end{eqnarray} 
%%%%%%%%%%%%%%%%%%%%%%%%%%%%%%%%%%%%%%%%%%%%%%%%%%%%%%%%%%%%%%%%%%% 
Furthermore, taking the Weyl BRST transformation of the third ETCR reads\footnote{We define 
the BRST transformation for the Weyl transformation as $\bar \delta_B \Phi \equiv [ i \bar Q_B, \Phi \}$ 
where $\Phi$ is a generic field and $[ \hspace{2mm}, \hspace{2mm} \}$ denotes the graded bracket. 
Of course, in case of the GCT BRST transformation, it is replaced by $\delta_B \Phi \equiv [ i Q_B, \Phi \}$.}
%**   Q-WB   %%%%%%%%%%%%%%%%%%%%%%%%%%%%%%%%%%%%%%%%%%%%%%%%%%%%%%%%%
\begin{eqnarray}
0 = \{ [ i \bar Q_B, B], \dot{\bar c}^\prime \} + [ B, \{ i \bar Q_B, \dot{\bar c}^\prime \} ]
= [ B, i \dot B^\prime ],
\label{Q-WB} 
\end{eqnarray} 
%%%%%%%%%%%%%%%%%%%%%%%%%%%%%%%%%%%%%%%%%%%%%%%%%%%%%%%%%%%%%%%%%%% 
where we have used the Weyl BRST transformation (\ref{Weyl-BRST}). As a result we have the ETCR:
%**   B-dotB   %%%%%%%%%%%%%%%%%%%%%%%%%%%%%%%%%%%%%%%%%%%%%%%%%%%%%%%%%
\begin{eqnarray}
[ B, \dot B^\prime ] = 0.
\label{B-dotB} 
\end{eqnarray} 
%%%%%%%%%%%%%%%%%%%%%%%%%%%%%%%%%%%%%%%%%%%%%%%%%%%%%%%%%%%%%%%%%%% 

Next, from $[ \pi_B, \bar c^\prime_\lambda ] = [ \pi_B, c^{\sigma\prime} ] = 0$,
we find  
%**   dot-phi-bar-c-lambda   %%%%%%%%%%%%%%%%%%%%%%%%%%%%%%%%%%%%%%%%%%%%%%%%%%%%%%%%%
\begin{eqnarray}
[ \dot \phi, \bar c^\prime_\lambda ] = [ \dot \phi, c^{\sigma\prime} ] = 0.
\label{dot-phi-bar-c-lambda} 
\end{eqnarray} 
%%%%%%%%%%%%%%%%%%%%%%%%%%%%%%%%%%%%%%%%%%%%%%%%%%%%%%%%%%%%%%%%%%% 
Similarly, from $[ \pi_B, \bar c^\prime ] = [ \pi_B, c^\prime ] = 0$,
we have
%**   dot-phi-bar-c   %%%%%%%%%%%%%%%%%%%%%%%%%%%%%%%%%%%%%%%%%%%%%%%%%%%%%%%%%
\begin{eqnarray}
[ \dot \phi, \bar c^\prime ] = [ \dot \phi, c^\prime ] = 0.
\label{dot-phi-bar-c} 
\end{eqnarray} 
%%%%%%%%%%%%%%%%%%%%%%%%%%%%%%%%%%%%%%%%%%%%%%%%%%%%%%%%%%%%%%%%%%% 
Using the field equation for $\bar c_\lambda$ in (\ref{q-field-eq}), i.e., $g^{\mu\nu} \partial_\mu 
\partial_\nu \bar c_\lambda = 0$, the ETCR, $[ \phi, \bar c^\prime_\lambda ] = 0$, the formula (\ref{identity}),
and Eq. (\ref{dot-phi-bar-c-lambda}), it is easy to derive the equations:
%**   phi-bar-c-eqs   %%%%%%%%%%%%%%%%%%%%%%%%%%%%%%%%%%%%%%%%%%%%%%%%%%%%%%%%%
\begin{eqnarray}
[ \phi, \ddot {\bar c}^\prime_\lambda ] = [ \dot \phi, \dot {\bar c}^\prime_\lambda ] 
= [ \ddot \phi, \bar c^\prime_\lambda ] = 0.
\label{phi-bar-c-eqs} 
\end{eqnarray} 
%%%%%%%%%%%%%%%%%%%%%%%%%%%%%%%%%%%%%%%%%%%%%%%%%%%%%%%%%%%%%%%%%%% 
Similar equations also hold when $\bar c^\prime_\lambda$ is replaced with $c^{\sigma\prime}$,
$c^\prime$ or $\bar c^\prime$. 

Now, using the equations obtained above, we are ready to evaluate the type of the ETCRs, $[ \dot \Phi, b^\prime_\rho ]$
where $\Phi$ is a generic field. First, let us focus on $[ \dot \phi, b^\prime_\rho ]$. To do that, 
we start with $[ \dot \phi, \bar c^\prime_\rho ] = 0$ in Eq. (\ref{dot-phi-bar-c-lambda}) 
and take its BRST variation for the GCT as follows:
%**   Qdot-phi-b   %%%%%%%%%%%%%%%%%%%%%%%%%%%%%%%%%%%%%%%%%%%%%%%%%%%%%%%%%
\begin{eqnarray}
0 &=& \{ i Q_B, [ \dot \phi, \bar c^\prime_\rho ] \} 
\nonumber\\
&=& \{ [ i Q_B, \dot \phi ], \bar c^\prime_\rho \} + [ \dot \phi, \{ i Q_B, \bar c^\prime_\rho \} ]
\nonumber\\
&=& \{ - \partial_0 ( c^\lambda \partial_\lambda \phi ), \bar c^\prime_\rho \} 
+ [ \dot \phi, i ( b^\prime_\rho + i c^{\lambda\prime} \partial_\lambda \bar c^\prime_\rho ) ]
\nonumber\\
&=& - \{ \dot c^\lambda,  \bar c^\prime_\rho \} \partial_\lambda \phi + i [ \dot \phi, b^\prime_\rho ].
\label{Qdot-phi-b} 
\end{eqnarray} 
%%%%%%%%%%%%%%%%%%%%%%%%%%%%%%%%%%%%%%%%%%%%%%%%%%%%%%%%%%%%%%%%%%% 
Using Eq. (\ref{gh-antigh}), we are able to obtain
%**   dot-phi-b-eq   %%%%%%%%%%%%%%%%%%%%%%%%%%%%%%%%%%%%%%%%%%%%%%%%%%%%%%%%%
\begin{eqnarray}
[ \dot \phi, b^\prime_\rho ] = - i \tilde f \phi^{-2} \partial_\rho \phi \delta^3.
\label{dot-phi-b-eq} 
\end{eqnarray} 
%%%%%%%%%%%%%%%%%%%%%%%%%%%%%%%%%%%%%%%%%%%%%%%%%%%%%%%%%%%%%%%%%%% 
It turns out that the ETCRs, $[ \pi_{c \lambda}, \pi_g^{\alpha 0\prime} ] 
= [ \pi^\sigma_{\bar c}, \pi_g^{\alpha 0\prime} ] = 0$ give rise to
%**   dot-c-b-eq   %%%%%%%%%%%%%%%%%%%%%%%%%%%%%%%%%%%%%%%%%%%%%%%%%%%%%%%%%
\begin{eqnarray}
[ \dot{\bar c}_\lambda, b^\prime_\rho ] = - i \tilde f \phi^{-2} \partial_\rho \bar c_\lambda \delta^3, \qquad
[ \dot c^\sigma, b^\prime_\rho ] = - i \tilde f \phi^{-2} \partial_\rho c^\sigma \delta^3.
\label{dot-c-b-eq} 
\end{eqnarray} 
%%%%%%%%%%%%%%%%%%%%%%%%%%%%%%%%%%%%%%%%%%%%%%%%%%%%%%%%%%%%%%%%%%% 
Similarly,  the ETCRs, $[ \pi_c, \pi_g^{\alpha 0\prime} ] = [ \pi_{\bar c}, \pi_g^{\alpha 0\prime} ] = 0$ give us
%**   dot-Wc-b-eq   %%%%%%%%%%%%%%%%%%%%%%%%%%%%%%%%%%%%%%%%%%%%%%%%%%%%%%%%%
\begin{eqnarray}
[ \dot{\bar c}, b^\prime_\rho ] = - i \tilde f \phi^{-2} \partial_\rho \bar c \delta^3, \qquad
[ \dot c, b^\prime_\rho ] = - i \tilde f \phi^{-2} \partial_\rho c \delta^3.
\label{dot-Wc-b-eq} 
\end{eqnarray} 
%%%%%%%%%%%%%%%%%%%%%%%%%%%%%%%%%%%%%%%%%%%%%%%%%%%%%%%%%%%%%%%%%%% 
In order to evaluate $[ \dot B, b^\prime_\rho ]$, we make use of $[\pi_c, b^\prime_\rho ] = 0$, which
can be easily proved. Taking its BRST transformation for the Weyl transformation leads to the equation:
%**   Q-Wpi-b   %%%%%%%%%%%%%%%%%%%%%%%%%%%%%%%%%%%%%%%%%%%%%%%%%%%%%%%%%
\begin{eqnarray}
[ \{ i \bar Q_B, \pi_c \}, b^\prime_\rho ] = 0,
\label{Q-Wpi-b} 
\end{eqnarray} 
%%%%%%%%%%%%%%%%%%%%%%%%%%%%%%%%%%%%%%%%%%%%%%%%%%%%%%%%%%%%%%%%%%% 
where $[ i \bar Q_B, b^\prime_\rho ] = 0$ was used. We can show that $\{ i \bar Q_B, \pi_c \} =
\tilde g^{0 \mu} \phi^2 \partial_\mu B$, so using (\ref{3-g-b}) and (\ref{phi-b}), we can calculate
%**   B-b-CR   %%%%%%%%%%%%%%%%%%%%%%%%%%%%%%%%%%%%%%%%%%%%%%%%%%%%%%%%%
\begin{eqnarray}
[ \dot B, b^\prime_\rho ] = - i \tilde f \phi^{-2} \partial_\rho B \delta^3.
\label{B-b-CR} 
\end{eqnarray} 
%%%%%%%%%%%%%%%%%%%%%%%%%%%%%%%%%%%%%%%%%%%%%%%%%%%%%%%%%%%%%%%%%%% 
Finally, the ETCR, $[ \dot g_{\mu\nu}, b_\rho^\prime ]$ (or equivalently, $[ g_{\mu\nu}, \dot b_\rho^\prime ]$)
can be obtained by using the method developed in our previous article \cite{Oda-Q}. Only the result is written out as
%**   dot g-b   %%%%%%%%%%%%%%%%%%%%%%%%%%%%%%%%%%%%%%%%%%%%%%%%%%%%%%%%%
\begin{eqnarray}
[ \dot g_{\mu\nu}, b_\rho^\prime ] &=& - i \Bigl\{ \tilde f \phi^{-2} ( \partial_\rho g_{\mu\nu} 
+ \delta_\mu^0 \dot g_{\rho\nu} + \delta_\nu^0 \dot g_{\rho\mu} ) \delta^3 
\nonumber\\
&+& [ ( \delta_\mu^k - 2 \delta_\mu^0 \tilde f \tilde g^{0 k} ) g_{\rho\nu}
+ (\mu\leftrightarrow \nu) ] \partial_k ( \tilde f \phi^{-2} \delta^3 ) \Bigr\},
\label{dot g-b}  
\end{eqnarray}
%%%%%%%%%%%%%%%%%%%%%%%%%%%%%%%%%%%%%%%%%%%%%%%%%%%%%%%%%%%%%%%%%%% 
or equivalently,
%**   g-dot b   %%%%%%%%%%%%%%%%%%%%%%%%%%%%%%%%%%%%%%%%%%%%%%%%%%%%%%%%%
\begin{eqnarray}
[ g_{\mu\nu}, \dot b_\rho^\prime ] &=& i \Bigl\{ [ \tilde f \phi^{-2} \partial_\rho g_{\mu\nu} 
- \partial_0 ( \tilde f \phi^{-2} ) ( \delta_\mu^0 g_{\rho\nu} + \delta_\nu^0 g_{\rho\mu} ) ] \delta^3 
\nonumber\\
&+& [ ( \delta_\mu^k - 2 \delta_\mu^0 \tilde f \tilde g^{0k} ) g_{\rho\nu} + (\mu \leftrightarrow \nu) ]
\partial_k (\tilde f \phi^{-2} \delta^3) \Bigr\}.
\label{g-dot b}  
\end{eqnarray}
%%%%%%%%%%%%%%%%%%%%%%%%%%%%%%%%%%%%%%%%%%%%%%%%%%%%%%%%%%%%%%%%%%% 
Following our previous calculation \cite{Oda-Q}, we can prove
%**   b-b   %%%%%%%%%%%%%%%%%%%%%%%%%%%%%%%%%%%%%%%%%%%%%%%%%%%%%%%%%
\begin{eqnarray}
&{}& [ b_\mu, b_\nu^\prime ] = 0,  
\nonumber\\
&{}& [ b_\mu, \dot b_\nu^\prime ] = i \tilde f \phi^{-2} ( \partial_\mu b_\nu + \partial_\nu b_\mu ) \delta^3. 
\label{b-b}  
\end{eqnarray}
%%%%%%%%%%%%%%%%%%%%%%%%%%%%%%%%%%%%%%%%%%%%%%%%%%%%%%%%%%%%%%%%%%% 

\section{Equal-time commutation relations in gravitational sector}

The remaining nontrivial ETCRs are related to the time derivative of the metric field, i.e.,
the ETCRs, $[ \dot g_{\mu\nu}, \Phi^\prime ]$ where $\Phi$ is a generic field. In this section,
we will evaluate such the ETCRs. 

First of all, let us start with the ETCR, $[ \pi_\phi, g_{\mu\nu}^\prime ] = 0$. From the
expression of $\pi_\phi$ in Eq. (\ref{CCM}), this ETCR can be described as
%**   pi-phi-g   %%%%%%%%%%%%%%%%%%%%%%%%%%%%%%%%%%%%%%%%%%%%%%%%%%%%%%%%%
\begin{eqnarray}
&{}& \tilde g^{00} [ \dot \phi, g_{\mu\nu}^\prime ] + \frac{1}{6} \phi ( \tilde g^{00} g^{\rho\sigma} 
- \tilde g^{0\rho} g^{0\sigma} ) [ \dot g_{\rho\sigma}, g_{\mu\nu}^\prime ]
+ \tilde g^{00} \phi [ \dot B, g_{\mu\nu}^\prime ] 
\nonumber\\
&{}& = - 4 i \tilde f \phi^{-1} \sqrt{-g} 
\delta^0_\mu \delta^0_\nu \delta^3. 
\label{pi-phi-g}  
\end{eqnarray}
%%%%%%%%%%%%%%%%%%%%%%%%%%%%%%%%%%%%%%%%%%%%%%%%%%%%%%%%%%%%%%%%%%% 
Next, the ETCR, $[ \pi_\phi, \phi^\prime ] = - i \delta^3$ produces the equation:
%**   pi-phi-phi   %%%%%%%%%%%%%%%%%%%%%%%%%%%%%%%%%%%%%%%%%%%%%%%%%%%%%%%%%
\begin{eqnarray}
( \tilde g^{00} g^{\rho\sigma} - \tilde g^{0\rho} g^{0\sigma} ) [ \dot g_{\rho\sigma}, \phi^\prime ] = 0.
\label{pi-phi-phi}  
\end{eqnarray}
%%%%%%%%%%%%%%%%%%%%%%%%%%%%%%%%%%%%%%%%%%%%%%%%%%%%%%%%%%%%%%%%%%% 
Moreover, the ETCR, $[ \pi_\phi, B^\prime ] = 0$ reads
%**   pi-phi-B   %%%%%%%%%%%%%%%%%%%%%%%%%%%%%%%%%%%%%%%%%%%%%%%%%%%%%%%%%
\begin{eqnarray}
( \tilde g^{00} g^{\rho\sigma} - \tilde g^{0\rho} g^{0\sigma} ) [ \dot g_{\rho\sigma}, B^\prime ] 
= 6 i \phi^{-2} \delta^3.
\label{pi-phi-B}  
\end{eqnarray}
%%%%%%%%%%%%%%%%%%%%%%%%%%%%%%%%%%%%%%%%%%%%%%%%%%%%%%%%%%%%%%%%%%% 

The extended de Donder gauge, $\partial_\mu ( \tilde g^{\mu\nu} \phi^2 ) = 0$, can be
rewritten as 
%**   Ext-de-Donder1   %%%%%%%%%%%%%%%%%%%%%%%%%%%%%%%%%%%%%%%%%%%%%%%%%%%%%%%%%
\begin{eqnarray}
{\cal D}^{\lambda\rho\sigma} \dot g_{\rho\sigma} + 4 \phi^{-1} g^{\lambda\rho} \partial_\rho \phi
= ( 2 g^{\lambda\rho} g^{\sigma k} - g^{\rho\sigma} g^{\lambda k} ) \partial_k g_{\rho\sigma},
\label{Ext-de-Donder1}  
\end{eqnarray}
%%%%%%%%%%%%%%%%%%%%%%%%%%%%%%%%%%%%%%%%%%%%%%%%%%%%%%%%%%%%%%%%%%% 
where ${\cal D}^{\lambda\rho\sigma} \equiv g^{0\lambda} g^{\rho\sigma} - 2 g^{\lambda\rho} g^{0\sigma}$.
Since the right-hand side (RHS) of Eq. (\ref{Ext-de-Donder1}) is independent of $\dot g_{\mu\nu}$, it
commutes with $g_{\mu\nu}, \phi$ or $B$. Thus, we have three identities:
%**   Ext-de-Donder-g   %%%%%%%%%%%%%%%%%%%%%%%%%%%%%%%%%%%%%%%%%%%%%%%%%%%%%%%%%
\begin{eqnarray}
{\cal D}^{\lambda\rho\sigma} [ \dot g_{\rho\sigma},  g_{\mu\nu}^\prime ] + 4 \phi^{-1} g^{\lambda 0} 
[ \dot \phi, g_{\mu\nu}^\prime ] = 0.
\label{Ext-de-Donder-g}  
\end{eqnarray}
%%%%%%%%%%%%%%%%%%%%%%%%%%%%%%%%%%%%%%%%%%%%%%%%%%%%%%%%%%%%%%%%%%% 
%**   Ext-de-Donder-phi   %%%%%%%%%%%%%%%%%%%%%%%%%%%%%%%%%%%%%%%%%%%%%%%%%%%%%%%%%
\begin{eqnarray}
{\cal D}^{\lambda\rho\sigma} [ \dot g_{\rho\sigma},  \phi^\prime ] = 0.
\label{Ext-de-Donder-phi}  
\end{eqnarray}
%%%%%%%%%%%%%%%%%%%%%%%%%%%%%%%%%%%%%%%%%%%%%%%%%%%%%%%%%%%%%%%%%%% 
%**   Ext-de-Donder-B   %%%%%%%%%%%%%%%%%%%%%%%%%%%%%%%%%%%%%%%%%%%%%%%%%%%%%%%%%
\begin{eqnarray}
{\cal D}^{\lambda\rho\sigma} [ \dot g_{\rho\sigma},  B^\prime ] = 4 i \tilde f \phi^{-2} g^{\lambda 0} \delta^3.
\label{Ext-de-Donder-B}  
\end{eqnarray}
%%%%%%%%%%%%%%%%%%%%%%%%%%%%%%%%%%%%%%%%%%%%%%%%%%%%%%%%%%%%%%%%%%% 
In Eqs. (\ref{Ext-de-Donder-phi}) and (\ref{Ext-de-Donder-B}), we have used Eq. (\ref{2-phi}).

Putting $\lambda = 0$ in Eq. (\ref{Ext-de-Donder-phi}) and using Eq. (\ref{pi-phi-phi}), we have
%**   Ext-de-Donder-phi2   %%%%%%%%%%%%%%%%%%%%%%%%%%%%%%%%%%%%%%%%%%%%%%%%%%%%%%%%%
\begin{eqnarray}
g^{\rho\sigma} [ \dot g_{\rho\sigma},  \phi^\prime ] = g^{0 \rho} g^{0 \sigma} [ \dot g_{\rho\sigma},  \phi^\prime ] = 0.
\label{Ext-de-Donder-phi2}  
\end{eqnarray}
%%%%%%%%%%%%%%%%%%%%%%%%%%%%%%%%%%%%%%%%%%%%%%%%%%%%%%%%%%%%%%%%%%% 
In general, from the argument of symmetry, $[ \dot g_{\rho\sigma}, \phi^\prime ]$ must be of form:
%**   d-g-p0   %%%%%%%%%%%%%%%%%%%%%%%%%%%%%%%%%%%%%%%%%%%%%%%%%%%%%%%%%
\begin{eqnarray}
[ \dot g_{\rho\sigma},  \phi^\prime ] = a_1 ( g_{\rho\sigma} + a_2 \delta^0_\rho \delta^0_\sigma ) \delta^3,
\label{d-g-p0}  
\end{eqnarray}
%%%%%%%%%%%%%%%%%%%%%%%%%%%%%%%%%%%%%%%%%%%%%%%%%%%%%%%%%%%%%%%%%%% 
where $a_1, a_2$ are constants. Eq. (\ref{Ext-de-Donder-phi2}) then requires us to take $a_1 = a_2 = 0$. Thus, we have
%**   d-g-p   %%%%%%%%%%%%%%%%%%%%%%%%%%%%%%%%%%%%%%%%%%%%%%%%%%%%%%%%%
\begin{eqnarray}
[ \dot g_{\rho\sigma},  \phi^\prime ] = 0.
\label{d-g-p}  
\end{eqnarray}
%%%%%%%%%%%%%%%%%%%%%%%%%%%%%%%%%%%%%%%%%%%%%%%%%%%%%%%%%%%%%%%%%%% 
Next, in a similar manner, we can set 
%**   d-g-B0   %%%%%%%%%%%%%%%%%%%%%%%%%%%%%%%%%%%%%%%%%%%%%%%%%%%%%%%%%
\begin{eqnarray}
[ \dot g_{\rho\sigma},  B^\prime ] = b_1 ( g_{\rho\sigma} + b_2 \delta^0_\rho \delta^0_\sigma ) \delta^3,
\label{d-g-B0}  
\end{eqnarray}
%%%%%%%%%%%%%%%%%%%%%%%%%%%%%%%%%%%%%%%%%%%%%%%%%%%%%%%%%%%%%%%%%%% 
where $b_1, b_2$ are constants. From Eq. (\ref{pi-phi-B}), $b_1$ is determined to be $2 i \tilde f \phi^{-2}$,
and then Eq. (\ref{Ext-de-Donder-B}) requires $b_2$ to be vanishing, so we can obtain
%**   d-g-B   %%%%%%%%%%%%%%%%%%%%%%%%%%%%%%%%%%%%%%%%%%%%%%%%%%%%%%%%%
\begin{eqnarray}
[ \dot g_{\rho\sigma},  B^\prime ] = 2 i \tilde f \phi^{-2} g_{\rho\sigma} \delta^3.
\label{d-g-B}  
\end{eqnarray}
%%%%%%%%%%%%%%%%%%%%%%%%%%%%%%%%%%%%%%%%%%%%%%%%%%%%%%%%%%%%%%%%%%% 

Finally, we wish to evaluate $[ \dot g_{\rho\sigma},  g^\prime_{\mu\nu} ]$, for which we need some calculations.
Before doing so, let us rewrite Eq. (\ref{pi-phi-g}) by means of Eqs. (\ref{d-g-p}) and (\ref{d-g-B}) into the form:
%**   pi-phi-g2   %%%%%%%%%%%%%%%%%%%%%%%%%%%%%%%%%%%%%%%%%%%%%%%%%%%%%%%%%
\begin{eqnarray}
( \tilde g^{00} g^{\rho\sigma} - \tilde g^{0\rho} g^{0\sigma} ) [ \dot g_{\rho\sigma}, g_{\mu\nu}^\prime ]
= - 12 i \phi^{-2} \left( g_{\mu\nu} + \frac{2}{g^{00}} \delta^0_\mu \delta^0_\nu \right) \delta^3. 
\label{pi-phi-g2}  
\end{eqnarray}
%%%%%%%%%%%%%%%%%%%%%%%%%%%%%%%%%%%%%%%%%%%%%%%%%%%%%%%%%%%%%%%%%%% 
Similarly, Eq. (\ref{Ext-de-Donder-g}) reduces to 
%**   Ext-de-Donder-g2   %%%%%%%%%%%%%%%%%%%%%%%%%%%%%%%%%%%%%%%%%%%%%%%%%%%%%%%%%
\begin{eqnarray}
( g^{0\lambda} g^{\rho\sigma} - 2 g^{\lambda\rho} g^{0\sigma} ) [ \dot g_{\rho\sigma}, g_{\mu\nu}^\prime ] = 0.
\label{Ext-de-Donder-g2}  
\end{eqnarray}
%%%%%%%%%%%%%%%%%%%%%%%%%%%%%%%%%%%%%%%%%%%%%%%%%%%%%%%%%%%%%%%%%%% 

We are now willing to evaluate the ETCR, $[ \dot g_{\rho\sigma}, g_{\mu\nu}^\prime ]$. This ETCR 
has a symmetry under the simultaneous exchange of $(\mu\nu) \leftrightarrow (\rho\sigma)$ and primed 
$\leftrightarrow$ unprimed in addition to the usual symmetry $\mu \leftrightarrow \nu$ and $\rho \leftrightarrow \sigma$. 
Then, we can write down its general expression like
%**   dot-G-G   %%%%%%%%%%%%%%%%%%%%%%%%%%%%%%%%%%%%%%%%%%%%%%%%%%%%%%%%%
\begin{eqnarray}
[ \dot g_{\rho\sigma}, g_{\mu\nu}^\prime ] &=& \biggl\{ c_1 g_{\rho\sigma} g_{\mu\nu} + c_2 ( g_{\rho\mu} g_{\sigma\nu}
+ g_{\rho\nu} g_{\sigma\mu} )
\nonumber\\
&+& h \tilde f \Bigl[ c_3 ( \delta_\rho^0 \delta_\sigma^0 g_{\mu\nu} + \delta_\mu^0 \delta_\nu^0 g_{\rho\sigma} )
+ c_4 ( \delta_\rho^0 \delta_\mu^0 g_{\sigma\nu} + \delta_\rho^0 \delta_\nu^0 g_{\sigma\mu} 
\nonumber\\
&+& \delta_\sigma^0 \delta_\mu^0 g_{\rho\nu} + \delta_\sigma^0 \delta_\nu^0 g_{\rho\mu} ) \Bigr]
+ ( h \tilde f )^2 c_5 \delta_\rho^0 \delta_\sigma^0 \delta_\mu^0 \delta_\nu^0 \biggr\} \delta^3,  
\label{dot-G-G}  
\end{eqnarray}
%%%%%%%%%%%%%%%%%%%%%%%%%%%%%%%%%%%%%%%%%%%%%%%%%%%%%%%%%%%%%%%%%%% 
where $c_i ( i = 1, \cdots, 5)$ are some coefficients. 
Imposing Eq. (\ref{Ext-de-Donder-g2}) on (\ref{dot-G-G}) leads to relations among the coefficients:
%**   c-relation   %%%%%%%%%%%%%%%%%%%%%%%%%%%%%%%%%%%%%%%%%%%%%%%%%%%%%%%%%
\begin{eqnarray}
c_3 = 2 ( c_1 + c_2 ), \qquad c_4 = - c_2, \qquad c_5 = 4 ( c_1 + c_2 ).
\label{c-relation}  
\end{eqnarray}
%%%%%%%%%%%%%%%%%%%%%%%%%%%%%%%%%%%%%%%%%%%%%%%%%%%%%%%%%%%%%%%%%%% 
Furthermore, imposing Eq. (\ref{pi-phi-g2}), we can determine $c_2, c_3, c_4$ and $c_5$ via $c_1$ as
%**   Rel of c   %%%%%%%%%%%%%%%%%%%%%%%%%%%%%%%%%%%%%%%%%%%%%%%%%%%%%%%%%
\begin{eqnarray}
&{}& c_3 = - c_1 - 12 i \tilde f \phi^{-2}, \qquad
c_4 = - c_2 = \frac{3}{2} c_1 + 6 i \tilde f \phi^{-2}, 
\nonumber\\
&{}& c_5 = -2 c_1 - 24 i \tilde f \phi^{-2}.
\label{Rel of c}  
\end{eqnarray}
%%%%%%%%%%%%%%%%%%%%%%%%%%%%%%%%%%%%%%%%%%%%%%%%%%%%%%%%%%%%%%%%%%% 

In order to fix the coefficient $c_1$, we need to calculate the ETCR, $[ \dot g_{kl}, g_{mn}^\prime ]$ 
explicitly in terms of $[ \pi_g^{kl}, g_{mn}^\prime  ] = - i \frac{1}{2} ( \delta_m^k \delta_n^l 
+ \delta_m^l \delta_n^k) \delta^3$ in Eq. (\ref{CCRs}) and the concrete expression of $\pi_g^{kl}$ 
in Eq. (\ref{CCM}). To do that, from Eq. (\ref{CCM}), let us write
%**   Pi-G-CCM   %%%%%%%%%%%%%%%%%%%%%%%%%%%%%%%%%%%%%%%%%%%%%%%%%%%%%%%%%
\begin{eqnarray}
\pi_g^{kl} = \hat A^{kl} + \hat B^{kl\rho} b_\rho + \hat C^{klmn} \dot g_{mn} + \hat D^{kl} \dot \phi.
\label{Pi-G-CCM}  
\end{eqnarray}
%%%%%%%%%%%%%%%%%%%%%%%%%%%%%%%%%%%%%%%%%%%%%%%%%%%%%%%%%%%%%%%%%%% 
Here $\hat A^{kl}, \hat B^{kl\rho}, \hat C^{klmn}$ and $\hat D^{kl}$ commute with $g_{mn}$, and $\hat C^{klmn}$
and $\hat D^{kl}$ are defined as\footnote{It turns out that the concrete expressions of $\hat A^{kl}$ and 
$\hat B^{kl\rho}$ are irrelevant to the calculation of $[ \dot g_{kl}, g_{mn}^\prime ]$.}
%**   Pi-G-CCM2   %%%%%%%%%%%%%%%%%%%%%%%%%%%%%%%%%%%%%%%%%%%%%%%%%%%%%%%%%
\begin{eqnarray}
\hat C^{klmn} = \frac{1}{24} h \phi^2 K^{klmn}, \qquad
\hat D^{kl} = \frac{1}{6} \phi ( \tilde g^{00} g^{kl} - \tilde g^{0k} g^{0l} ),
\label{Pi-G-CCM2}  
\end{eqnarray}
%%%%%%%%%%%%%%%%%%%%%%%%%%%%%%%%%%%%%%%%%%%%%%%%%%%%%%%%%%%%%%%%%%% 
where the definition of $K^{klmn}$ and its property are given by
%**   Pi-G-CCM3   %%%%%%%%%%%%%%%%%%%%%%%%%%%%%%%%%%%%%%%%%%%%%%%%%%%%%%%%%
\begin{eqnarray}
&{}& K^{klmn} = \left|
\begin{array}{rrr}
g^{00} & g^{0l} & g^{0n} \\
g^{k0} & g^{kl} & g^{kn} \\
g^{m0} & g^{ml} & g^{mn} \\
\end{array}
\right|,
\nonumber\\
&{}& K^{klmn} \frac{1}{2} (g^{00})^{-1} ( g_{ij} g_{mn} - g_{im} g_{jn} - g_{in} g_{jm} )
= \frac{1}{2} ( \delta_i^k \delta_j^l + \delta_i^l \delta_j^k ).
\label{Pi-G-CCM3}  
\end{eqnarray}
%%%%%%%%%%%%%%%%%%%%%%%%%%%%%%%%%%%%%%%%%%%%%%%%%%%%%%%%%%%%%%%%%%% 

From Eq. (\ref{Pi-G-CCM}), we can calculate 
%**   Pi-G-CCM4   %%%%%%%%%%%%%%%%%%%%%%%%%%%%%%%%%%%%%%%%%%%%%%%%%%%%%%%%%
\begin{eqnarray}
[ \dot g_{kl}, g_{mn}^\prime ] &=& \hat C^{-1}_{klpq} \left( [ \pi_g^{pq}, g_{mn}^\prime ] 
-  \hat B^{pq\rho} [ b_\rho, g_{mn}^\prime ] -  \hat D^{pq} [ \dot \phi, g_{mn}^\prime ] \right).
\nonumber\\
&=& - i \frac{1}{2} \hat C^{-1}_{klpq} ( \delta^p_m \delta^q_n + \delta^q_m \delta^p_n ) \delta^3,
\label{Pi-G-CCM4}  
\end{eqnarray}
%%%%%%%%%%%%%%%%%%%%%%%%%%%%%%%%%%%%%%%%%%%%%%%%%%%%%%%%%%%%%%%%%%% 
where we have used Eqs. (\ref{CCRs}), (\ref{g-b}), and (\ref{d-g-p}). Since we can calculate 
%**   C-inverse   %%%%%%%%%%%%%%%%%%%%%%%%%%%%%%%%%%%%%%%%%%%%%%%%%%%%%%%%%
\begin{eqnarray}
\hat C^{-1}_{klpq} = 12 \tilde f \phi^{-2} ( g_{kl} g_{pq} - g_{kp} g_{lq} - g_{kq} g_{lp} ),
\label{C-inverse}  
\end{eqnarray}
%%%%%%%%%%%%%%%%%%%%%%%%%%%%%%%%%%%%%%%%%%%%%%%%%%%%%%%%%%%%%%%%%%% 
we can eventually arrive at the result:
%**   Pi-G-CCM-final   %%%%%%%%%%%%%%%%%%%%%%%%%%%%%%%%%%%%%%%%%%%%%%%%%%%%%%%%%
\begin{eqnarray}
[ \dot g_{kl}, g_{mn}^\prime ] = - 12 i \tilde f \phi^{-2} ( g_{kl} g_{mn} - g_{km} g_{ln} - g_{kn} g_{lm} )
\delta^3.
\label{Pi-G-CCM-final}  
\end{eqnarray}
%%%%%%%%%%%%%%%%%%%%%%%%%%%%%%%%%%%%%%%%%%%%%%%%%%%%%%%%%%%%%%%%%%% 
Meanwhile, from Eq. (\ref{dot-G-G}) we have the ETCR:
%**   Pi-G-CCM-com   %%%%%%%%%%%%%%%%%%%%%%%%%%%%%%%%%%%%%%%%%%%%%%%%%%%%%%%%%
\begin{eqnarray}
[ \dot g_{kl}, g_{mn}^\prime ] = \left[ c_1 g_{kl} g_{mn} + c_2 ( g_{km} g_{ln} + g_{kn} g_{lm} ) \right] \delta^3
\label{Pi-G-CCM-com}  
\end{eqnarray}
%%%%%%%%%%%%%%%%%%%%%%%%%%%%%%%%%%%%%%%%%%%%%%%%%%%%%%%%%%%%%%%%%%% 
Hence, comparing (\ref{Pi-G-CCM-final}) with (\ref{Pi-G-CCM-com}), we can obtain 
%**   c1-c2   %%%%%%%%%%%%%%%%%%%%%%%%%%%%%%%%%%%%%%%%%%%%%%%%%%%%%%%%%
\begin{eqnarray}
c_1 = - 12 i \tilde f \phi^{-2}, \qquad
c_2 = 12 i \tilde f \phi^{-2}.
\label{c1-c2}  
\end{eqnarray}
%%%%%%%%%%%%%%%%%%%%%%%%%%%%%%%%%%%%%%%%%%%%%%%%%%%%%%%%%%%%%%%%%%% 
Note that these values satisfy the relation in Eq. (\ref{Rel of c}), $- c_2 = \frac{3}{2} c_1 
+ 6 i \tilde f \phi^{-2}$, which gives us a nontrivial verification of our result.
In this way, we have succeeded in getting the following ETCR:
%**   ETCR-final   %%%%%%%%%%%%%%%%%%%%%%%%%%%%%%%%%%%%%%%%%%%%%%%%%%%%%%%%%
\begin{eqnarray}
[ \dot g_{\rho\sigma}, g_{\mu\nu}^\prime ] &=& -12 i \tilde f \phi^{-2} 
[ g_{\rho\sigma} g_{\mu\nu} - g_{\rho\mu} g_{\sigma\nu} - g_{\rho\nu} g_{\sigma\mu} 
+ h \tilde f ( \delta_\rho^0 \delta_\mu^0 g_{\sigma\nu} 
\nonumber\\
&+& \delta_\rho^0 \delta_\nu^0 g_{\sigma\mu}  + \delta_\sigma^0 \delta_\mu^0 g_{\rho\nu} 
+ \delta_\sigma^0 \delta_\nu^0 g_{\rho\mu} ) ] 
\delta^3.
\label{ETCR-final}  
\end{eqnarray}
%%%%%%%%%%%%%%%%%%%%%%%%%%%%%%%%%%%%%%%%%%%%%%%%%%%%%%%%%%%%%%%%%%% 

\section{Unitarity of physical S-matrix}

As in the conventional BRST formalism, the physical state $| \rm{phys} \rangle$ is defined by
imposing two subsidiary conditions \cite{Kugo-Ojima}:
%**   Phys-state   %%%%%%%%%%%%%%%%%%%%%%%%%%%%%%%%%%%%%%%%%%%%%%%%%%%%%%%%%
\begin{eqnarray}
Q_B | \rm{phys} \rangle = \bar Q_B | \rm{phys} \rangle = 0.
\label{Phys-state}  
\end{eqnarray}
%%%%%%%%%%%%%%%%%%%%%%%%%%%%%%%%%%%%%%%%%%%%%%%%%%%%%%%%%%%%%%%%%%% 
It is then well known that the physical S-matrix is unitary under the assumption that all the
BRST singlet states have positive norm. In this section, we would like to prove the unitarity of the
physical S-matrix. Since there is a ghost-like scalar field $\phi$ as well as timelike and longitudinal 
components of the metric field in our formalism, this is not a trivial problem.  

In analysing the unitarity, it is enough to take account of asymptotic fields of all the fundamental
fields and the free part of the Lagrangian. Let us first assume the asymptotic fields as
%**   Asmp-exp   %%%%%%%%%%%%%%%%%%%%%%%%%%%%%%%%%%%%%%%%%%%%%%%%%%%%%%%%%
\begin{eqnarray}
g_{\mu\nu} &=& \eta_{\mu\nu} + \varphi_{\mu\nu},  \qquad
\phi = \phi_0 + \tilde \phi, \qquad
b_\mu = \beta_\mu, \qquad
B = \beta,
\nonumber\\
c^\mu &=& \gamma^\mu, \qquad
\bar c_\mu = \bar \gamma_\mu, \qquad
c = \gamma, \qquad
\bar c = \bar \gamma,
\label{Asmp-exp}  
\end{eqnarray}
%%%%%%%%%%%%%%%%%%%%%%%%%%%%%%%%%%%%%%%%%%%%%%%%%%%%%%%%%%%%%%%%%%% 
where $\eta_{\mu\nu} ( = \eta^{\mu\nu} )$ is the flat Minkowski metric with the mostly positive signature 
and $\phi_0$ is a constant. In this section, the Minkowski metric is used to lower or raise the 
Lorentz indices. Using these asymptotic fields, the free part of the Lagrangian reads
%**   Free-Lag   %%%%%%%%%%%%%%%%%%%%%%%%%%%%%%%%%%%%%%%%%%%%%%%%%%%%%%%%%
\begin{eqnarray}
{\cal L}_q &=& \frac{1}{12} \phi_0^2 \left( \frac{1}{4} \varphi_{\mu\nu} \Box \varphi^{\mu\nu} 
- \frac{1}{4} \varphi \Box \varphi - \frac{1}{2} \varphi^{\mu\nu} \partial_\mu \partial_\rho \varphi_\nu{}^\rho
+ \frac{1}{2} \varphi^{\mu\nu} \partial_\mu \partial_\nu \varphi \right)
\nonumber\\
&+& \frac{1}{6} \phi_0 \tilde \phi \left( - \Box \varphi + \partial_\mu \partial_\nu \varphi^{\mu\nu} \right)
+ \frac{1}{2} \partial_\mu \tilde \phi \partial^\mu \tilde \phi
- i \phi_0^2 \partial_\mu \bar \gamma_\rho \partial^\mu \gamma^\rho
\nonumber\\
&-& \left( 2 \eta^{\mu\nu} \phi_0 \tilde \phi - \phi_0^2 \varphi^{\mu\nu} + \frac{1}{2} 
\phi_0^2 \eta^{\mu\nu} \varphi \right) \partial_\mu \beta_\nu
\nonumber\\
&+& \phi_0 \partial_\mu \beta \partial^\mu \tilde \phi - i \phi_0^2 \partial_\mu \bar \gamma \partial^\mu \gamma,
\label{Free-Lag}  
\end{eqnarray}
%%%%%%%%%%%%%%%%%%%%%%%%%%%%%%%%%%%%%%%%%%%%%%%%%%%%%%%%%%%%%%%%%%% 
where $\Box \equiv \eta^{\mu\nu} \partial_\mu \partial_\nu$ and $\varphi \equiv \eta^{\mu\nu} \varphi_{\mu\nu}$.
Based on this Lagrangian, it is easy to derive the linearized field equations: 
%**   Linear-Eq   %%%%%%%%%%%%%%%%%%%%%%%%%%%%%%%%%%%%%%%%%%%%%%%%%%%%%%%%%
\begin{eqnarray}
&{}& \frac{1}{12} \phi_0 \biggl( \frac{1}{2} \Box \varphi_{\mu\nu} - \frac{1}{2} \eta_{\mu\nu} \Box \varphi 
- \partial_\rho \partial_{(\mu} \varphi_{\nu)}{}^\rho + \frac{1}{2} \partial_\mu \partial_\nu \varphi
+ \frac{1}{2} \eta_{\mu\nu} \partial_\rho \partial_\sigma \varphi^{\rho\sigma} \biggr)
\nonumber\\
&{}& + \frac{1}{6} \left( - \eta_{\mu\nu} \Box + \partial_\mu \partial_\nu \right) \tilde \phi 
+\phi_0 \partial_{(\mu} \beta_{\nu)} - \frac{1}{2} \phi_0 \eta_{\mu\nu} \partial_\rho \beta^\rho = 0.
\label{Linear-Eq1}
\\
&{}& \frac{1}{6} ( \Box \varphi - \partial_\mu \partial_\nu \varphi^{\mu\nu} )
+ 2 \partial_\rho \beta^\rho + \Box \beta = 0.
\label{Linear-Eq2}
\\
&{}& \partial_\mu \tilde \phi - \frac{1}{2} \phi_0 ( \partial^\nu \varphi_{\mu\nu} 
- \frac{1}{2} \partial_\mu \varphi ) = 0.
\label{Linear-Eq3}
\\
&{}& \Box \tilde \phi = \Box \gamma^\mu =  \Box \bar \gamma_\mu = \Box \gamma 
= \Box \bar \gamma = 0. 
\label{Linear-Eq4}  
\end{eqnarray}
%%%%%%%%%%%%%%%%%%%%%%%%%%%%%%%%%%%%%%%%%%%%%%%%%%%%%%%%%%%%%%%%%%% 
Here we have introduced the symmetrization notation $A_{(\mu} B_{\nu)} \equiv \frac{1}{2} ( A_\mu
B_\nu + A_\nu B_\mu )$. 
Now, operating $\partial^\mu$ on Eq. (\ref{Linear-Eq3}) and using Eq. (\ref{Linear-Eq4}), we can obtain
%**   Linear-Eq5   %%%%%%%%%%%%%%%%%%%%%%%%%%%%%%%%%%%%%%%%%%%%%%%%%%%%%%%%%
\begin{eqnarray}
\partial_\mu \partial_\nu \varphi^{\mu\nu} - \frac{1}{2} \Box \varphi = 0.
\label{Linear-Eq5}  
\end{eqnarray}
%%%%%%%%%%%%%%%%%%%%%%%%%%%%%%%%%%%%%%%%%%%%%%%%%%%%%%%%%%%%%%%%%%% 
Next, taking the trace of Eq. (\ref{Linear-Eq1}) with the help of Eqs. (\ref{Linear-Eq4}) and 
(\ref{Linear-Eq5}) leads to
%**   Linear-Eq6   %%%%%%%%%%%%%%%%%%%%%%%%%%%%%%%%%%%%%%%%%%%%%%%%%%%%%%%%%
\begin{eqnarray}
\Box \varphi + 24 \partial_\rho \beta^\rho = 0.
\label{Linear-Eq6}  
\end{eqnarray}
%%%%%%%%%%%%%%%%%%%%%%%%%%%%%%%%%%%%%%%%%%%%%%%%%%%%%%%%%%%%%%%%%%% 
Then, with the help of Eqs. (\ref{Linear-Eq5}) ans (\ref{Linear-Eq6}), Eq. (\ref{Linear-Eq2}) can be rewritten as
%**   Linear-Eq7   %%%%%%%%%%%%%%%%%%%%%%%%%%%%%%%%%%%%%%%%%%%%%%%%%%%%%%%%%
\begin{eqnarray}
\Box \beta = 0.
\label{Linear-Eq7}  
\end{eqnarray}
%%%%%%%%%%%%%%%%%%%%%%%%%%%%%%%%%%%%%%%%%%%%%%%%%%%%%%%%%%%%%%%%%%% 
Moreover, acting $\partial^\mu$ on Eq. (\ref{Linear-Eq1}) yields
%**   Linear-Eq8   %%%%%%%%%%%%%%%%%%%%%%%%%%%%%%%%%%%%%%%%%%%%%%%%%%%%%%%%%
\begin{eqnarray}
\Box \beta_\mu = 0.
\label{Linear-Eq8}  
\end{eqnarray}
%%%%%%%%%%%%%%%%%%%%%%%%%%%%%%%%%%%%%%%%%%%%%%%%%%%%%%%%%%%%%%%%%%% 
Finally, using various equations obtained thus far, Eq. (\ref{Linear-Eq1}) is reduced to the form:
%**   Linear-Eq9   %%%%%%%%%%%%%%%%%%%%%%%%%%%%%%%%%%%%%%%%%%%%%%%%%%%%%%%%%
\begin{eqnarray}
\Box \varphi_{\mu\nu} + 24 \partial_{(\mu} \beta_{\nu)} = 0,
\label{Linear-Eq9}  
\end{eqnarray}
%%%%%%%%%%%%%%%%%%%%%%%%%%%%%%%%%%%%%%%%%%%%%%%%%%%%%%%%%%%%%%%%%%% 
which means that the field $\varphi_{\mu\nu}$ is a dipole field:
%**   Linear-Eq10   %%%%%%%%%%%%%%%%%%%%%%%%%%%%%%%%%%%%%%%%%%%%%%%%%%%%%%%%%
\begin{eqnarray}
\Box^2 \varphi_{\mu\nu} = 0.
\label{Linear-Eq10}  
\end{eqnarray}
%%%%%%%%%%%%%%%%%%%%%%%%%%%%%%%%%%%%%%%%%%%%%%%%%%%%%%%%%%%%%%%%%%% 
On the other hand, the other fields are all simple pole fields:
%**   Linear-Eq11   %%%%%%%%%%%%%%%%%%%%%%%%%%%%%%%%%%%%%%%%%%%%%%%%%%%%%%%%%
\begin{eqnarray}
\Box \tilde \phi = \Box \beta_\mu = \Box \beta =\Box \gamma^\mu =  \Box \bar \gamma_\mu 
= \Box \gamma = \Box \bar \gamma = 0. 
\label{Linear-Eq11}  
\end{eqnarray}
%%%%%%%%%%%%%%%%%%%%%%%%%%%%%%%%%%%%%%%%%%%%%%%%%%%%%%%%%%%%%%%%%%% 
Note that Eq. (\ref{Linear-Eq11}) corresponds to Eq. (\ref{X-M-eq}) in a curved space-time.

Following the standard technique, we can calculate the four-dimensional (anti-)commutation 
relations (4D CRs) between asymptotic fields. The point is that the simple pole fields, for instance, 
the Nakanishi-Lautrup field $\beta(x)$ can be expressed in terms of the invariant delta
function $D(x)$ as
%**   D-beta   %%%%%%%%%%%%%%%%%%%%%%%%%%%%%%%%%%%%%%%%%%%%%%%%%%%%%%%%%
\begin{eqnarray}
\beta_\mu (x) = \int d^3 z D(x-z) \overleftrightarrow{\partial}_0^z \beta_\mu (z),
\label{D-beta}  
\end{eqnarray}
%%%%%%%%%%%%%%%%%%%%%%%%%%%%%%%%%%%%%%%%%%%%%%%%%%%%%%%%%%%%%%%%%%% 
whereas the dipole field $\varphi_{\mu\nu}(x)$ can be done as
%**   E-varphi   %%%%%%%%%%%%%%%%%%%%%%%%%%%%%%%%%%%%%%%%%%%%%%%%%%%%%%%%%
\begin{eqnarray}
&{}& \varphi_{\mu\nu} (x) = \int d^3 z \left[ D(x-z) \overleftrightarrow{\partial}_0^z \varphi_{\mu\nu} (z)
+ E(x-z) \overleftrightarrow{\partial}_0^z \Box \varphi_{\mu\nu} (z) \right]
\nonumber\\
&{}& = \int d^3 z \left[ D(x-z) \overleftrightarrow{\partial}_0^z \varphi_{\mu\nu} (z)
- 24 E(x-z) \overleftrightarrow{\partial}_0^z \partial_{(\mu} \beta_{\nu)} (z) \right],
\label{E-varphi}  
\end{eqnarray}
%%%%%%%%%%%%%%%%%%%%%%%%%%%%%%%%%%%%%%%%%%%%%%%%%%%%%%%%%%%%%%%%%%% 
where in the last equality we have used Eq. (\ref{Linear-Eq9}).
Here the invariant delta function $D(x)$ for massless simple pole fields and its properties
are described as
%**   D-function   %%%%%%%%%%%%%%%%%%%%%%%%%%%%%%%%%%%%%%%%%%%%%%%%%%%%%%%%%
\begin{eqnarray}
&{}& D(x) = - \frac{i}{(2 \pi)^3} \int d^4 k \, \epsilon (k_0) \delta (k^2) e^{i k x}, \qquad
\Box D(x) = 0,
\nonumber\\
&{}& D(-x) = - D(x), \qquad D(0, \vec{x}) = 0, \qquad 
\partial_0 D(0, \vec{x}) = \delta^3 (x), 
\label{D-function}  
\end{eqnarray}
%%%%%%%%%%%%%%%%%%%%%%%%%%%%%%%%%%%%%%%%%%%%%%%%%%%%%%%%%%%%%%%%%%% 
where $\epsilon (k_0) \equiv \frac{k_0}{|k_0|}$. Similarly, the invariant delta function $E(x)$ 
for massless dipole fields and its properties are given by
%**   E-function   %%%%%%%%%%%%%%%%%%%%%%%%%%%%%%%%%%%%%%%%%%%%%%%%%%%%%%%%%
\begin{eqnarray}
&{}& E(x) = - \frac{i}{(2 \pi)^3} \int d^4 k \, \epsilon (k_0) \delta^\prime (k^2) e^{i k x}, \qquad  
\Box E(x) = D(x),
\nonumber\\
&{}& E(-x) = - E(x), \qquad 
E(0, \vec{x}) = \partial_0 E(0, \vec{x}) = \partial_0^2 E(0, \vec{x}) = 0, 
\nonumber\\ 
&{}& \partial_0^3 E(0, \vec{x}) = - \delta^3 (x). 
\label{E-function}  
\end{eqnarray}
%%%%%%%%%%%%%%%%%%%%%%%%%%%%%%%%%%%%%%%%%%%%%%%%%%%%%%%%%%%%%%%%%%%
It is easy to show that the RHS of Eqs. (\ref{D-beta}) and (\ref{E-varphi}) is independent of
$z^0$. Thus, for instance, when we evaluate the four-dimensional commutation relation
$[ \varphi_{\mu\nu} (x), \varphi_{\sigma\tau} (y) ]$, we can put $z^0 = y^0$ and use the 
three-dimensional commutation relations among asymptotic fields. The resultant 4D CRs 
are summarized as
%**   4D-CR   %%%%%%%%%%%%%%%%%%%%%%%%%%%%%%%%%%%%%%%%%%%%%%%%%%%%%%%%%
\begin{eqnarray}
&{}& [ \varphi_{\mu\nu} (x), \varphi_{\sigma\tau} (y) ] = 12 i \phi_0^{-2} [ ( \eta_{\mu\nu} \eta_{\sigma\tau}
- \eta_{\mu\sigma} \eta_{\nu\tau} - \eta_{\mu\tau} \eta_{\nu\sigma} ) D(x-y) 
\nonumber\\
&{}& + ( \eta_{\mu\sigma} \partial_\nu \partial_\tau + \eta_{\nu\sigma} \partial_\mu \partial_\tau +
\eta_{\mu\tau} \partial_\nu \partial_\sigma + \eta_{\nu\tau} \partial_\mu \partial_\sigma ) E(x-y) ],
\label{4D-CR1}
\\
&{}& [ \varphi_{\mu\nu} (x), \beta_\rho (y) ] = i \phi_0^{-2} ( \eta_{\mu\rho} \partial_\nu
+ \eta_{\nu\rho} \partial_\mu ) D(x-y). 
\label{4D-CR2}
\\
&{}& [ \varphi_{\mu\nu} (x), \beta (y) ] = - 2 i \phi_0^{-1} \eta_{\mu\nu} D(x-y). 
\label{4D-CR3}
\\
&{}& [ \tilde \phi (x), \beta (y) ] = i \phi_0^{-1} D(x-y). 
\label{4D-CR4}
\\
&{}& \{ \gamma^\sigma (x), \bar \gamma_\tau (y) \} = - \phi_0^{-2} \delta_\tau^\sigma D(x-y). 
\label{4D-CR5}
\\
&{}& \{ \gamma (x), \bar \gamma (y) \} = - \phi_0^{-2} D(x-y). 
\label{4D-CR6}  
\end{eqnarray}
%%%%%%%%%%%%%%%%%%%%%%%%%%%%%%%%%%%%%%%%%%%%%%%%%%%%%%%%%%%%%%%%%%% 
The other 4D CRs vanish identically.

Now we would like to discuss the issue of the unitarity of the theory in hand. To do that, it is
convenient to perform the Fourier transformation of Eqs. (\ref{4D-CR1})-(\ref{4D-CR6}).
However, for the dipole field we cannot use the three-dimensional Fourier expansion to define 
the creation and annihilation operators. We therefore make use of the four-dimensional 
Fourier expansion \cite{N-O-text}:\footnote{The Fourier transform of a field is denoted 
by the same field except for the argument $x$ or $p$, for simplicity.}
%**   FT-varphi   %%%%%%%%%%%%%%%%%%%%%%%%%%%%%%%%%%%%%%%%%%%%%%%%%%%%%%%%%
\begin{eqnarray}
\varphi_{\mu\nu} (x) = \frac{1}{(2 \pi)^{\frac{3}{2}}} \int d^4 p \, \theta (p_0) [ \varphi_{\mu\nu} (p) e^{i p x}
+ \varphi_{\mu\nu}^\dagger (p) e^{- i p x} ],
\label{FT-varphi}  
\end{eqnarray}
%%%%%%%%%%%%%%%%%%%%%%%%%%%%%%%%%%%%%%%%%%%%%%%%%%%%%%%%%%%%%%%%%%% 
where $\theta (p_0)$ is the step function. For any simple pole fields, we adopt the same Fourier expansion,
for instance, 
%**   FT-beta   %%%%%%%%%%%%%%%%%%%%%%%%%%%%%%%%%%%%%%%%%%%%%%%%%%%%%%%%%
\begin{eqnarray}
\beta_\mu (x) = \frac{1}{(2 \pi)^{\frac{3}{2}}} \int d^4 p \, \theta (p_0) [ \beta_\mu (p) e^{i p x}
+ \beta_\mu^\dagger (p) e^{- i p x} ].
\label{FT-beta}  
\end{eqnarray}
%%%%%%%%%%%%%%%%%%%%%%%%%%%%%%%%%%%%%%%%%%%%%%%%%%%%%%%%%%%%%%%%%%%
Incidentally, for a generic simple pole field $\Phi$, the three-dimensional Fourier expansion is defined as
%**   3D-FT   %%%%%%%%%%%%%%%%%%%%%%%%%%%%%%%%%%%%%%%%%%%%%%%%%%%%%%%%%
\begin{eqnarray}
\Phi (x) = \frac{1}{(2 \pi)^{\frac{3}{2}}} \int d^3 p \, \frac{1}{\sqrt{2 |\vec{p}|}}  
[ \Phi (\vec{p}) e^{i p x} + \Phi^\dagger (\vec{p}) e^{- i p x} ],
\label{3D-FT}  
\end{eqnarray}
%%%%%%%%%%%%%%%%%%%%%%%%%%%%%%%%%%%%%%%%%%%%%%%%%%%%%%%%%%%%%%%%%%% 
where the on-shell relation $p_0 = |\vec{p}|$ must be satisfied ($\vec{p}$ denotes the three-dimensional
momentum) whereas the four-dimensional Fourier expansion reads
%**   4D-FT   %%%%%%%%%%%%%%%%%%%%%%%%%%%%%%%%%%%%%%%%%%%%%%%%%%%%%%%%%
\begin{eqnarray}
\Phi (x) = \frac{1}{(2 \pi)^{\frac{3}{2}}} \int d^4 p \, \theta (p_0) [ \Phi (p) e^{i p x}
+ \Phi^\dagger (p) (p) e^{- i p x} ].
\label{4D-FT}  
\end{eqnarray}
%%%%%%%%%%%%%%%%%%%%%%%%%%%%%%%%%%%%%%%%%%%%%%%%%%%%%%%%%%%%%%%%%%%
Thus, the annihilation operator $\Phi (p)$ in the four-dimensional Fourier expansion has connection with 
the annihilation operator $\Phi (\vec{p})$ in the three-dimensional Fourier expansion like
%**   3D-4D   %%%%%%%%%%%%%%%%%%%%%%%%%%%%%%%%%%%%%%%%%%%%%%%%%%%%%%%%%
\begin{eqnarray}
\Phi (p) = \theta (p_0) \delta (p^2) \sqrt{2 |\vec{p}|} \Phi (\vec{p}).
\label{3D-4D}  
\end{eqnarray}
%%%%%%%%%%%%%%%%%%%%%%%%%%%%%%%%%%%%%%%%%%%%%%%%%%%%%%%%%%%%%%%%%%%
Based on these Fourier expansions, we can calculate the Fourier transform of Eqs. (\ref{4D-CR1})-(\ref{4D-CR6}):
%**   FT-4D-CR   %%%%%%%%%%%%%%%%%%%%%%%%%%%%%%%%%%%%%%%%%%%%%%%%%%%%%%%%%
\begin{eqnarray}
&{}& [ \varphi_{\mu\nu} (p), \varphi_{\sigma\tau}^\dagger (q) ] = 12 \phi_0^{-2} \theta (p_0) \delta^4 (p-q)
[ \delta(p^2) ( \eta_{\mu\nu} \eta_{\sigma\tau}- \eta_{\mu\sigma} \eta_{\nu\tau} - \eta_{\mu\tau} \eta_{\nu\sigma} ) 
\nonumber\\
&{}& - 3  \delta^\prime (p^2) ( \eta_{\mu\sigma} p_\nu p_\tau + \eta_{\nu\sigma} p_\mu p_\tau +
\eta_{\mu\tau} p_\nu p_\sigma + \eta_{\nu\tau} p_\mu p_\sigma ) ].
\label{FT-4D-CR1}
\\
&{}& [ \varphi_{\mu\nu} (p), \beta_\rho^\dagger (q) ] = i \phi_0^{-2} ( \eta_{\mu\rho} p_\nu + \eta_{\nu\rho} p_\mu ) 
\theta (p_0) \delta(p^2) \delta^4 (p-q). 
\label{FT-4D-CR2}
\\
&{}& [ \varphi_{\mu\nu} (p), \beta^\dagger (q) ] = - 2 \phi_0^{-1} \eta_{\mu\nu} \theta (p_0) \delta(p^2) \delta^4 (p-q). 
\label{FT-4D-CR3}
\\
&{}& [ \tilde \phi (p), \beta^\dagger (q) ] = \phi_0^{-1} \theta (p_0) \delta(p^2) \delta^4 (p-q). 
\label{FT-4D-CR4}
\\
&{}& \{ \gamma^\sigma (p), \bar \gamma^\dagger_\tau (q) \} = i \phi_0^{-2} \delta_\tau^\sigma \theta (p_0) 
\delta(p^2) \delta^4 (p-q). 
\label{FT-4D-CR5}
\\
&{}& \{ \gamma (p), \bar \gamma^\dagger (q) \} = i \phi_0^{-2} \theta (p_0) \delta(p^2) \delta^4 (p-q). 
\label{FT-4D-CR6}  
\end{eqnarray}
%%%%%%%%%%%%%%%%%%%%%%%%%%%%%%%%%%%%%%%%%%%%%%%%%%%%%%%%%%%%%%%%%%% 

Next, let us turn our attention to the linearized field equations. In the Fourier transformation, 
Eq. (\ref{Linear-Eq3}) takes the form:
%**   FT-Linear-Eq3   %%%%%%%%%%%%%%%%%%%%%%%%%%%%%%%%%%%%%%%%%%%%%%%%%%%%%%%%%
\begin{eqnarray}
p^\nu \varphi_{\mu\nu} - \frac{1}{2} p_\mu \varphi = 2 \phi_0^{-1} p_\mu \tilde \phi.
\label{FT-Linear-Eq3}
\end{eqnarray}
%%%%%%%%%%%%%%%%%%%%%%%%%%%%%%%%%%%%%%%%%%%%%%%%%%%%%%%%%%%%%%%%%%% 
If we fix the degree of freedom associated with $\tilde \phi$, which will be discussed later,
this equation gives us four independent relations in ten components of $\varphi_{\mu\nu} (p)$,
thereby reducing the independent components of $\varphi_{\mu\nu} (p)$ to be six. To deal with
six independent components of $\varphi_{\mu\nu} (p)$, it is convenient to take a specific Lorentz 
frame such that $p_1 = p_2 = 0$ and $p_3 > 0$, and choose the six components as follows:
%**   Lorentz   %%%%%%%%%%%%%%%%%%%%%%%%%%%%%%%%%%%%%%%%%%%%%%%%%%%%%%%%%
\begin{eqnarray}
&{}& \varphi_1 (p) = \frac{1}{2} [ \varphi_{11} (p) - \varphi_{22} (p) ],  \qquad
\varphi_2 (p) = \varphi_{12} (p),  \qquad
\omega_0 (p) = - \frac{1}{2 p_0} \varphi_{00} (p),   
\nonumber\\
&{}& \omega_I (p) = - \frac{1}{p_0} \varphi_{0I} (p),  \qquad
\omega_3 (p) = - \frac{1}{2 p_3} \varphi_{33} (p), 
\label{Lorentz}  
\end{eqnarray}
%%%%%%%%%%%%%%%%%%%%%%%%%%%%%%%%%%%%%%%%%%%%%%%%%%%%%%%%%%%%%%%%%%% 
where the index $I$ takes the transverse components $I = 1, 2$. 

In this respect, it is worthwhile to consider the GCT BRST transformation for these components.
First, let us write down the GCT BRST transformation for the Fourier expansion of the asymptotic fields, 
which reads
%**   Q_B-FT   %%%%%%%%%%%%%%%%%%%%%%%%%%%%%%%%%%%%%%%%%%%%%%%%%%%%%%%%%
\begin{eqnarray}
&{}& \delta_B \varphi_{\mu\nu} (p) = - i [ p_\mu \gamma_\nu (p) + p_\nu \gamma_\mu (p) ], \quad
\delta_B \gamma^\mu (p) = 0, \quad 
\delta_B \bar \gamma_\mu (p) = i \beta_\mu (p),
\nonumber\\
&{}& \delta_B \tilde \phi (p) = \delta_B \beta_\mu (p) = \delta_B \beta (p) 
= \delta_B \gamma (p) = \delta_B \bar \gamma (p) = 0. 
\label{Q_B-FT}  
\end{eqnarray}
%%%%%%%%%%%%%%%%%%%%%%%%%%%%%%%%%%%%%%%%%%%%%%%%%%%%%%%%%%%%%%%%%%% 
Using this BRST transformation, the GCT BRST transformation for the components
in (\ref{Lorentz}) takes the form:
%**   Q_B-Comp   %%%%%%%%%%%%%%%%%%%%%%%%%%%%%%%%%%%%%%%%%%%%%%%%%%%%%%%%%
\begin{eqnarray}
&{}& \delta_B \varphi_I (p) = 0, \qquad
\delta_B \omega_\mu (p) = i \gamma_\mu (p),
\nonumber\\
&{}& \delta_B \bar \gamma_\mu (p) = i \beta_\mu (p), \qquad 
\delta_B \gamma_\mu (p) = \delta_B \beta_\mu (p) = 0,
\label{Q_B-Comp}  
\end{eqnarray}
%%%%%%%%%%%%%%%%%%%%%%%%%%%%%%%%%%%%%%%%%%%%%%%%%%%%%%%%%%%%%%%%%%% 
where $p_1 = p_2 = 0$ was used. This BRST transformation implies that $\varphi_I (p)$
could be the physical observable while a set of fields, $\{ \omega_\mu (p), \beta_\mu (p), 
\gamma_\mu (p), \bar \gamma_\mu (p) \}$ might belong to the BRST quartet, which are dropped
from the physical state by the Kugo-Ojima subsidiary condition, $Q_B | \rm{phys} \rangle = 0$ \cite{Kugo-Ojima}.  
However, note that $\beta_\mu (p), \gamma_\mu (p)$ and $\bar \gamma_\mu (p)$ are
simple pole fields obeying $p^2 \beta_\mu (p) = p^2 \gamma_\mu (p) = p^2 \bar \gamma_\mu (p) = 0$,
but $\varphi_{\mu\nu} (p)$ is a dipole field satisfying $( p^2 )^2 \varphi_{\mu\nu} (p) = 0$, 
so that a naive Kugo-Ojima's quartet mechanism does not work. 

To clarify the BRST quartet mechanism, let us calculate their 4D CRs. 
From Eqs. (\ref{FT-4D-CR1})-(\ref{FT-4D-CR6}) and the definition (\ref{Lorentz}), it is straightforward to 
derive the following 4D CRs:
%**   Lor-4D-CR   %%%%%%%%%%%%%%%%%%%%%%%%%%%%%%%%%%%%%%%%%%%%%%%%%%%%%%%%%
\begin{eqnarray}
&{}& [ \varphi_I (p), \varphi_J^\dagger (q) ] = - 12 \phi_0^{-2} \delta_{IJ} \theta (p_0) \delta(p^2) \delta^4 (p-q).
\label{Lor-4D-CR1}
\\
&{}& [ \varphi_I (p), \omega_\mu^\dagger (q) ] = [ \varphi_I (p), \beta_\mu^\dagger (q) ]
= [ \beta_\mu (p), \beta_\nu^\dagger (q) ] = 0. 
\label{Lor-4D-CR2}
\\
&{}& [ \omega_\mu (p), \beta_\nu^\dagger (q) ] = - i \phi_0^{-2} \eta_{\mu\nu} \theta (p_0) 
\delta(p^2) \delta^4 (p-q). 
\label{Lor-4D-CR3}
\\
&{}& \{ \gamma_\mu (p), \bar \gamma_\nu^\dagger (q) \} = i \phi_0^{-2} \eta_{\mu\nu} \theta (p_0) 
\delta(p^2) \delta^4 (p-q). 
\label{Lor-4D-CR4}
\end{eqnarray}
%%%%%%%%%%%%%%%%%%%%%%%%%%%%%%%%%%%%%%%%%%%%%%%%%%%%%%%%%%%%%%%%%%%
In addition to them, we have a bit complicated expression for $[ \omega_\mu (p), \omega_\nu^\dagger (q) ]$
because $\varphi_{\mu\nu} (p)$ is a dipole field, but luckily enough this expression is not necessary
for our aim \cite{Kugo-Ojima}. 
It is known how to take out a simple pole field from a dipole field, which amounts to
using an operator defined by \cite{Kugo-Ojima} 
%**   D-ope   %%%%%%%%%%%%%%%%%%%%%%%%%%%%%%%%%%%%%%%%%%%%%%%%%%%%%%%%%
\begin{eqnarray}
{\cal D}_p = \frac{1}{2 |\vec{p}|^2} p_0 \frac{\partial}{\partial p_0} + c, 
\label{D-ope}
\end{eqnarray}
%%%%%%%%%%%%%%%%%%%%%%%%%%%%%%%%%%%%%%%%%%%%%%%%%%%%%%%%%%%%%%%%%%% 
where $c$ is a constant. Using this operator, we can define a simple pole field
$\hat \varphi_{\mu\nu} (p)$ from the dipole field $\varphi_{\mu\nu} (p)$, which
obeys $(p^2)^2 \varphi_{\mu\nu} (p) = 0$, as
%**   Simple-field   %%%%%%%%%%%%%%%%%%%%%%%%%%%%%%%%%%%%%%%%%%%%%%%%%%%%%%%%%
\begin{eqnarray}
\hat \varphi_{\mu\nu} (p) &\equiv& \varphi_{\mu\nu} (p) - {\cal D}_p p^2 \varphi_{\mu\nu} (p)
\nonumber\\
&=& \varphi_{\mu\nu} (p) - 24 i {\cal D}_p p_{(\mu} \beta_{\nu)} (p),
\label{Simple-field}
\end{eqnarray}
%%%%%%%%%%%%%%%%%%%%%%%%%%%%%%%%%%%%%%%%%%%%%%%%%%%%%%%%%%%%%%%%%%% 
where in the last equality we have used the Fourier transform of the linearized field equation (\ref{Linear-Eq9}).
It is then easy to verify the equation:
%**   Simple-field2   %%%%%%%%%%%%%%%%%%%%%%%%%%%%%%%%%%%%%%%%%%%%%%%%%%%%%%%%%
\begin{eqnarray}
p^2 \hat \varphi_{\mu\nu} (p) = 0.
\label{Simple-field2}
\end{eqnarray}
%%%%%%%%%%%%%%%%%%%%%%%%%%%%%%%%%%%%%%%%%%%%%%%%%%%%%%%%%%%%%%%%%%% 
Then,  in (\ref{Lorentz}) we replace $\varphi_{\mu\nu}$ of $\omega_\mu$ with $\hat \varphi_{\mu\nu}$,
and we redefine $\omega_\mu$ by $\hat \omega_\mu$ as
%**   hat-omega   %%%%%%%%%%%%%%%%%%%%%%%%%%%%%%%%%%%%%%%%%%%%%%%%%%%%%%%%%
\begin{eqnarray}
\hat \omega_0 (p) = - \frac{1}{2 p_0} \hat \varphi_{00} (p),  \quad
\hat \omega_I (p) = - \frac{1}{p_0} \hat \varphi_{0I} (p),  \quad
\hat \omega_3 (p) = - \frac{1}{2 p_3} \hat \varphi_{33} (p).
\label{hat-omega}  
\end{eqnarray}
%%%%%%%%%%%%%%%%%%%%%%%%%%%%%%%%%%%%%%%%%%%%%%%%%%%%%%%%%%%%%%%%%%% 
The key point is that with this redefinition from $\omega_\mu$ to $\hat \omega_\mu$,
the BRST transformation and the 4D CRs remain unchanged owing to
$\delta_B \beta_\mu = 0$ and $[ \beta_\mu (p), \beta^\dagger_\nu (q) ] = [ \varphi_I (p), 
\beta^\dagger_\mu (q) ] = 0$, those are,
%**   point   %%%%%%%%%%%%%%%%%%%%%%%%%%%%%%%%%%%%%%%%%%%%%%%%%%%%%%%%%
\begin{eqnarray}
&{}& \delta_B  \hat \omega_\mu (p) = i \gamma_\mu (p),  \quad
[ \hat \omega_\mu (p), \beta^\dagger_\nu (q) ] = [ \omega_\mu (p), \beta^\dagger_\nu (q) ],
\nonumber\\
&{}& [ \varphi_I (p), \hat \omega^\dagger_\mu (q) ] = [ \varphi_I (p), \omega^\dagger_\mu (q) ].
\label{point}  
\end{eqnarray}
%%%%%%%%%%%%%%%%%%%%%%%%%%%%%%%%%%%%%%%%%%%%%%%%%%%%%%%%%%%%%%%%%%% 

Now it turns out that all the fields, $\{ \varphi_I, \hat \omega_\mu, \beta_\mu,
\gamma_\mu, \bar \gamma_\mu \}$ are simple pole fields.\footnote{Without the redefinition,
$\varphi_I (p)$ is already a simple pole field as can be seen in Eq. (\ref{Lor-4D-CR1}).}
Since all the fields become simple pole fields, we can obtain the standard creation and
annihilation operators in the three-dimensional Fourier expansion from those in the four-dimensional
one through the relation (\ref{3D-4D}). As a result, the three-dimensional (anti-)commutation
relations, which are denoted as $[ \Phi (\vec{p}), \Phi^\dagger (\vec{q}) \}$ with
$\Phi (\vec{p}) \equiv \{ \varphi_I (\vec{p}), \hat \omega_\mu (\vec{p}), \beta_\mu (\vec{p}), 
\gamma_\mu (\vec{p}), \bar \gamma_\mu (\vec{p}) \}$, are given by\footnote{The bracket $[ A, B \}$ 
is the graded commutation relation denoting either commutator or anti-commutator, according to 
the Grassmann-even or odd character of $A$ and $B$, i.e., $[ A, B \} = A B - (-)^{|A| |B|} B A$.}  
%**   G-3D-CRs   %%%%%%%%%%%%%%%%%%%%%%%%%%%%%%%%%%%%%%%%%%%%%%%%%%%%%%%%%
\begin{eqnarray}
[ \Phi (\vec{p}), \Phi^\dagger (\vec{q}) \} &=&
\left(
\begin{array} {cc|cc|cc}
-12 \phi_0^{-2}  \delta_{IJ}              &     &    &    &         \\ 
\hline
    &        &   [ \hat \omega_\mu (\vec{p}), \hat \omega_\nu^\dagger (\vec{q}) ] &  - i \phi_0^{-2} \eta_{\mu\nu}  & \\ 
    &        &     i \phi_0^{-2} \eta_{\mu\nu}   &  0       &              \\
\hline   
    &        &     &          &       &    i \phi_0^{-2} \eta_{\mu\nu} \\  
    &        &     &          &   -i \phi_0^{-2} \eta_{\mu\nu} &            \\
\end{array}
\right) 
\nonumber\\
&{}& \times \delta ( \vec{p} - \vec{q} )_.
\label{G-3D-CRs}  
\end{eqnarray}
%%%%%%%%%%%%%%%%%%%%%%%%%%%%%%%%%%%%%%%%%%%%%%%%%%%%%%%%%%%%%%%%%%% 
The (anti-)commuatation relations (\ref{G-3D-CRs}) have in essence the same structure as those 
of the Yang-Mills theory \cite{Kugo-Ojima}. Hence, we find that $\varphi_I$ could 
be the physical observable while a set of fields $\{ \hat \omega_\mu, \beta_\mu, \gamma_\mu, 
\bar \gamma_\mu \}$ belongs to the BRST quartet.

Next, let us move on to another BRST transformation, which is the BRST transformation for the Weyl 
transformation. The Weyl BRST transformation for the asymptotic fields is of form:
%**   W-Q_B-Asym   %%%%%%%%%%%%%%%%%%%%%%%%%%%%%%%%%%%%%%%%%%%%%%%%%%%%%%%%%
\begin{eqnarray}
&{}& \bar \delta_B \varphi_{\mu\nu} = 2 c \eta_{\mu\nu}, \quad
\bar \delta_B \tilde \phi = - \phi_0 \gamma, \quad 
\bar \delta_B \gamma = 0, \quad 
\bar \delta_B \bar \gamma = i \beta,
\nonumber\\
&{}& \bar \delta_B \beta = \bar \delta_B \beta_\mu = \bar \delta_B \gamma_\mu 
= \bar \delta_B \bar \gamma_\mu = 0. 
\label{W-Q_B-Asym}  
\end{eqnarray}
%%%%%%%%%%%%%%%%%%%%%%%%%%%%%%%%%%%%%%%%%%%%%%%%%%%%%%%%%%%%%%%%%%% 
The Weyl BRST transformation of $\varphi_I$ is vanishing:
%**   W-Q_B-Obs   %%%%%%%%%%%%%%%%%%%%%%%%%%%%%%%%%%%%%%%%%%%%%%%%%%%%%%%%%
\begin{eqnarray}
\bar \delta_B \varphi_I = 0, 
\label{W-Q_B-Obs}  
\end{eqnarray}
%%%%%%%%%%%%%%%%%%%%%%%%%%%%%%%%%%%%%%%%%%%%%%%%%%%%%%%%%%%%%%%%%%% 
which means that together with $\delta_B \varphi_I = 0$, $\varphi_I$ is truely the physical observable.
The four-dimensional (anti-)commutation relations among the fields $\{ \tilde \phi, \beta,
\gamma, \bar \gamma \}$ read
%**   W-4D-CRs   %%%%%%%%%%%%%%%%%%%%%%%%%%%%%%%%%%%%%%%%%%%%%%%%%%%%%%%%%
\begin{eqnarray}
&{}& [ \tilde \phi (p), \tilde \phi^\dagger (q) ] = 0, 
\nonumber\\
&{}& [ \tilde \phi (p), \beta^\dagger (q) ] = \phi_0^{-1} \theta (p_0) \delta (p^2) \delta^4 (p-q), 
\nonumber\\
&{}& \{ \gamma (p), \bar \gamma^\dagger (q) ] = i \phi_0^{-2} \theta (p_0) \delta (p^2) \delta^4 (p-q).
\label{W-4D-CRs}  
\end{eqnarray}
%%%%%%%%%%%%%%%%%%%%%%%%%%%%%%%%%%%%%%%%%%%%%%%%%%%%%%%%%%%%%%%%%%% 
As can be also seen in these 4D CRs, all the fields $\{ \varphi_I, \tilde \phi, \beta, \gamma, \bar \gamma \}$   
are massless simple pole fields. Via the relation (\ref{3D-4D}), the three-dimensional (anti-)commutation
relations $[ \Phi (\vec{p}), \Phi^\dagger (\vec{q}) \}$ with $\Phi (\vec{p}) \equiv \{ \varphi_I (\vec{p}), 
\tilde \phi (\vec{p}), \beta (\vec{p}), \gamma (\vec{p}), \bar \gamma (\vec{p}) \}$, are of form  
%**   W-3D-CRs   %%%%%%%%%%%%%%%%%%%%%%%%%%%%%%%%%%%%%%%%%%%%%%%%%%%%%%%%%
\begin{eqnarray}
[ \Phi (\vec{p}), \Phi^\dagger (\vec{q}) \} =
\left(
\begin{array}{cc|cc|cc}
-12 \phi_0^{-2}  \delta_{IJ}              &     &    &    &         \\ 
\hline
    &        &               0   &    \phi_0^{-1}     &            \\ 
    &        &               \phi_0^{-1}    &  0       &              \\
\hline   
    &        &     &    &       &    i \phi_0^{-2}  \\  
    &        &     &    &   -i \phi_0^{-2} &            \\
\end{array}
\right) \delta ( \vec{p} - \vec{q} )_.
\label{W-3D-CRs}  
\end{eqnarray}
%%%%%%%%%%%%%%%%%%%%%%%%%%%%%%%%%%%%%%%%%%%%%%%%%%%%%%%%%%%%%%%%%%% 
Thus, $\varphi_I$ is the physical observable while a set of fields, $\{ \tilde \phi, \beta, \gamma, \bar \gamma \}$ 
consists of the BRST quartet and is the unphysical mode by the Kugo-Ojima's subsidiary condition \cite{Kugo-Ojima}. 
Here it is worth mentioning that the ghost-like scalar field $\phi$ belongs to the unphysical mode 
so together with the result obtained in the analysis of the GCT BRST cohomology, the physical S-matrix is found to 
be unitary.

\section{Choral symmetry}

As mentioned in Section 3, a set of fields (including the space-time coordinates $x^\mu$)
$X^M \equiv \{ x^\mu, b_\mu, \sigma, B, c^\mu, \bar c_\mu, c, \bar c \}$ obeys a very simple equation:
%**   d'Alemb-eq   %%%%%%%%%%%%%%%%%%%%%%%%%%%%%%%%%%%%%%%%%%%%%%%%%%%%%%%%%
\begin{eqnarray}
g^{\mu\nu} \partial_\mu \partial_\nu X^M = 0.
\label{d'Alemb-eq}  
\end{eqnarray}
%%%%%%%%%%%%%%%%%%%%%%%%%%%%%%%%%%%%%%%%%%%%%%%%%%%%%%%%%%%%%%%%%%% 
This equation holds if and only if we adopt the extended de Donder gauge and the new scalar
gauge as gauge-fixing conditions for the GCT and the Weyl transformation, respectively. The existence of 
this simple equation suggests that there could be many of conserved currents defined in Eq. (\ref{Cons-currents}).  
In this section, we shall show explicitly that there exist such currents and we have a huge global symmetry called 
choral symmetry, which is the $IOSp(10|10)$ in the present theory.

Let us start with the Lagrangian (\ref{ST-q-Lag3}), which can be cast to the form:
%**   Choral-Lag   %%%%%%%%%%%%%%%%%%%%%%%%%%%%%%%%%%%%%%%%%%%%%%%%%%%%%%%%%
\begin{eqnarray}
{\cal L}_q = \tilde g^{\mu\nu} \phi^2 \left( \frac{1}{12} R_{\mu\nu} - \frac{1}{2} \hat E_{\mu\nu} \right).
\label{Choral-Lag}  
\end{eqnarray}
%%%%%%%%%%%%%%%%%%%%%%%%%%%%%%%%%%%%%%%%%%%%%%%%%%%%%%%%%%%%%%%%%%% 
Here note that $\tilde g^{\mu\nu} \phi^2$ is a Weyl invariant metric and the Ricci tensor is invariant
under only a global scale transformation. We can further rewrite it into the form:
%**   Choral-OSp-Lag   %%%%%%%%%%%%%%%%%%%%%%%%%%%%%%%%%%%%%%%%%%%%%%%%%%%%%%%%%
\begin{eqnarray}
{\cal L}_q &=& \tilde g^{\mu\nu} \phi^2 \left( \frac{1}{12} R_{\mu\nu} - \frac{1}{2} \eta_{NM} 
\partial_\mu X^M \partial_\nu X^N \right)
\nonumber\\
&=& \tilde g^{\mu\nu} \phi^2 \left( \frac{1}{12} R_{\mu\nu} - \frac{1}{2}  
\partial_\mu X^M \tilde \eta_{MN} \partial_\nu X^N \right),
\label{Choral-OSp-Lag}  
\end{eqnarray}
%%%%%%%%%%%%%%%%%%%%%%%%%%%%%%%%%%%%%%%%%%%%%%%%%%%%%%%%%%%%%%%%%%% 
where we have introduced an $OSp(10|10)$ metric $\eta_{NM} = \eta_{MN}^T \equiv \tilde \eta_{MN}$ 
defined as \cite{Kugo}
%**   OSp-metric   %%%%%%%%%%%%%%%%%%%%%%%%%%%%%%%%%%%%%%%%%%%%%%%%%%%%%%%%%
\begin{eqnarray}
\eta_{NM} = \tilde \eta_{MN} =
\left(
\begin{array}{cc|cc|cc|cc}
     &                \delta_\mu^\nu &     &   &     &  \\ 
\delta^\mu_\nu  &                    &    &    &    &   \\ 
\hline
    &        &               -1    &   -1      &     &    &      \\ 
    &        &               -1    &  0       &       &    &    \\
\hline   
    &        &     &    &       &   -i\delta_\mu^\nu  &   & \\  
    &        &     &    &   i\delta^\mu_\nu &   &    & \\
\hline
    &        &                &        &       &     &         &  -i \\  
    &        &                &        &       &     &          i     & \\
\end{array}
\right)_.
\label{OSp-metric}  
\end{eqnarray}
%%%%%%%%%%%%%%%%%%%%%%%%%%%%%%%%%%%%%%%%%%%%%%%%%%%%%%%%%%%%%%%%%%% 
Let us note that this $OSp(10|10)$ metric $\eta_{NM}$, which is a c-number quantity, has the symmetry 
property such that 
%**   Prop-OSp-metric  %%%%%%%%%%%%%%%%%%%%%%%%%%%%%%%%%%%%%%%%%%%%%%%%%%%%%%%%%
\begin{eqnarray}
\eta_{MN}=(-)^{|M| \cdot |N|} \eta_{NM} = (-)^{|M|} \eta_{NM}=(-)^{|N|} \eta_{NM},
\label{Prop-OSp-metric}  
\end{eqnarray}
%%%%%%%%%%%%%%%%%%%%%%%%%%%%%%%%%%%%%%%%%%%%%%%%%%%%%%%%%%%%%%%%%%% 
where the statistics index $|M|$ is 0 or 1 when $X^M$ is Grassmann-even or 
Grassmann-odd, respectively. This property comes from the fact that $\eta_{MN}$ is `diagonal' 
in the sense that its off-diagonal, Grassmann-even and Grassmann-odd, and vice versa, matrix elements 
vanish, i.e., $\eta_{MN} = 0$ when $|M| \neq |N|$, thereby being $|M| = |N| = |M| \cdot| N|$ in front of 
$\eta_{MN}$ \cite{Kugo}. 

Now that the quantum Lagrangian (\ref{Choral-OSp-Lag}) is expressed in a manifestly
$IOSp(10|10)$ invariant form except for the Weyl invariant metric $\tilde g^{\mu\nu} \phi^2$, which
will be discussed later, there could exist an $IOSp(10|10)$ as a global symmetry in our theory. Let us show 
this fact first. The infinitesimal $OSp$ rotation is defined by
%**   OSp-rot  %%%%%%%%%%%%%%%%%%%%%%%%%%%%%%%%%%%%%%%%%%%%%%%%%%%%%%%%%
\begin{eqnarray}
\delta X^M = \eta^{ML} \varepsilon_{LN} X^N \equiv \varepsilon^M{}_N X^N,
\label{OSp-rot}  
\end{eqnarray}
%%%%%%%%%%%%%%%%%%%%%%%%%%%%%%%%%%%%%%%%%%%%%%%%%%%%%%%%%%%%%%%%%%% 
where $\eta^{MN}$ is the inverse matrix of $\eta_{MN}$, and the infinitesimal parameter
$\varepsilon_{MN}$ has the following properties:
%**   varepsilon  %%%%%%%%%%%%%%%%%%%%%%%%%%%%%%%%%%%%%%%%%%%%%%%%%%%%%%%%%
\begin{eqnarray}
\varepsilon_{MN} = (-)^{1 + |M| \cdot |N|} \varepsilon_{NM}, \qquad
\varepsilon_{MN} X^L = (-)^{|L| (|M| + |N|)} X^L \varepsilon_{MN}.
\label{varepsilon}  
\end{eqnarray}
%%%%%%%%%%%%%%%%%%%%%%%%%%%%%%%%%%%%%%%%%%%%%%%%%%%%%%%%%%%%%%%%%%% 
Moreover, in order to find the conserved current, we assume that the infinitesimal parameter
$\varepsilon_{MN}$ depends on the space-time coordinates $x^\mu$, i.e., 
$\varepsilon_{MN} = \varepsilon_{MN} (x^\mu)$.

Assuming for a while that the metric $\tilde g^{\mu\nu} \phi^2$ and $R_{\mu\nu}$ is invariant, 
the infinitesimal variation of the quantum Lagrangian (\ref{Choral-OSp-Lag}) under the $OSp$ 
rotation (\ref{OSp-rot}) is given by
%**   Var-OSp-Lag   %%%%%%%%%%%%%%%%%%%%%%%%%%%%%%%%%%%%%%%%%%%%%%%%%%%%%%%%%
\begin{eqnarray}
\delta {\cal L}_q = - \tilde g^{\mu\nu} \phi^2 \left( \partial_\mu \varepsilon_{NM} 
X^M \partial_\nu X^N + \varepsilon_{NM} \partial_\mu X^M \partial_\nu X^N \right).
\label{Var-OSp-Lag}  
\end{eqnarray}
%%%%%%%%%%%%%%%%%%%%%%%%%%%%%%%%%%%%%%%%%%%%%%%%%%%%%%%%%%%%%%%%%%% 
It is easy to prove that the second term on the RHS vanishes owing to the first property in 
Eq. (\ref{varepsilon}). Thus, ${\cal L}_q$ is invariant under the infinitesimal $OSp$ rotation.
The conserved current is then calculated as 
%**   OSp-current   %%%%%%%%%%%%%%%%%%%%%%%%%%%%%%%%%%%%%%%%%%%%%%%%%%%%%%%%%
\begin{eqnarray}
\delta {\cal L}_q &=& - \tilde g^{\mu\nu} \phi^2 \partial_\mu \varepsilon_{NM} X^M \partial_\nu X^N
\nonumber\\
&=& - \frac{1}{2} \tilde g^{\mu\nu} \phi^2 \partial_\mu \varepsilon_{NM} \left[ X^M \partial_\nu X^N
- (-)^{|M| \cdot |N|} X^N \partial_\nu X^M \right]
\nonumber\\
&=& - \frac{1}{2} \tilde g^{\mu\nu} \phi^2 \partial_\mu \varepsilon_{NM} \left( X^M \partial_\nu X^N
- \partial_\nu X^M X^N  \right)
\nonumber\\
&=& - \frac{1}{2} \tilde g^{\mu\nu} \phi^2 \partial_\mu \varepsilon_{NM} 
X^M \overset{\leftrightarrow}{\partial}_\nu X^N
\nonumber\\
&\equiv& - \frac{1}{2} \partial_\mu \varepsilon_{NM} {\cal M}^{\mu MN},
\label{OSp-current}  
\end{eqnarray}
%%%%%%%%%%%%%%%%%%%%%%%%%%%%%%%%%%%%%%%%%%%%%%%%%%%%%%%%%%%%%%%%%%% 
from which the conserved current ${\cal M}^{\mu MN}$ for the $OSp$ rotation takes the form: 
%**   OSp-current-M   %%%%%%%%%%%%%%%%%%%%%%%%%%%%%%%%%%%%%%%%%%%%%%%%%%%%%%%%%
\begin{eqnarray}
{\cal M}^{\mu MN} = \tilde g^{\mu\nu} \phi^2 X^M \overset{\leftrightarrow}{\partial}_\nu X^N.
\label{OSp-current-M}  
\end{eqnarray}
%%%%%%%%%%%%%%%%%%%%%%%%%%%%%%%%%%%%%%%%%%%%%%%%%%%%%%%%%%%%%%%%%%% 

In a similar way, we can derive the conserved current for the infinitesimal translation
%**   transl  %%%%%%%%%%%%%%%%%%%%%%%%%%%%%%%%%%%%%%%%%%%%%%%%%%%%%%%%%
\begin{eqnarray}
\delta X^M = \varepsilon^M,
\label{transl}  
\end{eqnarray}
%%%%%%%%%%%%%%%%%%%%%%%%%%%%%%%%%%%%%%%%%%%%%%%%%%%%%%%%%%%%%%%%%%% 
where $\varepsilon^M$ is the infinitesimal parameter and assume that it is a local one
for deriving the corresponding conserved current. Indeed, assuming again that 
the metric $\tilde g^{\mu\nu} \phi^2$ and $R_{\mu\nu}$ are invariant under the translation,
we can show that ${\cal L}_q$ is invariant under the infinitesimal translation as follows:
%**   transl-current   %%%%%%%%%%%%%%%%%%%%%%%%%%%%%%%%%%%%%%%%%%%%%%%%%%%%%%%%%
\begin{eqnarray}
\delta {\cal L}_q &=& - \tilde g^{\mu\nu} \phi^2 \eta_{NM} \partial_\mu \varepsilon^M \partial_\nu X^N
\nonumber\\
&=& - \tilde g^{\mu\nu} \phi^2 \partial_\mu \varepsilon_N \partial_\nu X^N
\nonumber\\
&\equiv& - \partial_\mu \varepsilon_M {\cal P}^{\mu M},
\label{transl-current}  
\end{eqnarray}
%%%%%%%%%%%%%%%%%%%%%%%%%%%%%%%%%%%%%%%%%%%%%%%%%%%%%%%%%%%%%%%%%%% 
which implies that the conserved current ${\cal P}^{\mu M}$ for the translation reads 
%**   transl-current-P   %%%%%%%%%%%%%%%%%%%%%%%%%%%%%%%%%%%%%%%%%%%%%%%%%%%%%%%%%
\begin{eqnarray}
{\cal P}^{\mu M} = \tilde g^{\mu\nu} \phi^2 \partial_\nu X^M
= \tilde g^{\mu\nu} \phi^2 \left( 1 \overset{\leftrightarrow}{\partial}_\nu X^M \right).
\label{transl-current-P}  
\end{eqnarray}
%%%%%%%%%%%%%%%%%%%%%%%%%%%%%%%%%%%%%%%%%%%%%%%%%%%%%%%%%%%%%%%%%%% 

The above proofs make sense only under the assumption that the metric $\tilde g^{\mu\nu} \phi^2$ 
and $R_{\mu\nu}$ are invariant under the $IOSp(10|10)$. So the problem reduces to a question: 
Is this assumption correct? The answer is obviously not true, but the non-invariant terms 
can be compensated by a suitable Weyl transformation. To show this fact, let us consider only the case of 
the infinitesimal $OSp$ rotation since we can treat the case of the translation in a perfectly 
similar manner. Under the infinitesimal $OSp$ rotation (\ref{OSp-rot}), the dilaton $\sigma(x)$, 
which is defined as $\phi = e^\sigma$, transforms as
%**   Dilaton-OSp   %%%%%%%%%%%%%%%%%%%%%%%%%%%%%%%%%%%%%%%%%%%%%%%%%%%%%%%%%
\begin{eqnarray}
\delta \sigma = \eta^{\sigma L} \varepsilon_{LN} X^N = - \varepsilon_{BN} X^N,
\label{Dilaton-OSp}  
\end{eqnarray}
%%%%%%%%%%%%%%%%%%%%%%%%%%%%%%%%%%%%%%%%%%%%%%%%%%%%%%%%%%%%%%%%%%% 
where we have used (\ref{OSp-metric}) and
%**   Matrix   %%%%%%%%%%%%%%%%%%%%%%%%%%%%%%%%%%%%%%%%%%%%%%%%%%%%%%%%%
\begin{eqnarray} 
\begin{pmatrix}
   -1 & -1 \\
   -1 & 0
\end{pmatrix}^{-1}
= \begin{pmatrix}
   0 & -1 \\
   -1 & 1
\end{pmatrix},
\label{Matrix}  
\end{eqnarray}
%%%%%%%%%%%%%%%%%%%%%%%%%%%%%%%%%%%%%%%%%%%%%%%%%%%%%%%%%%%%%%%%%%% 
where recall that the matrix $\eta^{ML}$ is the inverse matrix of $\eta_{ML}$. As for the scalar field $\phi(x)$, 
this transformation for the dilaton can be interpreted as a Weyl transformation:
%**   Weyl-phi   %%%%%%%%%%%%%%%%%%%%%%%%%%%%%%%%%%%%%%%%%%%%%%%%%%%%%%%%%
\begin{eqnarray} 
\phi \rightarrow \phi^\prime = e^{\epsilon (x)} \phi, 
\label{Weyl-phi}  
\end{eqnarray}
%%%%%%%%%%%%%%%%%%%%%%%%%%%%%%%%%%%%%%%%%%%%%%%%%%%%%%%%%%%%%%%%%%% 
where the infinitesimal parameter is defined as $\epsilon (x) = - \varepsilon_{BN} X^N$.
This Weyl transformation induces the Weyl transformation for the metric tensor field
at the same time:
%**   Weyl-g   %%%%%%%%%%%%%%%%%%%%%%%%%%%%%%%%%%%%%%%%%%%%%%%%%%%%%%%%%
\begin{eqnarray} 
g_{\mu\nu} \rightarrow g_{\mu\nu}^\prime = e^{- 2 \epsilon (x)} g_{\mu\nu}. 
\label{Weyl-g}  
\end{eqnarray}
%%%%%%%%%%%%%%%%%%%%%%%%%%%%%%%%%%%%%%%%%%%%%%%%%%%%%%%%%%%%%%%%%%% 
Let us recall that the metric $\tilde g^{\mu\nu} \phi^2$ is the Weyl invariant metric
so that it is invariant under the Weyl transformation (\ref{Weyl-phi}) and (\ref{Weyl-g}).
This implies that $\tilde g^{\mu\nu} \phi^2$ is essentially invariant under the $OSp$ rotation
if an appropriate Weyl transformation is achieved.

How about $R_{\mu\nu}$? Even if $R_{\mu\nu}$ is not invariant under the Weyl transformation
in itself, this object comes from the classical Lagrangian of the Weyl invariant scalar-tensor gravity
in (\ref{WIST-gravity}), so together with the metric tensor and the scalar field it essentially becomes 
invariant under the Weyl transformation (\ref{Weyl-phi}) and (\ref{Weyl-g}). Thus, in this sense,
$R_{\mu\nu}$ is also invariant under the $OSp$ rotation. In any case, it is worthwhile to stress
that in the present formulation, the choral symmetry $IOSp(10|10)$ is not a symmetry of only the FP ghosts 
and the Nakanishi-Latrup fields but closely related to classical fields $g_{\mu\nu}$ and $\phi$ 
which lie in the classical Lagrangian. 

An important remark is relevant to the expression of the conserved currents (\ref{OSp-current-M}) 
and (\ref{transl-current-P}). To make the quantum Lagrangian ${\cal L}_q$ be invariant 
under the choral symmetry $IOSp(10|10)$, it is necessary to perform the Weyl transformation 
(\ref{Weyl-phi}) and (\ref{Weyl-g}). Then, it is natural to ask if because of this associated 
Weyl transformation, the expression of the currents would be modified or not. Here a miracle happens. 
As shown in Refs. \cite{Jackiw, Oda-U}, the current for the Weyl transformation identically vanishes 
in the Weyl invariant scalar-tensor gravity. Thus, although we make the Weyl transformation 
(\ref{Weyl-phi}) and (\ref{Weyl-g}), the conserved currents (\ref{OSp-current-M}) and 
(\ref{transl-current-P}) are unchanged.   

From the conserved currents (\ref{OSp-current-M}) and (\ref{transl-current-P}), the corresponding
conserved charges become
%**   IOSp-charge   %%%%%%%%%%%%%%%%%%%%%%%%%%%%%%%%%%%%%%%%%%%%%%%%%%%%%%%%%
\begin{eqnarray}
M^{MN} &\equiv& \int d^3 x \, {\cal M}^{0 MN} = \int d^3 x \, \tilde g^{0 \nu} \phi^2  
X^M \overset{\leftrightarrow}{\partial}_\nu X^N,
\nonumber\\
P^M &\equiv& \int d^3 x \, {\cal P}^{0 M} = \int d^3 x \, \tilde g^{0 \nu} \phi^2 \partial_\nu X^M.
\label{IOSp-charge}  
\end{eqnarray}
%%%%%%%%%%%%%%%%%%%%%%%%%%%%%%%%%%%%%%%%%%%%%%%%%%%%%%%%%%%%%%%%%%% 
It then turns out that using various ETCRs obtained so far, the $IOSp(10|10)$ generators $\{ M^{MN}, P^M \}$ 
generate an $IOSp(10|10)$ algebra:
%**   IOSp-algebra   %%%%%%%%%%%%%%%%%%%%%%%%%%%%%%%%%%%%%%%%%%%%%%%%%%%%%%%%%
\begin{eqnarray}
&{}& [ P^M, P^N \} = 0, 
\nonumber\\
&{}& [ M^{MN}, P^R \} = i \bigl[ P^M \tilde \eta^{NR} - (-)^{|N| |R|} P^N \tilde \eta^{MR} \bigr],
\nonumber\\
&{}& [ M^{MN}, M^{RS} \} = i \bigl[ M^{MS} \tilde \eta^{NR} - (-)^{|N| |R|} M^{MR} \tilde \eta^{NS} 
- (-)^{|N| |R|} M^{NS} \tilde \eta^{MR} 
\nonumber\\
&{}& + (-)^{|M| |R| + |N| |S|} M^{NR} \tilde \eta^{MS} \bigr].
\label{IOSp-algebra}  
\end{eqnarray}
%%%%%%%%%%%%%%%%%%%%%%%%%%%%%%%%%%%%%%%%%%%%%%%%%%%%%%%%%%%%%%%%%%% 

As a final remark, it is worthwhile to point out that all the global symmetries existing in the 
present theory are expressed in terms of the generators of the choral symmetry. For instance,
the BRST charges for the GCT and Weyl transformation are respectively expressed as
%**   Choral-Symm   %%%%%%%%%%%%%%%%%%%%%%%%%%%%%%%%%%%%%%%%%%%%%%%%%%%%%%%%%
\begin{eqnarray}
&{}& Q_B \equiv M (b_\rho, c^\rho) = \int d^3 x \, \tilde g^{0 \nu} \phi^2 
b_\rho \overset{\leftrightarrow}{\partial}_\nu c^\rho, 
\nonumber\\
&{}& \bar Q_B \equiv M (B, c) = \int d^3 x \, \tilde g^{0 \nu} \phi^2 
B \overset{\leftrightarrow}{\partial}_\nu c.
\label{Choral-Symm}  
\end{eqnarray}
%%%%%%%%%%%%%%%%%%%%%%%%%%%%%%%%%%%%%%%%%%%%%%%%%%%%%%%%%%%%%%%%%%% 

\section{Gravitational conformal symmetry}

Even if we have already fixed the Weyl symmetry by the scalar gauge condition (\ref{Scalar-gauge}),
we still have its linearized, residual symmetries. In order to look for the residual symmetries, 
it is convenient to take the extended de Donder gauge (\ref{Ext-de-Donder}) into consideration 
simultaneously.\footnote{The same strategy has been adopted in different theories in Refs. 
\cite{Oda-R, Kamimura, Oda-RWS}.}   With the help of the extended de Donder gauge (\ref{Ext-de-Donder}), 
the scalar gauge condition (\ref{Scalar-gauge}) can be rewritten as 
%**   Scalar-gauge2   %%%%%%%%%%%%%%%%%%%%%%%%%%%%%%%%%%%%%%%%%%%%%%%%%%%%%%%%%
\begin{eqnarray}
0 = \partial_\mu ( \tilde g^{\mu\nu} \phi \partial_\nu \phi ) 
= \partial_\mu ( \tilde g^{\mu\nu} \phi^2 \partial_\nu \sigma ) 
= \tilde g^{\mu\nu} \phi^2 \partial_\mu \partial_\nu \sigma,
\label{Scalar-gauge2}  
\end{eqnarray}
%%%%%%%%%%%%%%%%%%%%%%%%%%%%%%%%%%%%%%%%%%%%%%%%%%%%%%%%%%%%%%%%%%% 
where we have used the relation between the scalar field and dilaton, $\phi = e^\sigma$.
Under the Weyl transformation (\ref{Weyl-transf}) with $\Omega(x) \equiv e^{\Lambda(x)}$,
the dilaton $\sigma$ transforms as 
%**   Dilaton-Weyl   %%%%%%%%%%%%%%%%%%%%%%%%%%%%%%%%%%%%%%%%%%%%%%%%%%%%%%%%%
\begin{eqnarray}
\sigma \rightarrow \sigma^\prime = \sigma - \log \Omega = \sigma - \Lambda,
\label{Dilaton-Weyl}  
\end{eqnarray}
%%%%%%%%%%%%%%%%%%%%%%%%%%%%%%%%%%%%%%%%%%%%%%%%%%%%%%%%%%%%%%%%%%% 
where we have used the Weyl transformation (\ref{Weyl-transf}) for the scalar field.
Since $\tilde g^{\mu\nu} \phi^2$ is a Weyl invariant quantity, the Weyl transformation
makes Eq. (\ref{Scalar-gauge2}) change to
%**   Scalar-gauge3   %%%%%%%%%%%%%%%%%%%%%%%%%%%%%%%%%%%%%%%%%%%%%%%%%%%%%%%%%
\begin{eqnarray}
0 = \tilde g^{\mu\nu} \phi^2 \partial_\mu \partial_\nu \sigma
\rightarrow 0 = \tilde g^{\mu\nu} \phi^2 \partial_\mu \partial_\nu \sigma^\prime
= \tilde g^{\mu\nu} \phi^2 \partial_\mu \partial_\nu ( \sigma - \Lambda).
\label{Scalar-gauge3}  
\end{eqnarray}
%%%%%%%%%%%%%%%%%%%%%%%%%%%%%%%%%%%%%%%%%%%%%%%%%%%%%%%%%%%%%%%%%%% 
This equation shows that when we use the extended de Donder gauge, the scalar gauge condition
is still invariant under the Weyl transformation as long as 
%**   Linear-equation   %%%%%%%%%%%%%%%%%%%%%%%%%%%%%%%%%%%%%%%%%%%%%%%%%%%%%%%%%
\begin{eqnarray}
g^{\mu\nu} \partial_\mu \partial_\nu \Lambda = 0,
\label{Linear-equation}  
\end{eqnarray}
%%%%%%%%%%%%%%%%%%%%%%%%%%%%%%%%%%%%%%%%%%%%%%%%%%%%%%%%%%%%%%%%%%% 
is satisfied, thereby implying the existence of the residual symmetries \cite{Oda-R, Kamimura, Oda-RWS}. 
Selecting the coefficients appropriately for later convenience, the solution to Eq. (\ref{Linear-equation}) 
is given by
%**   Sol-Linear-eq   %%%%%%%%%%%%%%%%%%%%%%%%%%%%%%%%%%%%%%%%%%%%%%%%%%%%%%%%%
\begin{eqnarray}
\Lambda = \lambda - 2 k_\mu x^\mu,
\label{Sol-Linear-eq}  
\end{eqnarray}
%%%%%%%%%%%%%%%%%%%%%%%%%%%%%%%%%%%%%%%%%%%%%%%%%%%%%%%%%%%%%%%%%%% 
where $\lambda, k_\mu$ are constants.\footnote{It is shown in Appendix B that the transformations associated with 
the parameters $\lambda$ and $k_\mu$, respectively, correpond to dilatation and special conformal transformation 
in a flat Minkowski background.}

We can also verify the invariance of the quantum Lagrangian under the residual symmetries more directly.  
To do that, let us assume that $\Lambda$, or equivalently, $\lambda$ and $k_\mu$, are the infinitesimal
parameters. It then turns out that the quantum Lagrangian (\ref{ST-q-Lag3}) is invariant under 
the residual symmetries:
%**   Res-symm   %%%%%%%%%%%%%%%%%%%%%%%%%%%%%%%%%%%%%%%%%%%%%%%%%%%%%%%%%
\begin{eqnarray}
&{}& \delta g_{\mu\nu} = 2 ( \lambda - 2 k_\rho x^\rho ) g_{\mu\nu},
\nonumber\\
&{}& \delta \sigma = - ( \lambda - 2 k_\rho x^\rho ), \qquad
\delta b_\mu = 2 k_\mu B,
\label{Res-symm}  
\end{eqnarray}
%%%%%%%%%%%%%%%%%%%%%%%%%%%%%%%%%%%%%%%%%%%%%%%%%%%%%%%%%%%%%%%%%%% 
where the other fields are unchanged. The generators corresponding to the transformation parameters
$\lambda$ and $k_\mu$ are respectively constructed out of those of the choral symmetry as
%**   Res-gen   %%%%%%%%%%%%%%%%%%%%%%%%%%%%%%%%%%%%%%%%%%%%%%%%%%%%%%%%%
\begin{eqnarray}
D_0 &\equiv& - P(B) = - \int d^3 x \, \tilde g^{0 \nu} \phi^2 \partial_\nu B,
\nonumber\\
K^\mu &\equiv& 2 M^\mu (x, B) = 2 \int d^3 x \, \tilde g^{0 \nu} \phi^2 x^\mu
\overset{\leftrightarrow}{\partial}_\nu B.
\label{Res-gen}  
\end{eqnarray}
%%%%%%%%%%%%%%%%%%%%%%%%%%%%%%%%%%%%%%%%%%%%%%%%%%%%%%%%%%%%%%%%%%% 
In addition to the generators $D_0$ and $K^\mu$, one can construct the translation generator $P_\mu$
and $GL(4)$ generator $G^\mu{}_\nu$ from those of the choral symmetry $IOSp(10|10)$ as
%**   Trans-GL   %%%%%%%%%%%%%%%%%%%%%%%%%%%%%%%%%%%%%%%%%%%%%%%%%%%%%%%%%
\begin{eqnarray}
P_\mu &\equiv& P_\mu (b) = \int d^3 x \, \tilde g^{0 \nu} \phi^2 \partial_\nu b_\mu,
\nonumber\\
G^\mu{}_\nu &\equiv& M^\mu{}_\nu (x, b) - i M^\mu{}_\nu (c^\tau, \bar c_\tau)
\nonumber\\
&=& \int d^3 x \, \tilde g^{0 \lambda} \phi^2 ( x^\mu \overset{\leftrightarrow}{\partial}_\lambda b_\nu
- i c^\mu \overset{\leftrightarrow}{\partial}_\lambda \bar c_\nu ).
\label{Trans-GL}  
\end{eqnarray}
%%%%%%%%%%%%%%%%%%%%%%%%%%%%%%%%%%%%%%%%%%%%%%%%%%%%%%%%%%%%%%%%%%% 

Now we would like to show that in our theory there is a gravitational conformal algebra which is slightly 
different from conformal algebra in a flat Minkowski space-time. For this aim, let us consider 
a set of generators, $\{ P_\mu, G^\mu{}_\nu, K^\mu, D_0 \}$. From these generators, we wish to
construct the generator $D$ for a scale transformation. Recall that in conformal field theory 
in the four-dimensional Minkowski space-time, the dilatation generator obeys the following algebra 
for an local operator $O_i (x)$ of conformal dimension $\Delta_i$ \cite{Gross, Nakayama}:\footnote{For clarity, 
we will call a global scale transformation in a flat Minkowski space-time ``dilatation''. Dilatation is usually 
interpreted as a subgroup of the general coordinate transformation in a such way that 
the space-time coordinates are transformed as $x^\mu \rightarrow \Omega x^\mu$ 
in the flat space-time where $\Omega$ is a constant scale factor, whereas the global scale 
transformation is a rescaling of all lengths by the same $\Omega$ by 
$g_{\mu\nu} \rightarrow \Omega^2 g_{\mu\nu}$. The two viewpoints are completely equivalent 
since all the lengths are defined via the line element $d s^2 = g_{\mu\nu} d x^\mu d x^\nu$.} 
%**   D-com   %%%%%%%%%%%%%%%%%%%%%%%%%%%%%%%%%%%%%%%%%%%%%%%%%%%%%%%%%
\begin{eqnarray}
[ i D, O_i (x) ] = x^\mu \partial_\mu O_i (x) + \Delta_i O_i (x).
\label{D-com}  
\end{eqnarray}
%%%%%%%%%%%%%%%%%%%%%%%%%%%%%%%%%%%%%%%%%%%%%%%%%%%%%%%%%%%%%%%%%%% 
Since the scalar field $\phi (x)$ has conformal dimension $1$, it must satisfy the equation:
%**   D-phi-com   %%%%%%%%%%%%%%%%%%%%%%%%%%%%%%%%%%%%%%%%%%%%%%%%%%%%%%%%%
\begin{eqnarray}
[ i D, \phi (x) ] = x^\mu \partial_\mu \phi (x) + \phi (x).
\label{D-phi-com}  
\end{eqnarray}
%%%%%%%%%%%%%%%%%%%%%%%%%%%%%%%%%%%%%%%%%%%%%%%%%%%%%%%%%%%%%%%%%%% 
 
To be consistent with this equation, we shall make a generator for the scale transformation.
From the definitions (\ref{Res-gen}) and (\ref{Trans-GL}), we find
%**   GD-phi-com   %%%%%%%%%%%%%%%%%%%%%%%%%%%%%%%%%%%%%%%%%%%%%%%%%%%%%%%%%
\begin{eqnarray}
[ i G^\mu{}_\nu, \phi (x) ] = x^\mu \partial_\mu \phi (x), \qquad
[ i D_0, \phi (x) ] = - \phi (x).
\label{GD-phi-com}  
\end{eqnarray}
%%%%%%%%%%%%%%%%%%%%%%%%%%%%%%%%%%%%%%%%%%%%%%%%%%%%%%%%%%%%%%%%%%% 
The following linear combination of $G^\mu{}_\nu$ and $D_0$ does the job:
%**   D-def   %%%%%%%%%%%%%%%%%%%%%%%%%%%%%%%%%%%%%%%%%%%%%%%%%%%%%%%%%
\begin{eqnarray}
D \equiv G^\mu{}_\mu - D_0.
\label{D-def}  
\end{eqnarray}
%%%%%%%%%%%%%%%%%%%%%%%%%%%%%%%%%%%%%%%%%%%%%%%%%%%%%%%%%%%%%%%%%%% 
As a consistency check, it is valuable to see how this operator $D$ acts on the metric field
whose result reads:
%**   D-g-com   %%%%%%%%%%%%%%%%%%%%%%%%%%%%%%%%%%%%%%%%%%%%%%%%%%%%%%%%%
\begin{eqnarray}
&{}& [ i D, g_{\sigma\tau} ] = [ i G^\mu{}_\mu, g_{\sigma\tau} ] - [ i D_0, g_{\sigma\tau} ] 
\nonumber\\
&{}& = ( x^\mu \partial_\mu g_{\sigma\tau} + 2 g_{\sigma\tau} ) - 2 g_{\sigma\tau}
=  x^\mu \partial_\mu g_{\sigma\tau},
\label{D-g-com}  
\end{eqnarray}
%%%%%%%%%%%%%%%%%%%%%%%%%%%%%%%%%%%%%%%%%%%%%%%%%%%%%%%%%%%%%%%%%%% 
which implies that the metric field has conformal dimension $0$ as desired and this result will be used 
later when discussing spontaneous symmetry breakdown.

Next, let us calculate an algebra among the generators $\{ P_\mu, G^\mu{}_\nu, K^\mu, D \}$.
After some calculations, we find that the algebra closes and takes the form:
%**   Grav-conf0   %%%%%%%%%%%%%%%%%%%%%%%%%%%%%%%%%%%%%%%%%%%%%%%%%%%%%%%%%
\begin{eqnarray}
&{}& [ P_\mu, P_\nu ] = 0, \quad 
[ P_\mu, G^\rho{}_\sigma ] = i P_\sigma \delta^\rho_\mu, \quad
[ P_\mu, K^\nu ] = - 2 i ( G^\rho{}_\rho - D ) \delta^\nu_\mu, \quad
\nonumber\\
&{}& [ P_\mu, D ] = i P_\mu, \quad 
[ G^\mu{}_\nu, G^\rho{}_\sigma ] = i ( G^\mu{}_\sigma \delta^\rho_\nu - G^\rho{}_\nu \delta^\mu_\sigma),
\nonumber\\
&{}& [ G^\mu{}_\nu, K^\rho ] = i K^\mu \delta^\rho_\nu, \quad
[ G^\mu{}_\nu, D ] = [ K^\mu, K^\nu ] = 0, 
\nonumber\\
&{}& [ K^\mu, D ] = - i K^\mu, \quad
[ D, D] = 0. 
\label{Grav-conf0}  
\end{eqnarray}
%%%%%%%%%%%%%%%%%%%%%%%%%%%%%%%%%%%%%%%%%%%%%%%%%%%%%%%%%%%%%%%%%%% 
To extract the gravitational conformal algebra in quantum gravity, it is necessary to introduce
the ``Lorentz'' generator, which can be contructed from the $GL(4)$ generator as
%**   Lor-gene   %%%%%%%%%%%%%%%%%%%%%%%%%%%%%%%%%%%%%%%%%%%%%%%%%%%%%%%%%
\begin{eqnarray}
M_{\mu\nu} \equiv - \eta_{\mu\rho} G^\rho{}_\nu + \eta_{\nu\rho} G^\rho{}_\mu. 
\label{Lor-gene}  
\end{eqnarray}
%%%%%%%%%%%%%%%%%%%%%%%%%%%%%%%%%%%%%%%%%%%%%%%%%%%%%%%%%%%%%%%%%%% 
In terms of the generator $M_{\mu\nu}$, the algebra (\ref{Grav-conf0}) can be cast to the form:
%**   Grav-conf   %%%%%%%%%%%%%%%%%%%%%%%%%%%%%%%%%%%%%%%%%%%%%%%%%%%%%%%%%
\begin{eqnarray}
&{}& [ P_\mu, P_\nu ] = 0, \quad 
[ P_\mu, M_{\rho\sigma} ] = i ( P_\rho \eta_{\mu\sigma} - P_\sigma \eta_{\mu\rho} ), 
\nonumber\\
&{}& [ P_\mu, K^\nu ] = - 2 i ( G^\rho{}_\rho - D ) \delta^\nu_\mu, \quad
[ P_\mu, D ] = i P_\mu, 
\nonumber\\
&{}& [ M_{\mu\nu}, M_{\rho\sigma} ] = - i ( M_{\mu\sigma} \eta_{\nu\rho} - M_{\nu\sigma} \eta_{\mu\rho}
+ M_{\rho\mu} \eta_{\sigma\nu} - M_{\rho\nu} \eta_{\sigma\mu}),
\nonumber\\
&{}& [ M_{\mu\nu}, K^\rho ] = i ( - K_\mu \delta^\rho_\nu + K_\nu \delta^\rho_\mu ), \quad
[ M_{\mu\nu}, D ] = [ K^\mu, K^\nu ] = 0, 
\nonumber\\
&{}& [ K^\mu, D ] = - i K^\mu, \quad
[ D, D] = 0. 
\label{Grav-conf}  
\end{eqnarray}
%%%%%%%%%%%%%%%%%%%%%%%%%%%%%%%%%%%%%%%%%%%%%%%%%%%%%%%%%%%%%%%%%%% 
where we have defined $K_\mu \equiv \eta_{\mu\nu} K^\nu$. It is of interest that
the the algebra (\ref{Grav-conf}) in quantum gravity, which we call ``gravitational conformal algebra'', 
formally resembles conformal algebra in the flat Minkowski space-time except for the expression
of $[ P_\mu, K^\nu ]$.\footnote{In case of conformal algebra in the flat space-time, $[ P_\mu, K^\nu ] 
= - 2 i ( \delta^\nu_\mu D + M_\mu{}^\nu )$.} This difference reflects from the difference of the definition 
of conformal dimension in both gravity and conformal field theory, for which the metric tensor field $g_{\mu\nu}$ 
has $2$ in gravity as seen in Eq. (\ref{Weyl-transf}) while it has $0$ in conformal field theory as seen 
in Eq. (\ref{D-g-com}).

\section{Spontaneous breakdown of symmetries}

In the theory in hand, there are huge global symmetries, which are $IOSp(10|10)$ supersymmetry, so it is valuable 
to investigate which symmetries are spontaneously broken or survive even in quantum regime.
In this section, we postulate the existence of a unique vacuum $| 0 \rangle$, which is normalized to be the unity:
%**   Vac-norm   %%%%%%%%%%%%%%%%%%%%%%%%%%%%%%%%%%%%%%%%%%%%%%%%%%%%%%%%%
\begin{eqnarray}
\langle 0 | 0 \rangle = 1.
\label{Vac-norm}  
\end{eqnarray}
%%%%%%%%%%%%%%%%%%%%%%%%%%%%%%%%%%%%%%%%%%%%%%%%%%%%%%%%%%%%%%%%%%% 
Furthermore, we assume that the vacuum is translation invariant:
%**   Trans-Vac   %%%%%%%%%%%%%%%%%%%%%%%%%%%%%%%%%%%%%%%%%%%%%%%%%%%%%%%%%
\begin{eqnarray}
P_\mu | 0 \rangle = 0,
\label{Trans-Vac}  
\end{eqnarray}
%%%%%%%%%%%%%%%%%%%%%%%%%%%%%%%%%%%%%%%%%%%%%%%%%%%%%%%%%%%%%%%%%%% 
and the vacuum expectation values (VEVs) of
the metric tensor $g_{\mu\nu}$ and the scalar field $\phi$ are respectively the Minkowski metric $\eta_{\mu\nu}$
and a non-zero constant $\phi_0 \neq 0$:
%**   VEV-A   %%%%%%%%%%%%%%%%%%%%%%%%%%%%%%%%%%%%%%%%%%%%%%%%%%%%%%%%%
\begin{eqnarray}
\langle 0 | g_{\mu\nu} | 0 \rangle = \eta_{\mu\nu}, \qquad
\langle 0 | \phi | 0 \rangle = \phi_0. 
\label{VEV-A}  
\end{eqnarray}
%%%%%%%%%%%%%%%%%%%%%%%%%%%%%%%%%%%%%%%%%%%%%%%%%%%%%%%%%%%%%%%%%%% 

By a straightforward calculation, we can obtain the following VEVs:
%**   Many-VEV   %%%%%%%%%%%%%%%%%%%%%%%%%%%%%%%%%%%%%%%%%%%%%%%%%%%%%%%%%
\begin{eqnarray}
&{}& \langle 0 | [ i P^\mu (x), b_\rho ] | 0 \rangle = - \delta^\mu_\rho, \quad 
\langle 0 | \{ i P^\mu (c^\tau), \bar c_\rho \} | 0 \rangle = i \delta^\mu_\rho, \quad
\nonumber\\
&{}& \langle 0 | \{ i P_\mu (\bar c_\tau), c^\rho \} | 0 \rangle = - i \delta^\rho_\mu, 
\nonumber\\
&{}& \langle 0 | [ i M^{\mu\nu} (x, x), \frac{1}{2} ( \partial_\lambda b_\rho 
- \partial_\rho b_\lambda ) ] | 0 \rangle = - ( \delta^\mu_\lambda \delta^\nu_\rho
- \delta^\nu_\lambda \delta^\mu_\rho ),
\nonumber\\
&{}& \langle 0 | \{ i M^{\mu\nu} (x, c^\tau), \partial_\lambda \bar c_\rho \} | 0 \rangle 
= i \delta^\mu_\lambda \delta^\nu_\rho,   \quad
\langle 0 | \{ i M^\mu{}_\nu (x, \bar c_\tau), \partial_\lambda c^\rho \} | 0 \rangle 
= - i \delta^\mu_\lambda \delta^\rho_\nu, 
\nonumber\\
&{}& \langle 0 | [ i P (\sigma), B ] | 0 \rangle = 1, \quad 
\langle 0 | \{ i P (c), \bar c \} | 0 \rangle = i, \quad
\langle 0 | \{ i P (\bar c), c \} | 0 \rangle = - i, 
\nonumber\\
&{}& \langle 0 | \{ i M (\sigma, c), \bar c \} | 0 \rangle = i \sigma_0,  \quad
\langle 0 | \{ i M (\sigma, \bar c), c \} | 0 \rangle = - i \sigma_0, 
\label{Many-VEV}  
\end{eqnarray}
%%%%%%%%%%%%%%%%%%%%%%%%%%%%%%%%%%%%%%%%%%%%%%%%%%%%%%%%%%%%%%%%%%% 
where $\langle 0 | \sigma(x) | 0 \rangle \equiv \sigma_0$. Eq. (\ref{Many-VEV}) shows that
the symmetries generated by the conserved charges 
%**   SSB-charge   %%%%%%%%%%%%%%%%%%%%%%%%%%%%%%%%%%%%%%%%%%%%%%%%%%%%%%%%%
\begin{eqnarray*}
&{}& \{ P^\mu (x), P^\mu (c^\tau), P_\mu (\bar c_\tau), 
M^{\mu\nu} (x, x), M^{\mu\nu} (x, c^\tau), M^\mu{}_\nu (x, \bar c_\tau), 
\nonumber\\
&{}& P (\sigma), P (c), P (\bar c), M (\sigma, c), M (\sigma, \bar c)\}
\label{SSB-charge}  
\end{eqnarray*}
%%%%%%%%%%%%%%%%%%%%%%%%%%%%%%%%%%%%%%%%%%%%%%%%%%%%%%%%%%%%%%%%%%% 
are necessarily broken spontaneously, thereby $b_\mu,
c^\mu, \bar c_\mu, B, c$ and $\bar c$ acquiring massless Nambu-Goldstone modes. Note that 
the exact masslessness of the dilaton $\sigma$ cannot be proved in this way.

Next, on the basis of the gravitational conformal symmetry, we will show that $GL(4)$, 
special conformal symmetry and scale symmetry are spontaneously broken down to the Poincar\'e
symmetry. We find that the VEV of a commutator between the $GL(4)$ generator 
and the metric field reads
%**   G-g-CM   %%%%%%%%%%%%%%%%%%%%%%%%%%%%%%%%%%%%%%%%%%%%%%%%%%%%%%%%%
\begin{eqnarray}
\langle 0 | [ i G^\mu{}_\nu, g_{\sigma\tau} ] | 0 \rangle 
= \delta^\mu_\sigma \eta_{\nu\tau} + \delta^\mu_\tau \eta_{\nu\sigma}.
\label{G-g-CM}  
\end{eqnarray}
%%%%%%%%%%%%%%%%%%%%%%%%%%%%%%%%%%%%%%%%%%%%%%%%%%%%%%%%%%%%%%%%%%% 
Thus, the Lorentz generator, which is defined in Eq. (\ref{Lor-gene}), has the vanishing VEV:
%**   M-g-CM   %%%%%%%%%%%%%%%%%%%%%%%%%%%%%%%%%%%%%%%%%%%%%%%%%%%%%%%%%
\begin{eqnarray}
\langle 0 | [ i M_{\mu\nu}, g_{\sigma\tau} ] | 0 \rangle = 0.
\label{M-g-CM}  
\end{eqnarray}
%%%%%%%%%%%%%%%%%%%%%%%%%%%%%%%%%%%%%%%%%%%%%%%%%%%%%%%%%%%%%%%%%%% 
On the other hand, the symmetric part, which is defined as $\bar M_{\mu\nu} \equiv \eta_{\mu\rho}
G^\rho{}_\nu + \eta_{\nu\rho} G^\rho{}_\mu$, has the non-vanishing VEV: 
%**   BM-g-CM   %%%%%%%%%%%%%%%%%%%%%%%%%%%%%%%%%%%%%%%%%%%%%%%%%%%%%%%%%
\begin{eqnarray}
\langle 0 | [ i \bar M_{\mu\nu}, g_{\sigma\tau} ] | 0 \rangle = 2 ( \eta_{\mu\sigma} \eta_{\nu\tau}
+ \eta_{\mu\tau} \eta_{\nu\sigma} ).
\label{BM-g-CM}  
\end{eqnarray}
%%%%%%%%%%%%%%%%%%%%%%%%%%%%%%%%%%%%%%%%%%%%%%%%%%%%%%%%%%%%%%%%%%% 
Thus, the $GL(4)$ symmetry is spontaneously broken to the Lorentz symmetry where the corresponding 
Nambu-Goldstone boson with ten independent components is nothing but the massless graviton \cite{NO}.
Here it is interesting that in a sector of the scalar field, the $GL(4)$ symmetry and of course the Lorentz
symmetry as well, do not give rise to a symmetry breaking as can be seen in the commutators:
%**   G-phi-CM   %%%%%%%%%%%%%%%%%%%%%%%%%%%%%%%%%%%%%%%%%%%%%%%%%%%%%%%%%
\begin{eqnarray}
\langle 0 | [ i G^\mu{}_\nu, \phi ] | 0 \rangle  = \langle 0 | [ i M_{\mu\nu}, \phi ] | 0 \rangle 
= \langle 0 | [ i \bar M_{\mu\nu}, \phi ] | 0 \rangle = 0.
\label{G-phi-CM}  
\end{eqnarray}
%%%%%%%%%%%%%%%%%%%%%%%%%%%%%%%%%%%%%%%%%%%%%%%%%%%%%%%%%%%%%%%%%%% 
 
Now we wish to clarify how the scale symmetry and special conformal symmetry are spontaneously broken and
what the corresponding Nambu-Goldstone bosons are. As for the scale symmetry, it is not the gravitational
field but the dilaton that gives rise to spontaneous symmetry breaking.
Indeed, Eq. (\ref{D-g-com}) gives us
%**   VEV-D-g   %%%%%%%%%%%%%%%%%%%%%%%%%%%%%%%%%%%%%%%%%%%%%%%%%%%%%%%%%
\begin{eqnarray}
\langle 0 | [ i D, g_{\sigma\tau} ] | 0 \rangle = 0.
\label{VEV-D-g}  
\end{eqnarray}
%%%%%%%%%%%%%%%%%%%%%%%%%%%%%%%%%%%%%%%%%%%%%%%%%%%%%%%%%%%%%%%%%%% 
On the other hand, for the dilaton, from Eq. (\ref{D-phi-com}) we have
%**   VEV-D-sigma   %%%%%%%%%%%%%%%%%%%%%%%%%%%%%%%%%%%%%%%%%%%%%%%%%%%%%%%%%
\begin{eqnarray}
\langle 0 | [ i D, \sigma ] | 0 \rangle = 1,
\label{VEV-D-sigma}  
\end{eqnarray}
%%%%%%%%%%%%%%%%%%%%%%%%%%%%%%%%%%%%%%%%%%%%%%%%%%%%%%%%%%%%%%%%%%% 
which elucidates the spontaneous symmetry breakdown of the scale symmetry whose
Nambu-Goldstone boson is just the massless dilaton $\sigma(x)$. 

Regarding the special conformal symmetry, we find
%**   VEV-K-phi   %%%%%%%%%%%%%%%%%%%%%%%%%%%%%%%%%%%%%%%%%%%%%%%%%%%%%%%%%
\begin{eqnarray}
\langle 0 | [ i K^\mu, \partial_\nu \sigma ] | 0 \rangle = 2 \delta^\mu_\nu.
\label{VEV-K-phi}  
\end{eqnarray}
%%%%%%%%%%%%%%%%%%%%%%%%%%%%%%%%%%%%%%%%%%%%%%%%%%%%%%%%%%%%%%%%%%% 
This equation means that the special conformal symmetry is certainly broken spontaneously
and its Nambu-Goldstone boson is the derivative of the dilaton. This interpretation can be
also verified from the gravitational conformal algebra. In the algebra (\ref{Grav-conf}), we have
a commutator between $P_\mu$ and $K^\nu$:
%**   P-K   %%%%%%%%%%%%%%%%%%%%%%%%%%%%%%%%%%%%%%%%%%%%%%%%%%%%%%%%%
\begin{eqnarray}
[ P_\mu, K^\nu ] = - 2 i ( G^\rho{}_\rho - D ) \delta^\nu_\mu.
\label{P-K}  
\end{eqnarray}
%%%%%%%%%%%%%%%%%%%%%%%%%%%%%%%%%%%%%%%%%%%%%%%%%%%%%%%%%%%%%%%%%%% 
Let us consider the Jacobi identity:
%**   Jacobi   %%%%%%%%%%%%%%%%%%%%%%%%%%%%%%%%%%%%%%%%%%%%%%%%%%%%%%%%%
\begin{eqnarray}
[ [ P_\mu, K^\nu ], \sigma ] + [ [ K^\nu, \sigma ], P_\mu ] + [ [ \sigma, P_\mu ], K^\nu ] = 0.
\label{Jacobi}  
\end{eqnarray}
%%%%%%%%%%%%%%%%%%%%%%%%%%%%%%%%%%%%%%%%%%%%%%%%%%%%%%%%%%%%%%%%%%% 
Using the translational invariance of the vacuum in Eq. (\ref{Trans-Vac}) and the equation
%**   Jacobi2   %%%%%%%%%%%%%%%%%%%%%%%%%%%%%%%%%%%%%%%%%%%%%%%%%%%%%%%%%
\begin{eqnarray}
[ P_\mu, \sigma ] =  - i \partial_\mu \sigma,
\label{Jacobi2}  
\end{eqnarray}
%%%%%%%%%%%%%%%%%%%%%%%%%%%%%%%%%%%%%%%%%%%%%%%%%%%%%%%%%%%%%%%%%%% 
and taking the VEV of the Jacobi identity (\ref{Jacobi}), we can obtain the VEV:
%**   VEV-Jacobi   %%%%%%%%%%%%%%%%%%%%%%%%%%%%%%%%%%%%%%%%%%%%%%%%%%%%%%%%%
\begin{eqnarray}
\langle 0 | [ K^\nu, \partial_\mu \sigma ] | 0 \rangle &=& - 2 \delta^\nu_\mu
\langle 0 | [ G^\rho{}_\rho - D, \sigma ] | 0 \rangle
\nonumber\\
&=& - 2 i \delta^\nu_\mu,
\label{VEV-Jacobi}  
\end{eqnarray}
%%%%%%%%%%%%%%%%%%%%%%%%%%%%%%%%%%%%%%%%%%%%%%%%%%%%%%%%%%%%%%%%%%% 
which coincides with Eq. (\ref{VEV-K-phi}) as promised. In other words, the $GL(4)$ 
symmetry is spontaneously broken to the Poincar\'e symmetry whose Nambu-Goldstone
boson is the graviton, the scale symmetry and the special conformal symmetry are also
spontaneously broken and the corresponding Nambu-Goldstone bosons are the dilaton
and the derivative of the dilaton, respectively. It is of interest that the Nambu-Goldstone
boson associated with the special conformal symmetry is not an independent field in 
quantum gravity as in conformal field theory \cite{Kobayashi}.

\section{Conclusion}

In this article, we have performed a manifestly covariant quantization and contructed a quantum 
theory of the Weyl invariant scalar-tensor gravity within the framework of the BRST formalism. 
In the past, Nakanishi has made a similar quantum gravitational theory of Einstein's general relativity
\cite{Nakanishi, N-O-text}, and the present work provides its natural generalization in the sense that 
the Weyl symmetry is treated on the same footing as the general coordinate symmetry. 

Since the Weyl invariant scalar-tensor gravity has been known to be equivalent to general relativity 
in the unitary gauge where the scalar field is gauge-fixed to be a constant, it is natural to expect 
that our present theory shares several characteristic features with the Nakanishi's quantum gravity. 
In particular, the both theories have a huge global symmetry called ``choral symmetry'', but our choral symmetry 
$ISOp(10|10)$ is larger than that of the Nakanishi's theory, which is an $ISOp(8|8)$, owing to the presence
of the Weyl symmetry in our formulation. Compared with the case of general relativity, one peculiar feature 
of our choral symmetry is that the choral symmetry needs the Weyl symmetry in proving its invariance 
of the quantum Lagrangian so that it is closely related to a gravitational sector while in the case of general 
relativity the choral symmetry is isolated from classical Lagrangian and comes from purely the Lagrangian 
involving the Nakanishi-Lautrup field and the FP ghosts. 

It is worth mentioning that in our quantum gravity there is a gravitational conformal algebra which
is relevant to conventional conformal algebra in a flat Minkowski space-time. According to
the Zumino theorem \cite{Zumino}, the theories which are invariant under the GCT and Weyl transformation
have conformal invariance in the flat Minkowski background at the classical level.  The present
study supports a conjecture that the Zumino theorem could be valid even in quantum gravity.  

Last but not least, we should comment on a Weyl anomaly. In this respect, let us recall that in the manifestly scale 
invariant regularization method \cite{Englert} - \cite{Ghilencea}, the scale invariance is free of scale anomaly. 
Though a completely satisfying formalism is still missing, we believe that in the Weyl invariant regularization
method, the Weyl invariance would be also kept in the operator level without the Weyl anomaly, and is spontaneously 
broken in considering states in the Hilbert space.

There are a lot of works to be done in future. First of all, we should make a manifestly Weyl invariant
regularization methods by introducing an additional scalar field which plays a role for the renormalization
mass scale $\mu$. Secondly, we should prove a quantum Zumino theorem in case that a classical
Lagrangian is an arbitrary Lagrangian which is invariant under the Weyl transformation. Thirdly,
we should add the Lagrangian of conformal gravity, that is, ${\cal L} \sim \sqrt{-g} \, C_{\mu\nu\rho\sigma}^2$
with conformal tensor $C_{\mu\nu\rho\sigma}$, and investigate if the similar analysis to the present
work could be done or not. Finally, it has been known that the Weyl invariant scalar-tensor gravity reduces to
the Weyl transverse gravity when the longitudinal general coordinate transformation is gauge-fixed 
\cite{Oda-S} - \cite{Oda-C}. The Weyl transverse gravity possesses the Weyl symmetry, to which we could apply 
the present formulation and investigate various quantum aspects. We hope to return these problems in the near future.

\section*{Acknowledgment}

This work is supported in part by the JSPS Kakenhi Grant No. 21K03539.

%%%%%%%%%%%%%%%%%%%%%%% Appendix %%%%%%%%%%%%%%%%%%%%%%%%%%%%%%%
%%%%%%%%%%%%%%%%%%%%%%%%%%%%%%%%%%%%%%%%%%%%%%%%%%%%%%%%%%%%%%%%%%
\appendix
\addcontentsline{toc}{section}{Appendix~\ref{app:scripts}: Training Scripts}
\section*{Appendix}
\label{app:scripts}
\renewcommand{\theequation}{A.\arabic{equation}}
\setcounter{equation}{0}

\section{Derivation of Eq. (\ref{b-rho-eq})}
\def\T{\text{T}}

In this appendix, we present a derivation of Eq. (\ref{b-rho-eq}). First of all, let us notice that
the scalar gauge condition (\ref{Scalar-gauge}) is equivalent to the equation:
%**   Box-phi   %%%%%%%%%%%%%%%%%%%%%%%%%%%%%%%%%%%%%%%%%%%%%%%%%%%%%%%%%
\begin{eqnarray}
\Box \phi^2 = 0.
\label{Box-phi}  
\end{eqnarray}
%%%%%%%%%%%%%%%%%%%%%%%%%%%%%%%%%%%%%%%%%%%%%%%%%%%%%%%%%%%%%%%%%%% 
Then, the Einstein equation in (\ref{q-field-eq}) reads
%**   Ein-eq   %%%%%%%%%%%%%%%%%%%%%%%%%%%%%%%%%%%%%%%%%%%%%%%%%%%%%%%%%
\begin{eqnarray}
G_{\mu\nu} - \phi^{-2} \nabla_\mu \nabla_\nu \phi^2  
-  6 \phi^{-2} ( E_{\mu\nu} - \frac{1}{2} g_{\mu\nu} E ) = 0.
\label{Ein-eq}  
\end{eqnarray}
%%%%%%%%%%%%%%%%%%%%%%%%%%%%%%%%%%%%%%%%%%%%%%%%%%%%%%%%%%%%%%%%%%% 
With the help of Eq. (\ref{Box-phi}), the trace part of this equation becomes
%**   Trace-Ein-eq   %%%%%%%%%%%%%%%%%%%%%%%%%%%%%%%%%%%%%%%%%%%%%%%%%%%%%%%%%
\begin{eqnarray}
R =  6 \phi^{-2} E.
\label{Trace-Ein-eq}  
\end{eqnarray}
%%%%%%%%%%%%%%%%%%%%%%%%%%%%%%%%%%%%%%%%%%%%%%%%%%%%%%%%%%%%%%%%%%% 
Inserting Eq. (\ref{Trace-Ein-eq}) to Eq. (\ref{Ein-eq}) leads to
%**   Ricci-eq   %%%%%%%%%%%%%%%%%%%%%%%%%%%%%%%%%%%%%%%%%%%%%%%%%%%%%%%%%
\begin{eqnarray}
R_{\mu\nu} = \phi^{-2} ( \nabla_\mu \nabla_\nu \phi^2 + 6 E_{\mu\nu} ).
\label{Ricci-eq}  
\end{eqnarray}
%%%%%%%%%%%%%%%%%%%%%%%%%%%%%%%%%%%%%%%%%%%%%%%%%%%%%%%%%%%%%%%%%%% 

Next, operating a covariant derivative $\nabla^\mu$ on Eq. (\ref{Ein-eq}) and using the
Bianchi identity $\nabla^\mu G_{\mu\nu} = 0$, we have
%**   Mod-Ein-eq   %%%%%%%%%%%%%%%%%%%%%%%%%%%%%%%%%%%%%%%%%%%%%%%%%%%%%%%%%
\begin{eqnarray}
&{}& 2 \nabla^\mu \phi \nabla_\mu \nabla_\nu \phi^2 - \phi R_\nu{}^\mu \nabla_\mu \phi^2 
+ 12 \nabla^\mu \phi \, ( E_{\mu\nu} - \frac{1}{2} g_{\mu\nu} E ) 
\nonumber\\
&{}& - 6 \phi \nabla^\mu ( E_{\mu\nu} - \frac{1}{2} g_{\mu\nu} E ) = 0,
\label{Mod-Ein-eq}  
\end{eqnarray}
%%%%%%%%%%%%%%%%%%%%%%%%%%%%%%%%%%%%%%%%%%%%%%%%%%%%%%%%%%%%%%%%%%%
where Eq. (\ref{Box-phi}) was used. Substituting Eq. (\ref{Ricci-eq}) into (\ref{Mod-Ein-eq}) produces
%**   Mod-Ein-eq2   %%%%%%%%%%%%%%%%%%%%%%%%%%%%%%%%%%%%%%%%%%%%%%%%%%%%%%%%%
\begin{eqnarray}
\nabla^\mu ( E_{\mu\nu} - \frac{1}{2} g_{\mu\nu} E ) + \phi^{-1} \nabla_\nu \phi \, E = 0.
\label{Mod-Ein-eq2}  
\end{eqnarray}
%%%%%%%%%%%%%%%%%%%%%%%%%%%%%%%%%%%%%%%%%%%%%%%%%%%%%%%%%%%%%%%%%%%

At this point, we make use of an identity which holds for any symmetric tensor $S^{\mu\nu} = S^{\nu\mu}$ 
\cite{Oda-Q}:
%**   S-ident   %%%%%%%%%%%%%%%%%%%%%%%%%%%%%%%%%%%%%%%%%%%%%%%%%%%%%%%%%
\begin{eqnarray}
\nabla_\nu S^\nu{}_\mu =  h^{-1} \partial_\nu ( h S^\nu{}_\mu ) 
+ \frac{1}{2} S_{\alpha\beta} \partial_\mu g^{\alpha\beta}.
\label{S-ident}  
\end{eqnarray}
%%%%%%%%%%%%%%%%%%%%%%%%%%%%%%%%%%%%%%%%%%%%%%%%%%%%%%%%%%%%%%%%%%%
Identifying $S^{\mu\nu}$ with $E^{\mu\nu}$, and using the relation (\ref{E vs h-E}), we can obtain
%**   S=E   %%%%%%%%%%%%%%%%%%%%%%%%%%%%%%%%%%%%%%%%%%%%%%%%%%%%%%%%%
\begin{eqnarray}
g^{\rho\nu} \partial_\rho \hat E_{\mu\nu} - \frac{1}{2} g^{\alpha\beta} \partial_\mu \hat E_{\alpha\beta} = 0,
\label{S=E}  
\end{eqnarray}
%%%%%%%%%%%%%%%%%%%%%%%%%%%%%%%%%%%%%%%%%%%%%%%%%%%%%%%%%%%%%%%%%%%
where we used the extended de Donder gauge condition (\ref{Ext-de-Donder}). Finally, when we calculate
the LHS of Eq. (\ref{S=E}) by using the definition of $\hat E_{\mu\nu}$ in (\ref{hat-E}), we can
arrive at the desired equation Eq. (\ref{b-rho-eq}).

\renewcommand{\theequation}{B.\arabic{equation}}
\setcounter{equation}{0}

\section{Residual symmetry and conformal symmetry}

In this Appendix, we would like to explain that the residual symmetries found in Eq. (\ref{Sol-Linear-eq}) 
in a curved space-time reduces to a dilatational invariance and special conformal invariance in a flat Minkowski 
space-time.

Before doing so, let us first recall that conformal transformation \cite{Gross, Nakayama} can be defined 
as the general coordinate transformation which can be undone by the Weyl transformation when the space-time 
metric is the flat Minkowski one. With this definition, the conformal transformation is described 
by the equation: 
%**   Conf-Killing   %%%%%%%%%%%%%%%%%%%%%%%%%%%%%%%%%%%%%%%%%%%%%%%%%%%%%%%%%
\begin{eqnarray}
\partial_\mu \epsilon_\nu + \partial_\nu \epsilon_\mu = 2 \Lambda(x) \eta_{\mu\nu},
\label{Conf-Killing}  
\end{eqnarray} 
%%%%%%%%%%%%%%%%%%%%%%%%%%%%%%%%%%%%%%%%%%%%%%%%%%%%%%%%%%%%%%%%%%%
where $\Lambda(x)$ is the infinitesimal transformation parameter of the Weyl transformation, i.e., 
$\Omega (x) \equiv e^{\Lambda(x)} \approx 1 + \Lambda(x)$.   

Taking the trace of Eq. (\ref{Conf-Killing}) enables us to determine $\Lambda(x)$ to be
%**   S=E   %%%%%%%%%%%%%%%%%%%%%%%%%%%%%%%%%%%%%%%%%%%%%%%%%%%%%%%%%
\begin{eqnarray}
\Lambda = \frac{1}{4} \partial^\rho \epsilon_\rho.
\label{lambda}  
\end{eqnarray} 
%%%%%%%%%%%%%%%%%%%%%%%%%%%%%%%%%%%%%%%%%%%%%%%%%%%%%%%%%%%%%%%%%%%
Inserting this $\Lambda$ to Eq. (\ref{Conf-Killing}) yields 
%**   S=E   %%%%%%%%%%%%%%%%%%%%%%%%%%%%%%%%%%%%%%%%%%%%%%%%%%%%%%%%%
\begin{eqnarray}
\partial_\mu \epsilon_\nu + \partial_\nu \epsilon_\mu = \frac{1}{2} \partial^\rho \epsilon_\rho \eta_{\mu\nu},
\label{Conf-Killing2}  
\end{eqnarray} 
%%%%%%%%%%%%%%%%%%%%%%%%%%%%%%%%%%%%%%%%%%%%%%%%%%%%%%%%%%%%%%%%%%%
which is often called the ``conformal Killing equation'' in the Minkowski space-time. It is worth stressing that
Eq. (\ref{Conf-Killing2}) implies the following fact: The flat Minkowski metric $g_{\mu\nu} = \eta_{\mu\nu}$
is invariant in the space of the metric functions under a suitable combination of the general coordinate transformation 
and the Weyl transformation in such a way that
%**   S=E   %%%%%%%%%%%%%%%%%%%%%%%%%%%%%%%%%%%%%%%%%%%%%%%%%%%%%%%%%
\begin{eqnarray}
\delta ( \epsilon_\mu ) = \delta_{GCT} ( \epsilon_\mu ) -  \delta_W ( \Lambda = \frac{1}{4} \partial^\rho \epsilon_\rho ),
\label{Comb-transf}  
\end{eqnarray} 
%%%%%%%%%%%%%%%%%%%%%%%%%%%%%%%%%%%%%%%%%%%%%%%%%%%%%%%%%%%%%%%%%%%
when the vector field $\epsilon_\mu(x)$ obeys the conformal Killing equation (\ref{Conf-Killing2}). To put it differently, 
the characteristic feature of the theory under consideration is that the Lagrangian (\ref{WIST-gravity}) possesses 
the conformal symmetry with 15 global parameters which is a subgroup of the general coordinate transformation 
and the Weyl transformation.

Multiplying it by $\partial^\mu \partial^\nu$, we obtain
%**   S=E   %%%%%%%%%%%%%%%%%%%%%%%%%%%%%%%%%%%%%%%%%%%%%%%%%%%%%%%%%
\begin{eqnarray}
\Box \partial^\mu \epsilon_\mu = 0.
\label{Conf-Killing3}  
\end{eqnarray} 
%%%%%%%%%%%%%%%%%%%%%%%%%%%%%%%%%%%%%%%%%%%%%%%%%%%%%%%%%%%%%%%%%%%
Moreover, multiplying Eq. (\ref{Conf-Killing2}) by $\partial^\mu \partial_\lambda$ and then symmetrizing the 
indices $\lambda$ and $\nu$ leads to the equation:
%**   S=E   %%%%%%%%%%%%%%%%%%%%%%%%%%%%%%%%%%%%%%%%%%%%%%%%%%%%%%%%%
\begin{eqnarray}
\partial_\nu \partial_\lambda \partial^\mu \epsilon_\mu = 0,
\label{Conf-Killing4}  
\end{eqnarray} 
%%%%%%%%%%%%%%%%%%%%%%%%%%%%%%%%%%%%%%%%%%%%%%%%%%%%%%%%%%%%%%%%%%%
where we have used Eqs. (\ref{Conf-Killing2}) and (\ref{Conf-Killing3}). It turns out that a general solution to 
Eq. (\ref{Conf-Killing4}) reads
%**   S=E   %%%%%%%%%%%%%%%%%%%%%%%%%%%%%%%%%%%%%%%%%%%%%%%%%%%%%%%%%
\begin{eqnarray}
\epsilon^\mu = a^\mu + \omega^{\mu\nu} x_\nu + \lambda x^\mu + k^\mu x^2 - 2 x^\mu k_\rho x^\rho,
\label{Conf-Killing-vector}  
\end{eqnarray} 
%%%%%%%%%%%%%%%%%%%%%%%%%%%%%%%%%%%%%%%%%%%%%%%%%%%%%%%%%%%%%%%%%%%
where $a^\mu, \omega^{\mu\nu} = - \omega^{\nu\mu}, \lambda$ and $k^\mu$ are all constant parameters and 
they correspond to the translation, the Lorentz transformation, the dilatation and the special conformal 
transformation, respectively.
 
At this point, it is useful to verify what expression the infinitesimal parameter $\Lambda$ generated by the 
``conformal Killing vector'' $\epsilon^\mu$ in Eq. (\ref{Conf-Killing-vector}) takes. Actually, substituting 
Eq. (\ref{Conf-Killing-vector}) into Eq. (\ref{lambda}), we have
%**   S=E   %%%%%%%%%%%%%%%%%%%%%%%%%%%%%%%%%%%%%%%%%%%%%%%%%%%%%%%%%
\begin{eqnarray}
\Lambda = \lambda - 2 k_\mu x^\mu.
\label{lambda2}  
\end{eqnarray} 
%%%%%%%%%%%%%%%%%%%%%%%%%%%%%%%%%%%%%%%%%%%%%%%%%%%%%%%%%%%%%%%%%%%
This is nothing but zero-mode solutions in Eq. (\ref{Sol-Linear-eq}). This result implies that finding the residual
symmetries (\ref{Sol-Linear-eq}) amounts to solving the conformal Killing equation in a flat Minkowski space-time.

To summarize, we have explicitly shown that in our quantum gravity the Weyl symmetry, together with 
the general coordinate invariance, generates the conformal symmetry in the flat Minkowski background. This result 
is a quantum-mechanical generalization of the well-known Zumino's theorem \cite{Zumino} which insists that 
the theories invariant under both the general coordinate transformation and the Weyl transformation (or local scale 
transformation) possess conformal symmetry in the flat Minkowski background. Even if we used a 
Weyl invariant classical Lagrangian (\ref{WIST-gravity}), we think that the result obtained here holds
for any theories which are invariant under the GCT and the Weyl transformation if we adopt the 
extended de Donder gauge and the scalar gauge for these invariances.

%%%%%%%%%%%%%%%%%%%%%%% reference %%%%%%%%%%%%%%%%%%%%%%%%%%%%%%%
%%%%%%%%%%%%%%%%%%%%%%%%%%%%%%%%%%%%%%%%%%%%%%%%%%%%%%%%%%%%%%%%%%

\end{document}